\documentclass{svjour3}                     
\smartqed  
\usepackage{txfonts}
\usepackage{graphicx}
\usepackage{amssymb}
\usepackage[below]{placeins}

\usepackage{graphicx}
\def \arcmin{$^{\prime}$}

\def \rosat{{\emph{ROSAT}}}
\def \xmm{{\emph{XMM-Newton}}}
\def \chandra{{\emph{Chandra}}}
\def \ixo{\emph{IXO}}
\def \asca{{\emph{ASCA}}}
\def \beppo{{\emph{BeppoSAX}}}
\def \suzaku{{\emph{Suzaku}}}

\def \leftp{{\emph{Left panel: }}}
\def \rightp{{\emph{Right panel: }}}

\def \sncc{SN$_{\mathrm{CC}}$}
\newcommand{\ion}[2]{#1\,{\sc{#2}}}
\usepackage{natbib}
\bibpunct{(}{)}{;}{a}{}{,}

\makeatletter
\renewcommand*{\@biblabel}[1]{\hfill}
\makeatother

\begin{document}

\title{X-ray Spectroscopy of Galaxy Clusters 
}
\subtitle{Studying astrophysical processes in the largest celestial laboratories}


\author{Hans B\"ohringer         \and
        Norbert Werner 
}


\institute{H. B\"ohringer \at
              Max-Planck-Institut f\"ur extraterrestrische Physik \\
              D-85748 Garching, Germany \\
              Tel.: +49 89 30000 3347\\
              Fax: +49 89 30000 3569\\
              \email{hxb@mpe.mpg.de}  \\
           \and
           N. Werner \at
              Kavli Institute for Particle Astrophysics and Cosmology\\
              Stanford University \\
              452 Lomita Mall, Stanford, CA 94305, USA
}

\date{Received: date / Accepted: date}

\maketitle

\maketitle

\begin{abstract}
Galaxy clusters, the largest clearly defined objects in our Universe, are ideal laboratories 
to study in detail the cosmic evolution of the intergalactic intracluster medium (ICM) 
and the cluster galaxy population. For the ICM, which is heated to X-ray radiating
temperatures, X-ray spectroscopy is the most important tool to obtain insight into the
structure and astrophysics of galaxy clusters. The ICM is also the hottest plasma that 
can be well studied under thermal equilibrium conditions. 

In this review we recall the basic principles of the interpretation of X-ray spectra from
a hot, tenuous plasma and we illustrate the wide range of scientific applications of X-ray 
spectroscopy. The determination of galaxy cluster masses, the most important prerequisite
for using clusters in cosmological studies, rest crucially on a precise spectroscopic determination
of the ICM temperature distribution. The study of the thermal structure of the ICM provides
a very interesting fossil record of the energy release during galaxy formation and evolution,
giving important constraints on galaxy formation models. The temperature and pressure distribution
of the ICM gives us important insight into the process of galaxy cluster merging
and the dissipation of the merger energy in form of turbulent motion.
Cooling cores in the centers of
about half of the cluster population are interesting laboratories to investigate the interplay
between gas cooling, star- and black hole formation and energy feedback, which is diagnosed
by means of X-ray spectroscopy. The element abundances deduced from X-ray spectra of the ICM
provide a cosmic history record of the contribution of different supernovae to the 
nucleosynthesis of heavy elements and their spatial distribution partly reflects important
transport processes in the ICM. 

Some discussion of plasma diagnostics for
conditions out of thermal equilibrium and an outlook on the future prospects of X-ray
spectroscopic cluster studies complete our review.

\end{abstract}

\section{Introduction}
\label{intro}
Galaxy clusters are contrary to their name more than just a collection of galaxies.
In the 1930s Fritz Zwicky \citep{Zwicky1937} discovered that it requires a large amount
of unseen matter to bind the fast moving galaxies in the Coma galaxy cluster into a long lasting
object. Today we have a clear cosmic scenario with galaxy clusters as an integral part of the 
large-scale structure of the Universe. They are the largest matter aggregates within the large-scale 
structure which have collapsed under their own gravity and are closely approaching a dynamical
equilibrium. Much theoretical effort has been spent to understand and characterize this
equilibrium structure. To the first order, galaxy clusters are now described as Dark Matter 
Halos with a characteristic universal shape of the Dark Matter potential 
\citep[e.g.][]{navarro1995,moore1999,gao2008}. 
Observations show indeed, that the cluster population can be described as constituting 
a nearly self-similar family with similar shapes of the matter distribution, where the small
scatter is due to the different formation histories and differently close approaches 
to the equilibrium configuration.  

Galaxy clusters are therefore very important giant astrophysical laboratories providing 
us with a well characterized physical environment in which we can study many interesting
astrophysical phenomena and cosmic processes on giant scales (Sarazin 1986). They also 
allow us to study large coeval galaxy populations and enable us to investigate 
their evolution in connection with the chemical and thermal evolution of the embedding 
intracluster medium (ICM) \citep[e.g.][]{dressler1980,dressler1997,poggianti1999,mei2006}. 

As tracers of the cosmic large-scale structure they
are also important probes for cosmology. It is the growth of structure in the matter 
distribution of the Universe that has a strong dependence on the cosmological model
parameters and in particular on the nature of Dark Matter and Dark Energy.
Since galaxy clusters are very sensitive tracers of structure growth, a census
of the cluster population as a function of redshift can be used to test
cosmological models \citep[e.g.][]{borgani2001,schuecker2003a,schuecker2003b,
henry2004b,vikhlinin2003,vikhlinin2009,henry2009}.

\begin{figure}
\begin{center}
  \includegraphics[height=9.0cm]{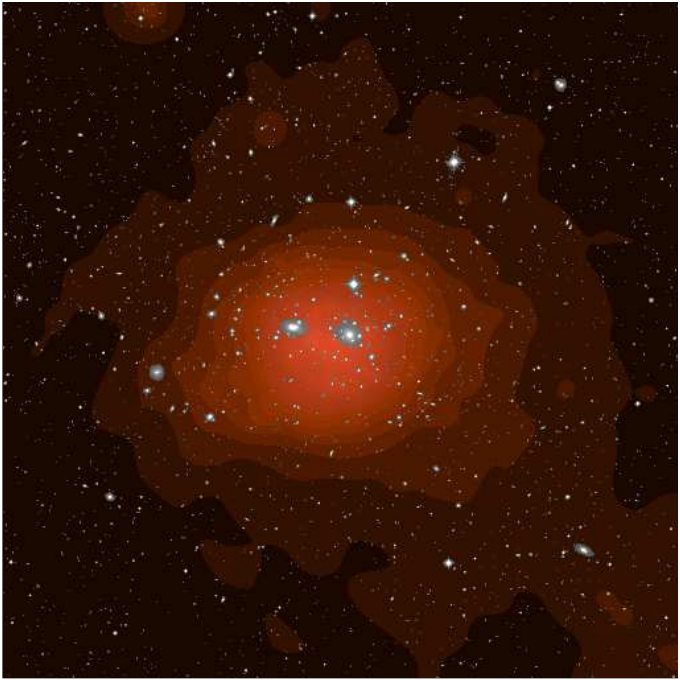}
\end{center}
\caption{ The Coma cluster of galaxies as seen in X-rays in the {\sl ROSAT} All-Sky
  Survey (underlaying red color) and the optically visible galaxy distribution
  in the Palomar Sky Survey Image (galaxy and stellar images from the
  digitized POSS plate superposed in grey).}
\label{fig:1}       
\end{figure}

The current most important limitation in using galaxy
cluster studies for cosmology is the calibration of the relation between various observables
and cluster mass. Therefore a lot of effort is currently being spent to improve 
the cluster mass determination and the understanding of cluster structure 
\citep[e.g.][]{arnaud2007,vikhlinin2006,pratt2006a,pratt2007}. 
X-ray spectroscopy of the cluster emission plays
a crucial technical role in this effort to characterize cluster structure precisely, to 
model the cluster population as a family of self-similar objects with explainable deviations,
and to establish scaling relations of global cluster parameters that allow to draw comprehensive
statistical conclusions on cluster properties from simple observables. 

Most of the detailed knowledge on galaxy clusters has been obtained in recent years through
X-ray astronomy. This is due to the fact that the intracluster medium (ICM) has been heated
to temperatures of tens of Millions of degrees (several keV per particle) which causes the hot
plasma to emit the bulk of the thermal energy in the regime of soft X-rays. Since this
is also the photon energy range where the well developed X-ray telescopes come into play,
galaxy clusters are among the most rewarding study objects for X-ray imaging and spectro-imaging
observations. Fig.~\ref{fig:1} shows a composite image of the Coma galaxy cluster, where an 
optical image from the Palomar Sky Survey showing the dense galaxy distribution of the Coma 
cluster is superposed in grey scale on top of an X-ray image from the ROSAT All-Sky Survey with 
X-ray brightness coded in red color. We clearly recognize that the X-ray image displays 
the cluster as one connected entity. This illustrates the fact that galaxy clusters are
well defined, fundamental building blocks of our Universe.

As largest cosmic objects they display several interesting astrophysical superlatives:
(i) the ICM is the hottest thermal equilibrium plasma that we can study in detail,
with temperatures up to two orders of magnitude larger than the temperature in the center
of the sun, (ii) the gravitational potential of clusters gives rise to the largest
effect of light deflection with deflection angles exceeding half an arcmin,
producing the most spectacular gravitational lensing effects \citep[e.g.][]{hattori1999}, 
(iii) the hot plasma cloud of the ICM casts the darkest shadows onto
the cosmic microwave background through the Sunyaev-Zeldovich effect (at wavelengths 
below about 1.4 mm they are seen as surface brightness enhancements) \citep[e.g.][]{birkinshaw1999}, 
and (iv) the merger of galaxy clusters produces the largest energy release in the Universe after
the big bang itself with energies up to orders of $10^{63}$ erg \citep[e.g.][]{feretti2002}.  

\begin{figure}
\begin{center}
  \includegraphics[height=6.5cm]{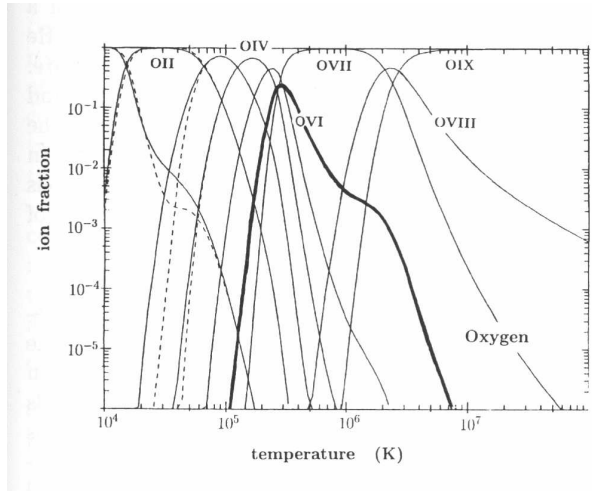}
\end{center}
\caption{Thermal equilibrium ionization structure of the oxygen ion family
 as a function of plasma temperature \citep{bohringer1998}.}
\label{fig:2}       
\end{figure}

X-ray observations and X-ray spectroscopy are the most important tools to obtain
detailed information on cluster properties and the processes occuring in their ICM
as will be illustrated in this article. In section 2 we will explain how observable X-ray
spectra can be understood and modeled. In section 3 we show how this spectral 
modeling is applied to study the thermal structure of the ICM that also provides
the tool for measuring cluster masses. Section 4 deals with the diagnostics of
the central regions of clusters where massive cooling is prevented by AGN heating
in those objects where the cooling time is short enough for effective cooling.
In section 5 we illustrate how the observed spectral lines can be used for the
chemical analysis of the ICM and what can be learned from these observations. In 
section 6 we take a look into the physics of plasma under non-equilibrium ionization
conditions and in the last section, 7, we provide an outlook on the capabilities
and potential of X-ray instruments planned for the future.

\section{X-ray Spectra of Hot Tenuous Plasma}
\label{sec:2}

The plasma of the ICM is very tenuous, with densities of $10^{-5}$ to $10^{-1}$ 
cm$^{-3}$ from the cluster outskirts to the densest regions of cool core clusters.
This low plasma density makes the cluster X-ray spectra modeling simple and enables 
a very straightforward interpretation. Three fundamental emission processes involving
electronic ``transitions'' contribute to the radiation: free-free or bremsstrahlung 
radiation caused by the deflection of an electron at close fly-by of an ion, free-bound
or recombination radiation caused by the capture of an electron by an ion following
ionization, and bound-bound or deexcitation radiation of an electron changing the 
quantum level in an ion. The first two processes give rise to continuum radiation
and the latter to line radiation. An exception in bound-bound transitions is the radiative
transition from the 2s to the 1s state, which is completely
forbidden by angular momentum conservation, but can happen as very slow two-photon process.  
This ``two-photon radiation'' involves a distribution function of the branching ratio
of the energies of the two electrons \citep{spitzer1951} and thus gives
rise to continuum radiation.

\begin{figure}
\begin{center}
  \includegraphics[height=8.0cm]{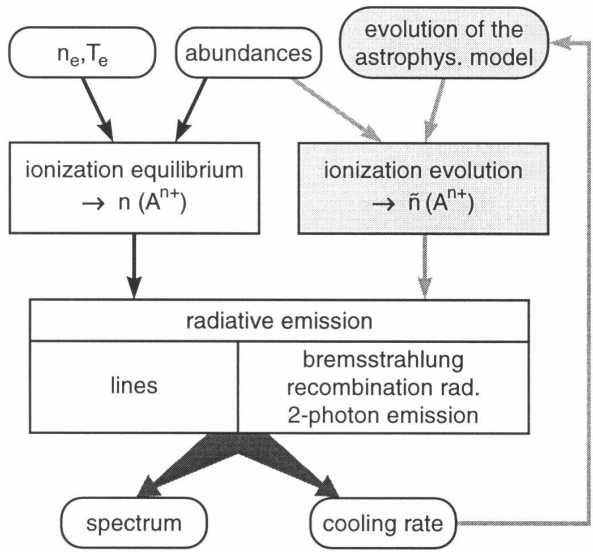}
\end{center}
\caption{Scheme of the calculation of the radiation spectrum and cooling function
of a hot, tenuous plasma in thermal equilibrium. The shaded boxes include the 
 situation of non-equilibrium ionization \citep[from][]{bohringer1998}.}
\label{fig:3}       
\end{figure}

We note that all these radiative processes depend on the collision (or close fly-by) of an 
electron and an ion. Due to the very low density of the plasma all the ions excited by collisions
have sufficient time for radiative deexcitation before a second deexciting collision 
occures. Thus contrary to laboratory plasmas,
where slow transitions are ``forbidden'' and the corresponding excited states are
much more rapidly deexcited by electronic collision, all ``forbidden'' transitions
actually happen in the ICM plasma. This leads to a scenario where all exciting,
recombining, and bremsstrahlung causing collisions lead to the radiation of a photon,
which is referred to as the thin plasma radiation limit (or ``coronal limit'',
as similar conditions prevail in the solar corona). The modeling of the thermal
plasma spectrum is therefore a book keeping exercise of all the electron ion collisions
rates and -- in the case of several deexcitation channels -- their branching ratios. These
collision rates are in general a function of temperature (for thermal plasma)
and the outcome is directly proportional to the electron density and the respective ion
density. The shape of the resulting spectrum is therefore a function of the temperature
and chemical composition
and its normalization is directly proportional to the electron density and the ion density.
The latter is true for the low plasma density limit that pertains in the ICM.

We further note that in general all photons so created leave the ICM plasma due
to its low density. An exception is discussed in section 5.5.3 for the strongest lines
in dense cluster cores. Thus no radiative transfer calculation is necessary for the
interpretation of the X-ray spectra of the ICM. This means in particular that the spectrum
we observe from a galaxy cluster provides an account of the entire ICM plasma, which
is very different from e.g. the spectra of stars, that provide information on merely a very
thin skin on their surface. This is another reason why cluster X-ray spectra are so
informative and straightforward to interpret.

\begin{figure}
\begin{center}
  \includegraphics[height=8.0cm]{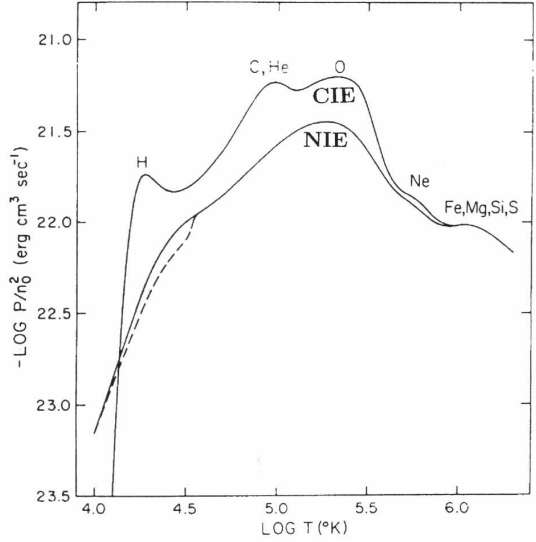}
\end{center}
\caption{Cooling coefficient of hot plasma as a function of the electron temperature
from \citep{shapiro1976}. Two scenarios of the cooling of plasma from an initially
hot phase of $T \sim 10^7$ K are compared. In the CIE (collisional ionization equilibrium) 
case the ionization structure is kept in full thermal equilibrium, in the NIE (non-ionization
equilibrium) case on the contrary, the ionization structure passively follows the 
cooling plasma and recombination can lag behind the cooling. The latter implies higher
ionization stages at the same temperature, that is less shell electrons and reduced line cooling.}
\label{fig:4}       
\end{figure}

As the radiation contribution specific to an ion species is proportional to the density
of this species in the plasma, the modeling
of the overall spectrum also requires the knowledge of the chemical composition
of the plasma and its ionization structure. For thermal equilibrium plasma, the ionization
structure is characterized by the balance of the ionization and recombination rates
for all the ionization steps of an element:
\begin{eqnarray}
[I^{n+}] n_e \left( C_{ci}+C_{ai}\right)~  
+~ [I^{n+}]\left( [H^{+}] X_{ce1} +[He^+,He^{++}]~ X_{ce2}\right) \\ \nonumber
= [I^{(n+1)+}] n_e \left( C_{rr}+C_{dr}\right)~  
+~ [I^{(n+1)+}]\left( [H] X_{ce3} +[He]~ X_{ce4}\right).
\end{eqnarray}
Several processes contribute to the ionization and recombination rate. The 
ionization comprises direct electron impact ionization, $C_{ci}$, electron
impact excitation into an auto-ionization state with subsequent auto-ionization,
$C_{ai}$, and ionizing charge exchange with $H^+$, $He^+$, and $He^{++}$ with
the rates $X_{ce1}$ and $X_{ce2}$, respectively. Direct ionization is the dominant
process. Recombination rates include radiative recombination, $C_{rr}$, 
dielectronic recombination, $C_{dr}$, and electron capture charge exchange
with $H$ and $He$,with the rates $X_{ce3}$ and $X_{ce4}$. 
In the dielectronic recombination process the recombination
collision complex is energetically stabilized by the excitation of a second 
bound electron which subsequently radiates the excess energy. This is a resonant
process which exhibits a strong energy dependence which is most important
for ions with many electrons (e.g. Fe ions). The ionization structure is calculated
from the complete set of linear equations for all important ions and elements
\citep[e.g.][]{arnaud1985,arnaud1992,mazzotta1998}. Fig.~\ref{fig:2} shows for example
the thermal equilibrium ionization structure of oxygen in the temperature range 
$10^4$ to $10^8$ K \citep{bohringer1998}. 

The collision rates that lead to the emission of radiation are then all of the form:
\begin{equation}
R = n_e n_I C_{Ix}  =  n_e^2 \left[ {n_I \over n_E} \right] \left[{n_E \over n_H} \right]
\left[{n_H \over n_e} \right]  ~C_{Ix}(T),  
\end{equation}
where $n_E$, $n_I$, and $n_H$ are the number densities of the elements, ions, and hydrogen nuclei, 
respectively,
and $C_{Ix}$ are the relevant collision rate coefficients for ion $I$.
Consequently, $\left[ {n_I \over n_E} \right]$ represents the fractional abundance
of ionization stage $I$, $\left[{n_E \over n_H} \right]$ the relative elemental abundance,
and $\left[{n_H \over n_e} \right]$ the Hydrogen nuclei to electron ratio, which is about 1.2.
Thus all rates are proportional to the squared plasma density, $n_e^2$. The normalization
of the spectrum is therefore determined from the emission measure, E, given by:
\begin{equation}
E  =  \int  n_e^2~~ dV.
\end{equation}
We note that this definition of emission measure is not unique. It conforms to most definitions
in text books, but the major public plasma radiation codes use the definition $E  =  \int  n_e n_H dV$.
The strategy of calculating the radiation spectrum of hot, thin plasma in
thermal equilibrium is thus summarized in Fig.~\ref{fig:3}. The notion of thermal equilibrium
involves a thermal Maxwellian equilibrium of the electrons and the thermal ionization
equilibrium of the ions. A deviation of the temperature of the electrons and the ions
will not affect the components of the radiation spectrum, but will be detected
through thermal line broadening. 

\begin{figure}
\begin{center}
\includegraphics[height=8.0cm]{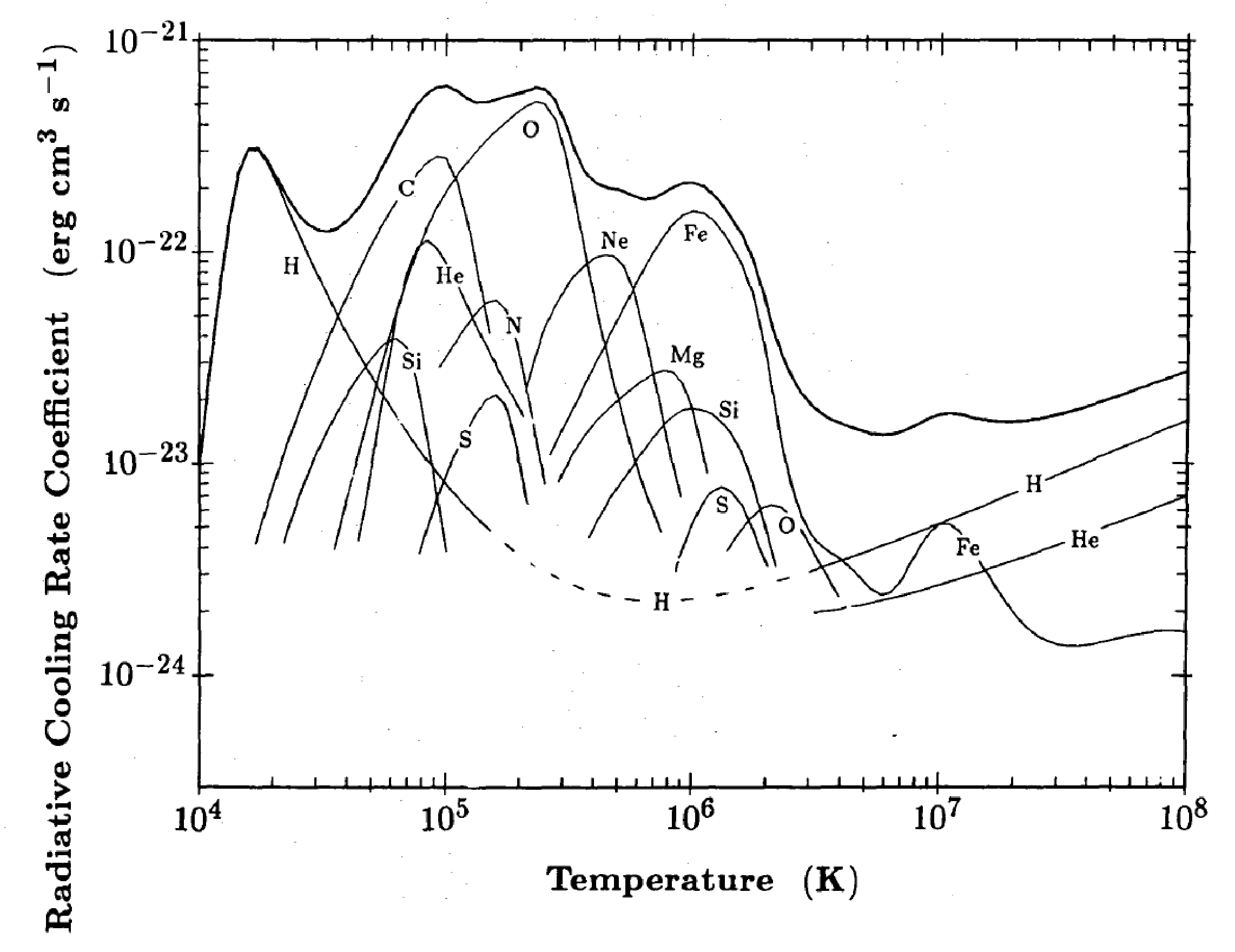}
\end{center}
\caption{Cooling rate of hot plasma as a function of the plasma temperature. 
The contribution to the cooling by the ions of different important abundant elements
is indicated \citep{bohringer1989}. 
Most of this contribution is in the form of line radiation, which is
by far the dominant form of radiation in the temperature range from about $10^4$ to
$2 \times 10^6$ K. At higher temperatures over most of the regime of interest
for the ICM the Bremsstrahlung contribution dominates.}
\label{fig:5}       
\end{figure}

With a change of the temperature
of the plasma the ionization structure is adjusted on the time scale of the inverse
ionization and recombination rates and thus lags behind the temperature change. 
In certain cases this time lag is significant and has to be included in the modeling, as sketched
in Fig.~\ref{fig:3} by explicitly following the time evolution of the ionization structure.
As an example, Fig.~\ref{fig:4} shows the calculation of a passively cooling plasma starting
at high temperature \citep{shapiro1976} where the instantaneous ionization 
equilibrium assumption is compared to the full time dependent ionization structure
calculations. We see that strong deviations occur at temperatures of the order of 
$10^5$ K, but no significant deviation from the ionization equilibrium calculations
are observed above $10^6$~K, which is the temperature range in clusters. So
far no observed X-ray spectrum of the ICM has required a non-ionization equilibrium
treatment. Fig.~\ref{fig:5} shows the contributions of continuum and line emission and the 
respective contributions of the most important elements (assuming solar abundances) to the 
total radiation power. The regime in which the recombination lags behind cooling can
easily be identified with the temperature range where line radiation from the important,
most abundant elements (particularly Fe) is strongly boosting the line
cooling. We note that continuum bremsstrahlung with a temperature dependence
of roughly $\propto T^{1/2}$ is the dominant radiation process at ICM temperatures.

Early modeling of the emission spectrum of hot, thin plasma include:
\citet{cox1969,tucker1971,cox1971,cox1972,shapiro1976,shapiro1977,kato1976,raymond1977,
landini1970,landini1990,landini1991,gaetz1983,raymond1988,masai1984,masai1987,
bohringer1989,schmutzler1993,sutherland1993,brickhouse1995,raymond1996}. Now the two most 
frequently used plasma radiation codes
are the MEKAL code
\citep{gronenschild1978,mewe1981,mewe1985,mewe1990,mewe1991,
kaastra1994,mewe1995} and the APEC code (Smith et al. 2001).  
While the older literature contains all the necessary information about the radiation codes including
all the details of the rate calculations and much detail on the underlying physical
processes, the modern codes are too complex to be easily described in the literature. Thus,
unfortunately very little documentation about the new development of the radiation codes
is available in written form after the mid 1990s.

\begin{figure}
\begin{center}
\hbox{
  \includegraphics[height=5.8cm]{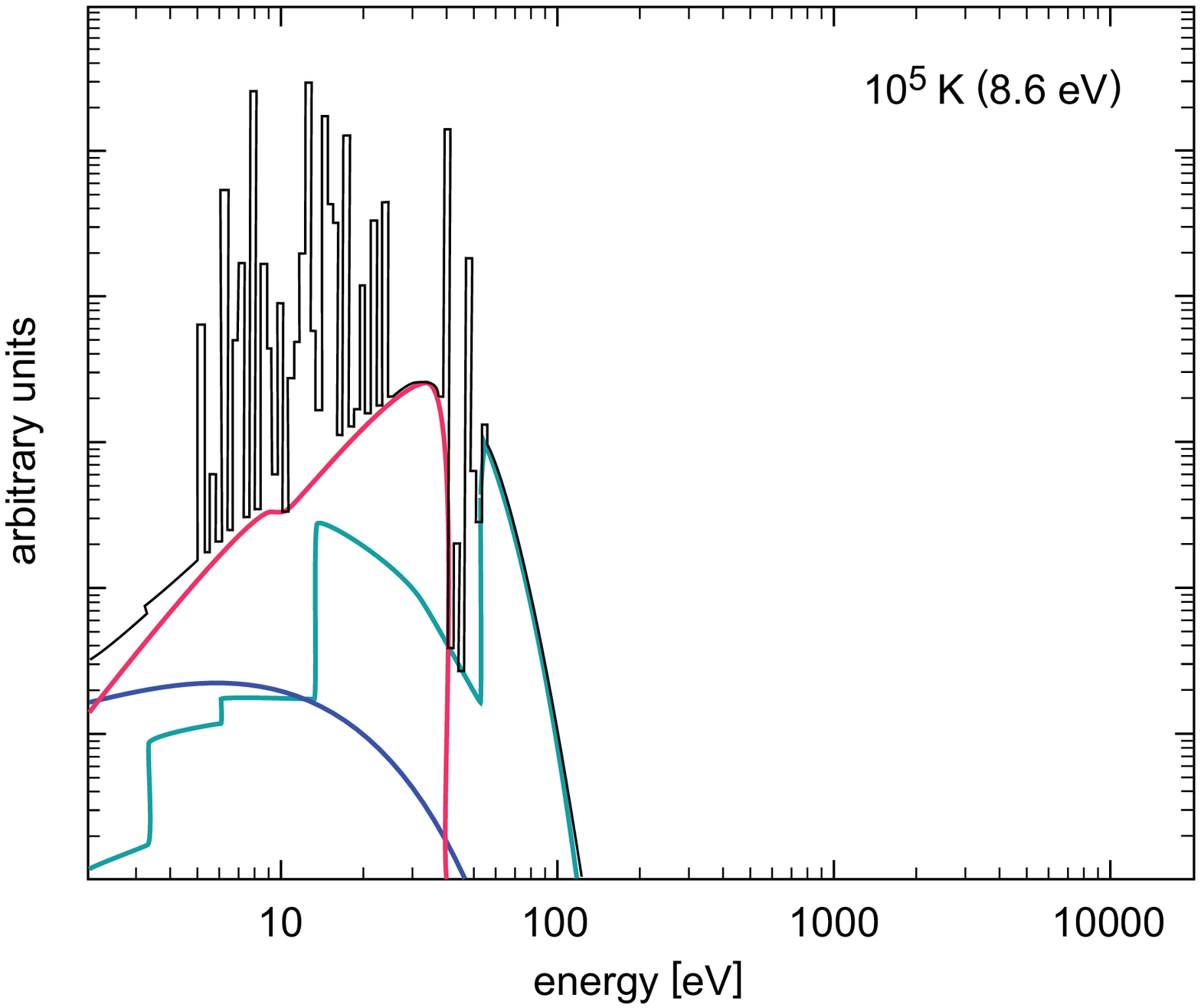}
\hspace{0.1cm}
  \includegraphics[height=5.8cm]{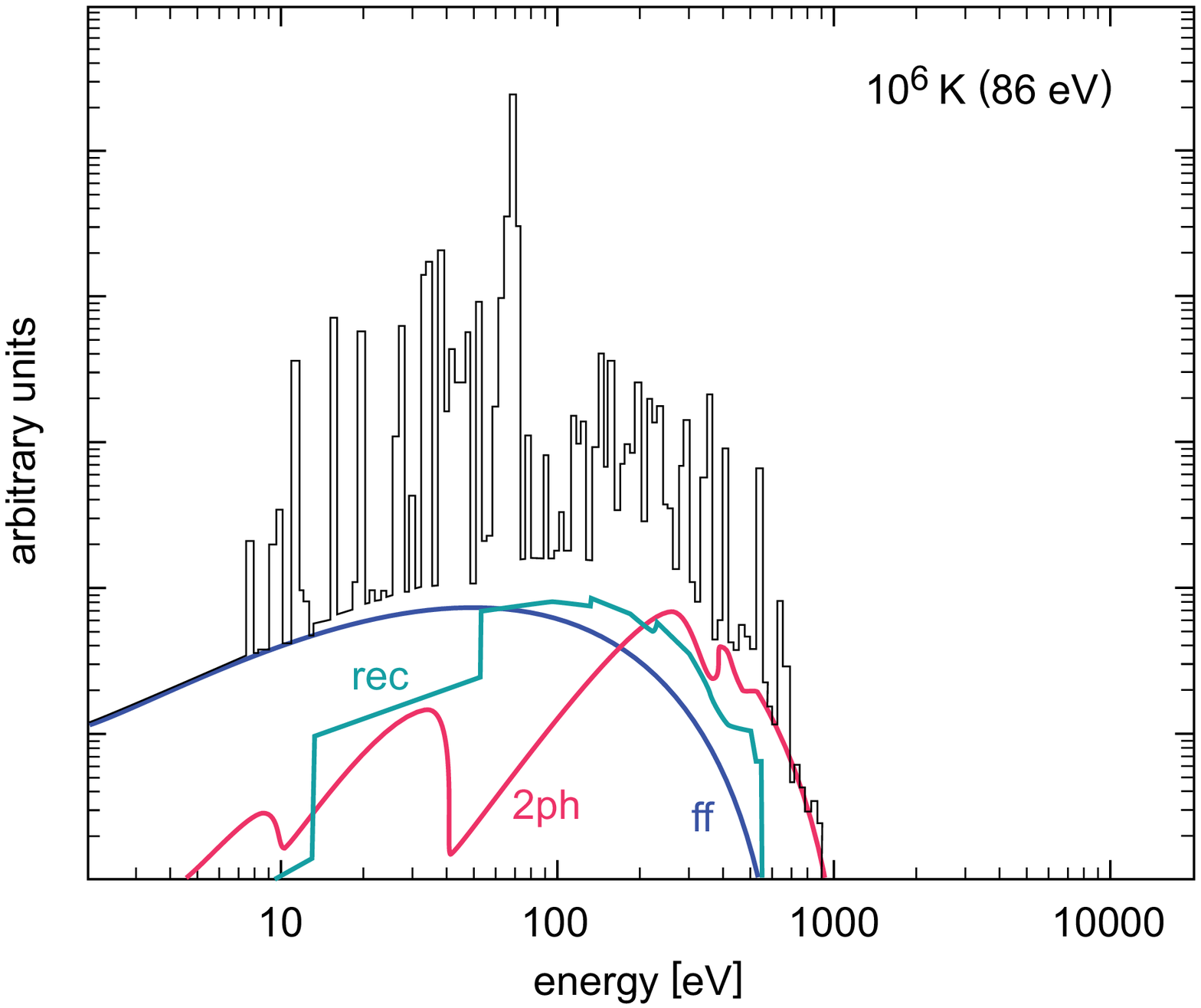}
}
\end{center}
\begin{center}
\hbox{
  \includegraphics[height=5.8cm]{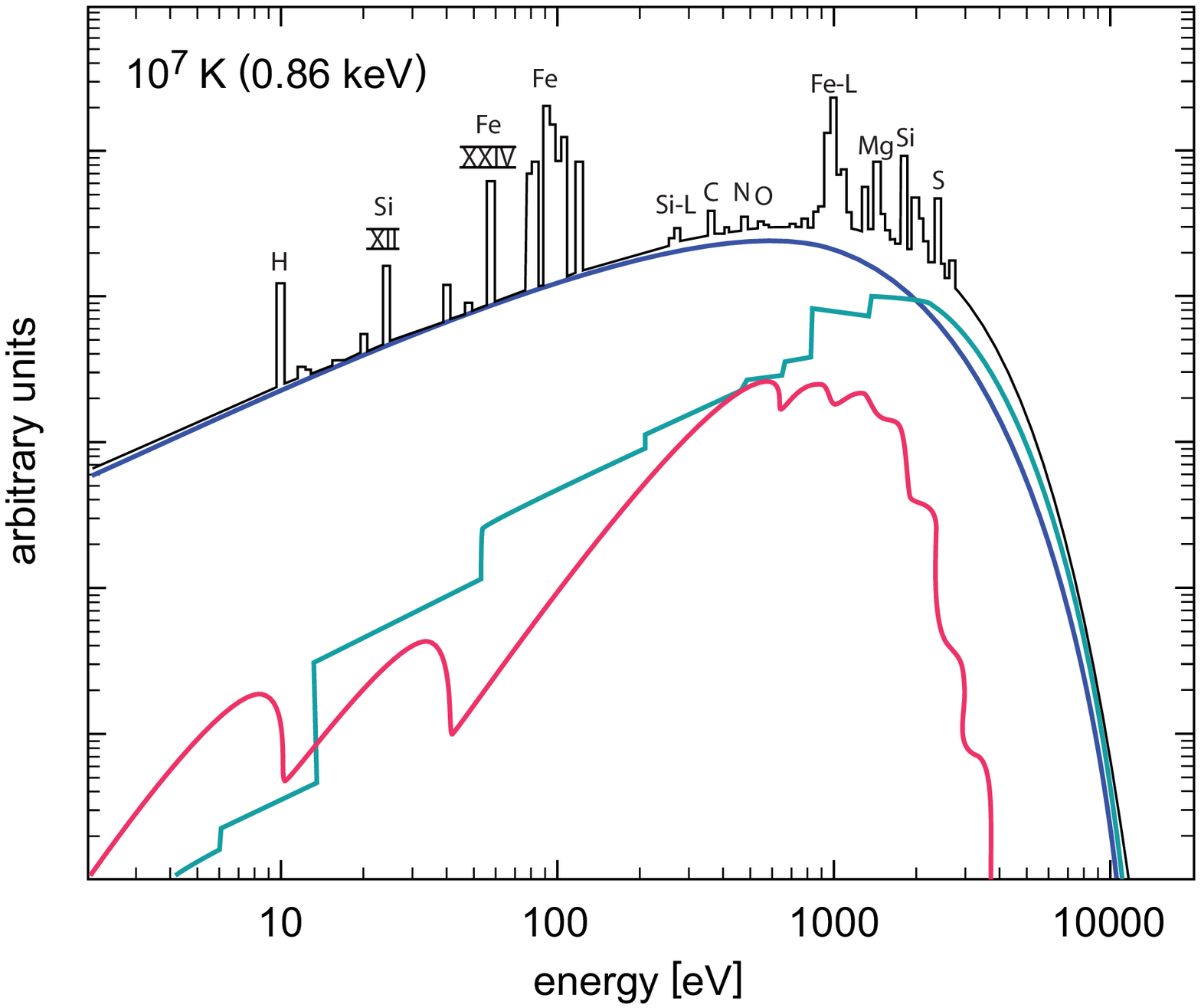}
\hspace{0.1cm}
  \includegraphics[height=5.8cm]{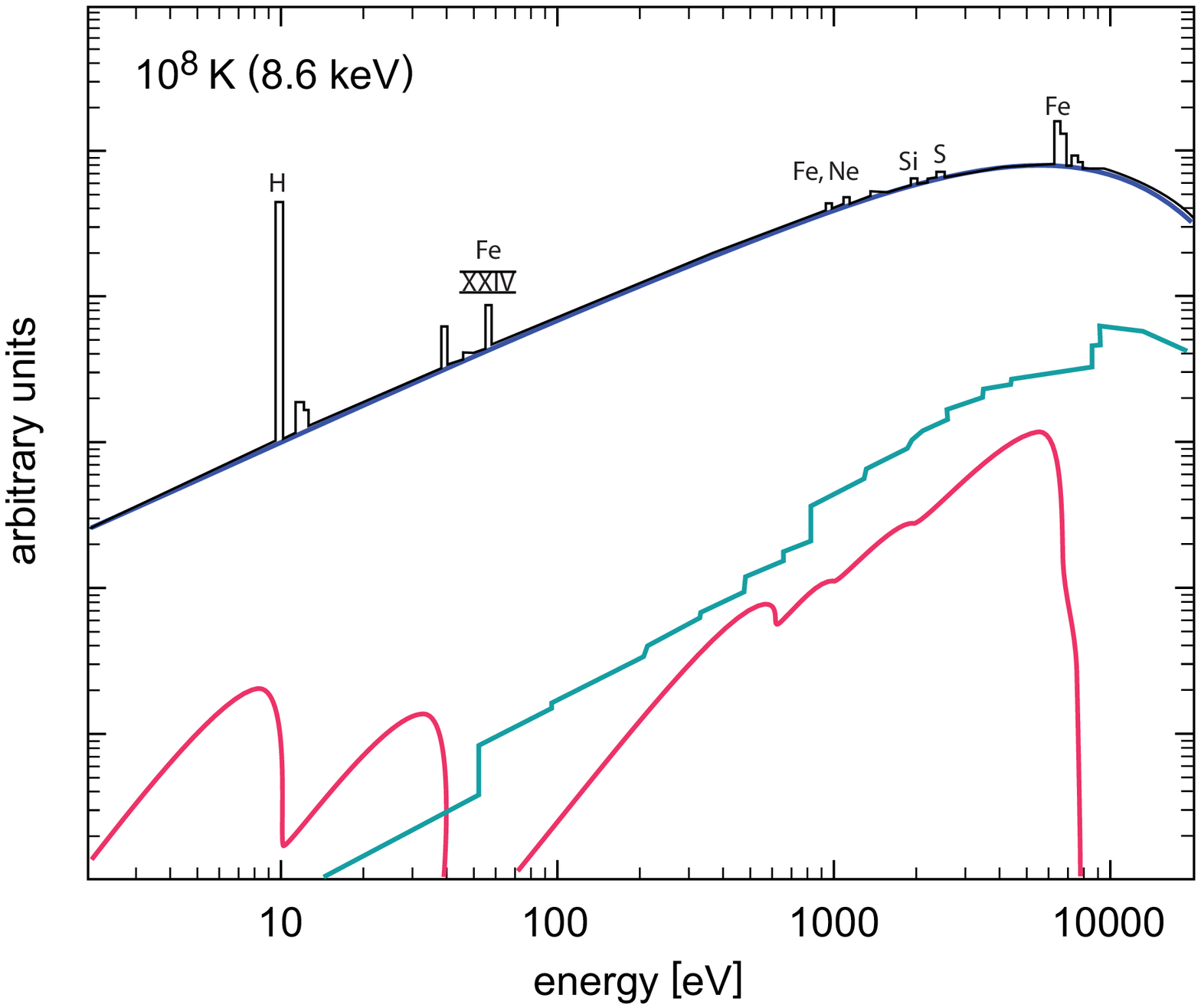}
}
\end{center}
\caption{X-ray spectra for solar abundance at different plasma temperatures.
  The continuum contributions from bremsstrahlung (blue), recombination
  radiation, characterized by the sharp ionization edges (green), and 2-photon
radiation (red) are indicated. At the highest temperatures relevant for
  massive clusters of galaxies bremsstrahlung is the dominant radiation
 process \citep[from the work described in][]{bohringer1989}. 
 The major emission lines in the panels for the higher temperatures 
relevant for galaxy clusters are designated by the elements from which they 
originate (The labels Fe-L ans Si-L
refer to transitions into the L-shell in ions of Fe and Si, respectively, and
two other lines with roman numbers carry the designation of the ions from which
they originate involving transitions within the L-shells.}

\label{fig:6}       
\end{figure}

Fig.~\ref{fig:6} shows typical spectra at temperatures of $10^5$, $10^6$, $10^7$, 
and $10^8$ K with solar element abundances, indicating the contribution
by line radiation and the continuum emission from bremsstrahlung, recombination and
two-photon transitions. We clearly see the increasing dominance of bremsstrahlung with
increasing temperature, which reflects the fact that fewer ions retain electrons
and the plasma is almost completely ionized at the higher temperatures.

\section{The Study of the thermal structure of the intracluster medium}
\label{sec:3}

\begin{figure}
\begin{center}
  \includegraphics[height=7.5cm]{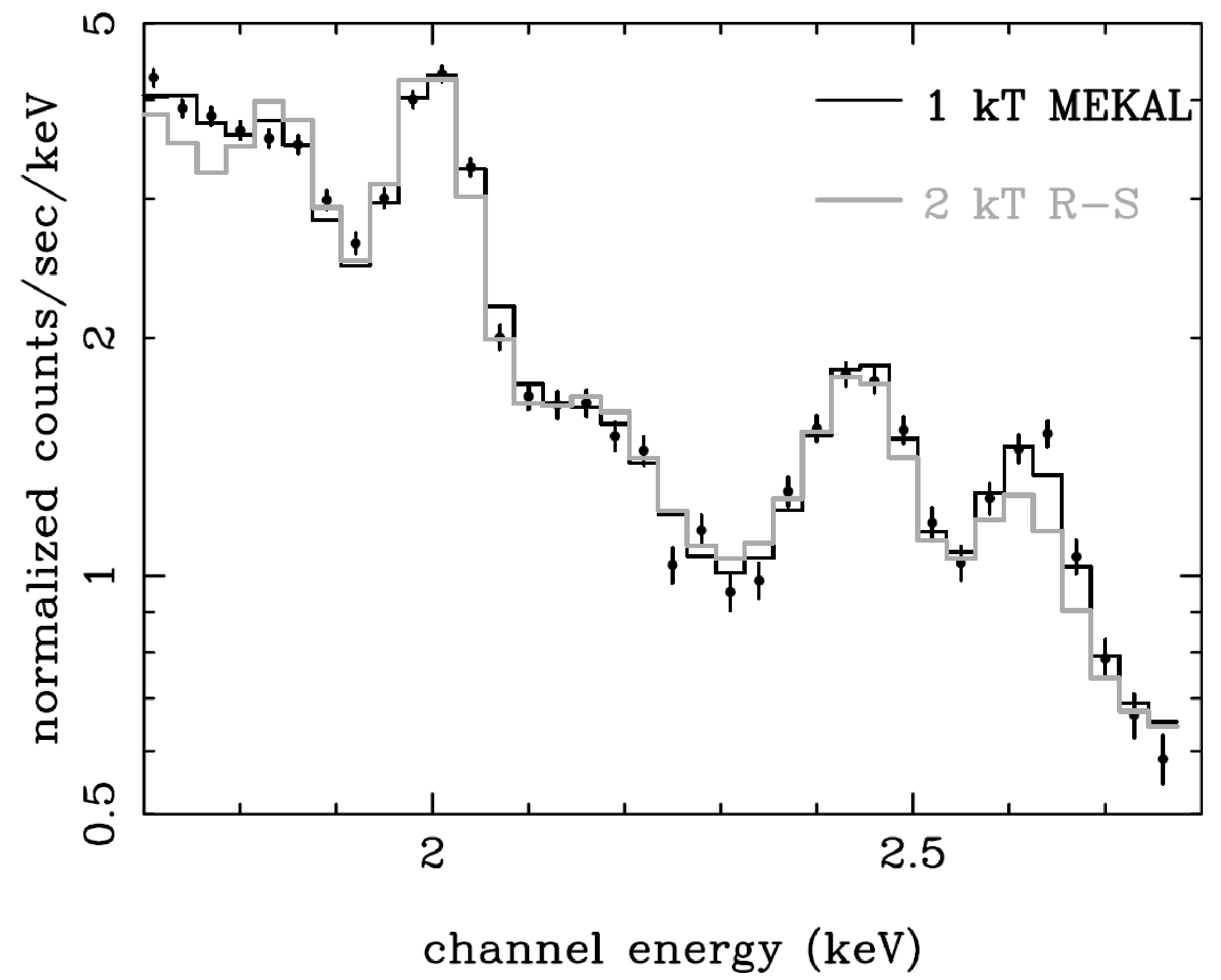}
\end{center}
\caption{X-ray spectrum of the ICM plasma in the halo of M87 at a temperature
around 1.7 keV observed with \xmm EMOS, showing exclusively the spectral range with the H- and He-like lines of
Si and S \citep{matsushita2002}. The spectrum is fitted with a one-temperature MEKAL
model (black line) and a two-temperature model with the temperature components 1.4 keV and
3.3 keV (grey line). The one-temperature model provides a better fit, supporting the view
that the temperature distribution is very narrow. For more details see the original publication.}
\label{fig:7}       
\end{figure}

We have seen in the previous section that the shape of the spectrum for a thermal 
equilibrium plasma is determined by the plasma temperature and the elemental
abundances. This is therefore the basic information we derive from the spectral analysis
of the ICM radiation: a temperature measurement and a chemical analysis. We consequently
illustrate in this, and the next chapter, the scientific insights gained from temperature
measurements from the state-of-the-art spectral analysis, and in section 5 the
lessons learned from the chemical analysis of the ICM.

It is mostly the shape of the continuum spectrum, dominated by 
bremsstrahlung, that provides information on the temperature. The spectral
energy distribution for the thermal bremsstrahlung spectrum for the collision
of an electron with ion, $i$, is given by  \citep[e.g.][]{gronenschild1978}:
\begin{equation}
\epsilon(\nu ) = {16~ e^6 \over 3~ m_e~ c^2} \left({2\pi \over 3 m_e ~k_B T_X}\right)^{1/2}
n_e n_i~ Z^2~ g_{ff}(Z,T_X,\nu)~ \mathrm{exp}\left({-h \nu \over k_B T_X}\right), 
\end{equation}
where $m_e$ and $n_e$ are the electron mass and density, respectively, $n_i$ is the respective ion 
density, $Z$ is the effective charge of the ion, and $g_{ff}$ is the gaunt factor, a quantity
close to unity which must be calculated numerically through a quantum-mechanical treatment.
The most prominent spectral signature of $T_X$ is the sharp cut-off of the spectrum
at high energies, due to the exponential term with the argument~ $- h\nu / k_B T$.
As long as this cut-off is seen in the energy window of the telescope, one has
a good handle on the temperature measurement. The element abundances are mostly
reflected in the intensity of the spectral lines. Only in well resolved spectra
of very high quality, where we can observe several lines of the same element, is the 
temperature also constrained by line strength ratios. A prominent such case
occurs at temperatures around 2--3 keV where we can observe the 
K-shell and L-shell lines of iron simultaneously in high photon count spectra. Another 
example of temperature diagnostics based on the study of the hydrogen and helium like 
Si and S lines is shown in Fig.~\ref{fig:7} \citep[taken from][]{matsushita2002}. This work 
provides a further nice illustrations of how different temperature indicators
can be used to check the consistency of the temperature measurement and to test if
the plasma has a distribution of temperature phases.

\begin{figure}
\begin{center}
  \includegraphics[height=7.4cm]{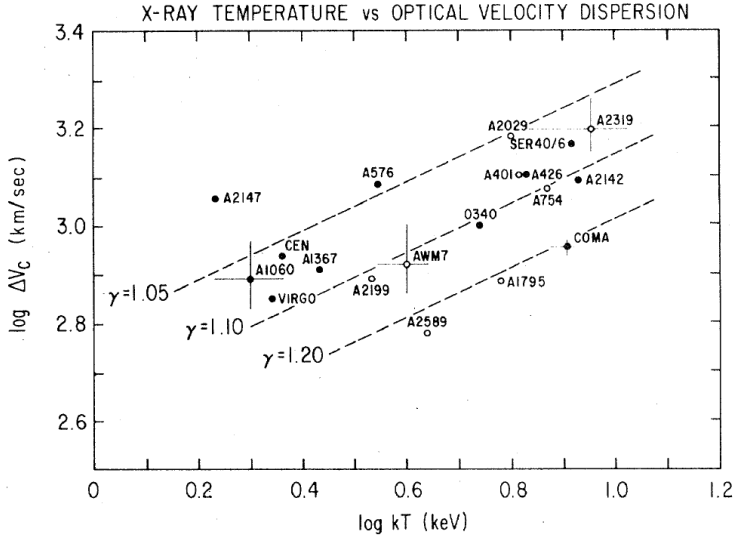}
\end{center}
\caption{Correlation of the ICM temperature with the velocity dispersion of the galaxies
in clusters \citep{mushotzky1984}. The temperatures have been derived from X-ray
spectra taken with the HEAO-1 satellite A-2 experiment. The dots are data for non-cD clusters
and the open symbols for cD clusters (clusters with a central dominant, cD type galaxy). The
lines show predictions of polytropic models with various indices \citep[see][]{mushotzky1984}.}
\label{fig:8}       
\end{figure}

The first X-ray spectral observations of clusters lacked sufficient angular resolution 
and thus involved the total ICM emission of the targets
and only provided us with information on the bulk temperature of the cluster ICM. What
does this parameter tell us about a cluster?
Galaxy clusters form from the gravitational collapse of overdense regions in the 
matter density distribution in the Universe and subsequently approach an equilibrium 
configuration characterized by a virial relation:
\begin{equation}
  E_{kinetic} = - 2 E_{potential}~~  \propto~~ {GM \over R},
\end{equation}
where the mass, $M$, refers to the total mass of the galaxy clusters including
the dark matter. Analogously to the virial equilibrium of galaxies and dark matter 
particles, the ICM plasma thermalizes and attains a ``virial temperature'' 
which reflects the depth of the gravitational potential of the cluster. In the 
collapse process the potential energy of the ICM is converted to internal heat. 
If the gravitational potentials of clusters
of different mass have a self-similar shape, as implied by numerical simulations
of gravitational collapse \citep[e.g.][]{navarro1995,moore1999}, then one 
finds the following self-similar relation between cluster mass and ICM temperature:
\begin{equation}
T \propto \sigma_{DM}^2 \propto {M/R}~~  \propto~~  M^{2/3},
\end{equation}
where $\sigma_{DM}$ is the velocity dispersion of the dark matter particles.
(The rightmost relation is due to $M \propto R^3$ and $\rho_{DM} = const.$~ in
the self-similar model).

The analysis of the first galaxy cluster spectra indirectly confirmed this trend
by showing that the ICM temperature increases with the galaxy velocity dispersion.
Fig.~\ref{fig:8} shows one of the first such relations \citep{mushotzky1984} with X-ray spectral
observations from the HEAO-1 A2 satellite experiment.
Modern versions of these relations - given here directly as $M - T_X$ relations - 
are shown in Fig.~\ref{fig:9}
for selected regular clusters from \citet{arnaud2007} and for
cluster sample from \citet{kotov2006}, respectively. The spectroscopically
determined ICM temperature thus turns out to be one of the best mass proxies as single
observable parameter \citep[e.g.][]{kravtsov2006}. The tightest relations are obtained,
if the core regions are excluded in the global temperature measurement, due to the
disproportionate influence of the central cool cores as will become aparant below, and
it has thus become standard to quote the mean temperature in the radial region
$r = 0.15 - 1 \times r_{500}$ as the most reliable single observable mass proxy for 
clusters \citep[e.g.][]{arnaud2005,pratt2008}\footnote{$r_{500}$ is the radius where the
mean total density of the cluster is 500 times the mass density of a Universe with critical
density. This radius describes the same relative scale of clusters of different mass 
in the self-similar scenario. See also explanation later in the text.}.

\begin{figure}
\begin{center}
\hbox{
  \includegraphics[height=5.8cm]{Boehringer_f9a.ps}
\hspace{0.1cm}
  \includegraphics[height=5.8cm]{Boehringer_f9b.ps}
}
\end{center}
\caption{{\bf Left:} The mass - temperature relation of a sample of regular clusters 
from \citep{arnaud2007}. The best fitting slope (solid line) is $1.71 \pm 0.09$ slightly 
steeper than the self similar relation.
{\bf Right :} The mass - temperature relation of galaxy clusters and its evolution
with redshift derived from two cluster samples at $<z> \sim 0.05$ and $<z> \sim 0.55$
\citep{kotov2006}. The relations have been displaced by a factor of 10 for
better visibility. The model predicted redshift evolution (clusters of equal mass get hotter
with look-back time) has been accounted for by scaling with the factor $h(z)$. This predicted
evolutionary trend is well supported by the data.}
\label{fig:9}       
\end{figure}

The advanced X-ray observatories \chandra\ and \xmm\ now routinely also provide
localized measurements of the ICM temperature. First
attempts had already been made using observations with the \rosat\ observatory
e.g. to produce temperature maps featuring still large temperature uncertainties 
\citep{briel1994, henry1995} and later
by means of the \asca\ observatory, but due to its comparably
low angular resolution most of the results remained somewhat ambiguous and depended
on the analysis method used \citep[e.g.][]{ikebe1997,markevitch1998,white2000}. 
With some studies by means of the Italian-Dutch X-ray mission \beppo\
some overview on the temperature structure of clusters was obtained, with decreasing
temperature profiles outside $\sim 0.2r_{500}$ and a diversity of the behaviour
in the core region depending on the thermal history of the cluster \citep{degrandi2002}. 
This picture has become very precise with systematic studies with \xmm\ and \chandra. 
Fig.~\ref{fig:10} shows results on the temperature profiles of cluster samples studied 
with both observatories \citep{vikhlinin2006,pratt2007}. 

To better understand these results, we have to 
make a brief excursion to scaling relations. For the comparison of the 
structure of galaxy clusters of different mass
based on the model of self-similarity mentioned above, we need a fiducial radius
which identifies corresponding scales in clusters of different mass. A useful
definition of such a scale is the radius at which the mean mass density of the
cluster is larger by a certain factor than the critical density of the Universe at
the redshift of the cluster.
One of the justifications of this picture comes from the spherical collapse model
in a critical density universe. A good practical choice for the overdensity factor
is 500 (see e.g. Evrard et al 1996 who show that inside a region of $\sim r_{500}$
randomized galaxy and mass particle orbits dominate clearly over infalling material), 
not least from an X-ray observers point
of view, since now a larger number of high quality \chandra\ and \xmm\ data
cover the cluster ICM spectroscopically out to this radius \citep[e.g.][]{vikhlinin2006,pratt2007}. 
We will most often make use of this scaling radius of  
$r_{500}$, but also use other values of overdensity depending on the literature
from which we draw the examples.

In Fig.~\ref{fig:10} the temperature profiles shown are scaled by the overdensity radii 
(with different overdensity values, which has no importance here). We note that there 
is a large diversity of the profiles at small radii. There are
two classes of galaxy clusters: the clusters with dense ICM cores show temperature
profiles with $T_X$ decreasing towards the center in the core regions, while 
in clusters with moderate central densities (typically below $10^{-2}$ cm$^{-3}$)
the temperature profiles are found to be flat or even slightly increasing
towards the cluster center. The clusters
will be called cool core and non-cool core clusters in the following. Due to
the logarithmic display of the radial range, the inner regions appear more 
prominent in Fig.~\ref{fig:10}.
Most of the cluster size and volume is in the outer region, where the temperature
profiles become very similar and appear squeezed into a narrow zone in scaled radius. 
This shows that on large scale the thermal structure of clusters can be viewed as following closely
a self-similar model. Characterizing this model and understanding the scatter around
the mean model is one of the current important observational goals of X-ray cluster 
research.

\begin{figure}
\begin{center}
\hbox{
  \includegraphics[height=6.0cm]{Boehringer_f10a.ps}
\hspace{0.5cm}
  \includegraphics[height=6.0cm]{Boehringer_f10b.ps}
}
\end{center}
\caption{Scaled temperature profiles of galaxy clusters derived from spectroscopic
observations with the \chandra\ satellite \citep[left,][]{vikhlinin2006} and results obtained
for a sample of clusters observed with \xmm\ \citep[right,][]{pratt2007}.}
\label{fig:10}       
\end{figure}

One of the difficulties in deriving the temperature distribution of the ICM
in clusters from X-ray spectroscopic observations is the fact that the observed
radiation is the result of an integral of radiative emission along the entire
line-of-sight through the cluster. Therefore one observational task is to ``deproject''
the cluster spectra, which can only be done by assuming a certain three-dimensional geometrical
shape of the cluster, in general spherical symmetry. The requirements for successful
spectral deprojection are very good photon statistics and an angular resolution of the data
(instrument point spread function) that is better than the corresponding radial binning used. In 
general the deprojected spectra are much more noisy than the observed spectra, but most
often the deprojection results show only small specific differences to the unprojected
data (because most of the emission shaping the spectrum comes from the center-most region).
 
A further problem in deriving the temperature structure from spectroscopy arises if the
plasma has more than one temperature phase locally, that is within the observed patch from
which the spectrum is extracted and within the deprojection bin. The question is then:
what determines the resulting temperature when this composite spectrum is fit with a single-phase
spectral model? A first intuitive guess is to assume that it is an average of the temperatures
weighted by the radiative emission contributions of the different phases, the so-called
emission weighted temperature, $T_{ew}$ \citep[e.g.][]{mathiesen2001}. However, as shown
by \citet{mazzotta2004} the resulting fitted temperature is generally biased low compared
to $T_{ew}$. \citet{mazzotta2004} derive a simple analytical approximation to calculate the
fitting result from a temperature mixture with an accuracy of a few percent for ICM temperatures
above 3 keV. The resulting fitted temperature they call spectroscopic-like temperature, 
$T_{sl}$. This approach is based on the fact that the temperature dependence of the
spectra according to Equ. 4 can be expressed as:

\begin{equation}
  \epsilon(E,T) \sim n_e^2~ \zeta(T,m)~ T^{-1/2}~ exp\left(-{E\over kT}\right) \sim  
n_e^2~ \zeta(T,m)~ T^{-1/2}~ \left(1-{E\over kT}\right)
\end{equation}

where $E$ is the photon energy, $m$ is the metallicity (heavy element abudance) of the plasma,
and $\zeta(T,m)$ includes the temperature dependence of the gaunt factor and some effect
of emission lines. 
The right hand side is a Taylor series expansion for low enough energies as 
covered by the energy window of X-ray telescopes (for $k_B T > 3$keV). Then it is easy to show that
temperature phase addition with subsequent single temperature fitting leads to the following
expression for the spectroscopic like temperature:

\begin{equation}
T_{sl} = { \int n^2 T^{\alpha}/T^{1/2} dV \over \int n^2 T^{\alpha}/T^{3/2} dV }   
\end{equation}
 
with tests showing that $\alpha = 0.75$ is a good approximation.
\citet{vikhlinin2006b} has generalized this approach also to low temperatures 
using a method that requires some more complex numerical simulations that take into account
the specifics of the spectroscopic instrument, energy band and galactic absorption
providing satisfying results with errors of a few percent. 

The latter approach allows us to derive $T_{sl}$ for a known temperature distribution,
but provides no means to correct for a temperature bias, if the distribution is unknown.
For example the determination of cluster masses (as explained below) rests on a local
average of the pressure which requires a mass weighted temperature average for known density.
Since T$_{sl}$ is always biased low with respect to mass averaging, the mass will be underestimated
in the presence of unresolved multi-temperature structure \citep[e.g.][]{rasia2005}.
A study by \citet{rasia2006} showed that the temperature bias can easily lead
to a mass underestimate by 10\% on average. With current instrumentation it is difficult to unveil
multi-temperature structure in the temperature range above 3 keV. A simple exercise that
can be performed with {\sl XSPEC} is to take an equal emission measure of 4 keV and 8 keV
plasma and to simulate an analysis with the instruments of \xmm. Even with a simulated
exposure that gives several ten million photons, a single temperature model with $T \sim 5.5$ keV
gives a perfect fit and we are unable to recover the original input phases. At lower
temperatures and very good photon statistics more diagnostics can be performed as
seen e.g. in the study shown in Fig.~\ref{fig:7} \citep{matsushita2002} and similar studies in
other cool core regions e.g. by \citet{fabian2005} or \citet{simionescu2008c} shown
below, due to the emisson lines characteristic to the different temperature phases. 

\begin{figure}
\begin{center}
\hbox{
  \includegraphics[height=5.5cm]{Boehringer_f11a.ps}
\hspace{0.01cm}
  \includegraphics[height=5.5cm]{Boehringer_f11b.ps}
}
\includegraphics[height=6.5cm]{Boehringer_f11c.ps}
\end{center}
\caption{{\bf Upper Left:} Mass profiles for a sample of regular clusters with a range of temperatures
derived from observations with \xmm\ \citep{pointecouteau2005}. {\bf Upper Right:} 
Same mass profiles as in the left panel scaled in radius by $r_{200}$ 
and in mass by $M_{200}$, the mass at the radius of $r_{200}$. The solid line shows 
the fit of a NFW mass distribution model which provides a good description of
the data. {\bf Lower Panel:} Scaled gas density profiles and scaled total density profiles
for a sample of galaxy clusters observed with \chandra\ \citep{vikhlinin2006}. The yellow line
shows the best fitting NFW-model for the total density profile.}
\label{fig:11}       
\end{figure}

\subsection{Cluster Mass Determination}

One of the most important applications of the knowledge of the temperature structure
in the ICM is the mass determination of galaxy clusters. Assuming the ICM is in hydrostatic
equilibrium and the cluster has approximately spherical symmetry, the distribution
of the total mass in the cluster is given by the density and temperature profiles
by the following equation:
\begin{equation}
M(r) = - {G~ k_B T_X \over \mu m_p} r \left({log \rho_g \over log r} + {log T_X \over log r}\right), 
\end{equation}
where $G$ and $k_B$ are the gravitational and Boltzmann's constant, respectively, $\mu$
is the mean particle mass ($\sim 0.6$), $m_p$ the mass of the proton, and $\rho_g$ the
gas density. 

One of the generally interesting results from recent efforts of precise mass determination 
is the conclusion that the "NFW" dark matter halo model proposed by Navarro, Frenk \& White (1995,1997)
gives a consistent description
of the mass distribution of seemingly relaxed clusters (clusters that show a high degree
of symmetry in projection and no obvious signs of recent merger activity) as shown 
in Fig.~\ref{fig:11} (\citet{pratt2006}, \citet{vikhlinin2005}, see also \citet{pratt2002},
\citet{pointecouteau2005},\citet{Buote2007},
\citet{voigt2006}). First attempts have been made of a rigorous comparison of the cluster 
masses determined from an X-ray analysis and from gravitational lensing studies \citep{zhang2007, 
mahdavi2008} showing that the X-ray mass determination provides a much smaller
individual uncertainty. We also have first indications that the lensing masses are generally
higher by about $12 (\pm 15) \%$ which may partly reflect the above mentioned temperature bias
and additional unaccounted turbulent pressure of the ICM if we assume that the lensing masses
are essentially unbiased. This mass calibration by combining several methods will constitute an
important effort in the coming years.

\subsection{Fossil Record of Structure and Galaxy Formation in the ICM Thermal Structure}

The temperature distribution in the ICM is also the key to the 
characterization of the thermal history and thermal structure, most conveniently expressed 
by the entropy structure of the cluster's ICM. The definition of ``entropy'', $S$, as used in the
astrophysical literature of galaxy clusters deviates from the general definition
in physics:
\begin{equation}
S = {k_B T_X \over n_e^{2/3} }.
\end{equation}
This definition is related to the general definition of entropy, $s$, by $ s = k_B~ ln\left(S^{3/2} 
(\mu m_p)^{5/2}\right) + s_0$ \citep{voit2005a}. The $S$ value can be seen as a parameter
to lable adiabates, and thus $S$ stays fixed in any hydrodynamic evolution of the ICM in which
all processes are adiabatic; $S$ increases e.g. in shock waves or with dissipation of turbulent
motion and descreases with radiative 
cooling. 

\begin{figure}
\begin{center}
  \includegraphics[height=7.0cm]{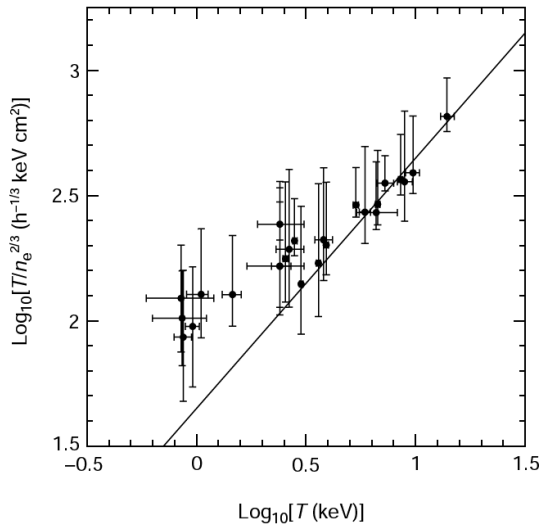}
\end{center}
\caption{Entropy in groups and clusters measured at a fiducial radius of 
$0.1~r_{virial}$ as a function of the system's ICM temperature \citep{ponman1999}. It was 
inferred from this plot that the entropy appears to converge at low $T_X$ to a characteristic floor 
value of $\sim 100~ h^{-1/3}~$ keV cm$^2$.}
\label{fig:12}       
\end{figure}

The most important application of studying the entropy structure in galaxy clusters is the 
distinction between heating by gravitational and non-gravitational processes. 
The term gravitational processes refers here to heat that comes from
conversion of potential to heat energy in the formation of structure and the collapse of clusters.
This form of heat is determined by structure formation processes that are mainly driven 
by the dark matter in the Universe. Non-gravitational processes are connected to energy input into
the ICM in the form of star formation driven galactic winds (which at the same time lead to
chemical enrichment of the ICM as discussed in section 5) and by AGN providing energy input via
jets of relativistic plasma.

The way to distinguish the two forms of heating is the study of scaling relations of thermal
cluster properties. The gravitational heating of the ICM during cluster formation is proportional
to the depth of the gravitational potential well of the cluster 
and thus the entropy gain is proportional
to the ICM temperature. Heating processes connected to feedback from the galaxy population
are expected to be related (proportional) to the total stellar mass in the cluster. Assuming
to first order a constant stellar to gas mass ratio of the baryonic cluster component, the
amount of non-gravitational energy input per unit gas mass is than also to first order
constant. Therefore in this most simple consideration the gravitational entropy contribution is 
proportional to the virial and ICM temperature, while the non-gravitational entropy contribution
is an additive constant.
The evidence for non-gravitational entropy characterized in this way was provided by
\citet{ponman1999} as illustrated in Fig.~\ref{fig:12} (see also \citet{david1996}
who showed with \rosat\ observations of cooler systems that groups have flatter 
entropy profiles than clusters). The results imply a so-called 
entropy floor of about $ 135$ keV cm$^2$ \citep{ponman1999,lloyd2000}. 
More modern results do not show the sudden appearance of a step, 
but rather a slope of the entropy - temperature 
relation which is shallower than the slope of 1 expected from purely gravitational models:
instead of the gravitational slope of $S(r_{scaled}) \propto T$, relations around $\propto T^{2/3}$
are found. Purely gravitational models of smooth and cold accretion of gas into the cluster
\citep[e.g.][]{tozzi2001} predict entropy profiles in clusters of the form $S(r) \propto r^{1.1}$.
Observations show profiles which can be as steep as this prediction for the most massive
systems, but they are shallower for smaller clusters, which is again a manifestation of the
fact that non-gravitational processes have a larger effect on systems with a smaller mass
and thus a shallower gravitational potential.
As a representative result we show the study of \citet{pratt2006}
involving 10 relaxed appearing clusters (Fig.~\ref{fig:13}), which illustrates the radial and ICM 
temperature scaling behaviour of the cluster ICM entropy. The diagnostics of entropy profiles
has now been pursued in detail for larger cluster samples by \citet{cavagnolo2009} and representative
cluster samples by Prat et al. (2009, in preparation) providing a more detailed picture, the
proper interpretation of which has to be assisted by simulations.

\begin{figure}
\begin{center}
\hbox{
  \includegraphics[height=6.0cm]{Boehringer_f13a.ps}
\hspace{0.2cm}
  \includegraphics[height=6.0cm]{Boehringer_f13b.ps}
}
\end{center}
\caption{{\bf Left:} The entropy - temperature relation for 10 regular galaxy clusters
determined at four different fiducial radii \citep{pratt2006}. The slopes of the relations
as determined from the orthogonal BCES method \citep{akritas1996} are for increasing
radii: $0.49 \pm 0.15$, $0.62 \pm 0.11$, $0.64 \pm 0.11$, and $0.62 \pm 0.08$. 
{\bf Right:} Scaled radial entropy profiles for the same 10 clusters. 
The radius is scaled by $r_{200}$. The entropy, $S$, has been 
scaled with an entropy-temperature relation proportional to $T^{0.65}$ using the global
cluster temperature. The grey
shaded area corresponds to the 1$\sigma$ standard deviation. The dashed line shows radial dependence  
$S \propto  r^{1.08}$\citep{pratt2006}.}
\label{fig:13}       
\end{figure}

Theoretical modeling to explain the quantitative entropy scaling behaviour involves ``preheating''
(early entropy increase of the intergalactic medium due to star formation before the formation
of the cluster), cooling and condensing of the low entropy material which increases $S$ in the remaining
gas phase, and feedback heating processes in the cluster. A complicated balance of all these processes
seems to be necessary to reproduce the entropy profiles, the scaling relations, and the amount of
baryons that are converted into stars in the cluster volume and efforts to obtain a completely
consistent and satisfying model are still ongoing 
\citep[e.g.][]{borgani2004,borgani2005,voit2003,voit2005a,mccarthy2008}.

\subsection{At and beyond the virial radius}
\begin{figure}
\begin{center}
  \includegraphics[height=11.0cm]{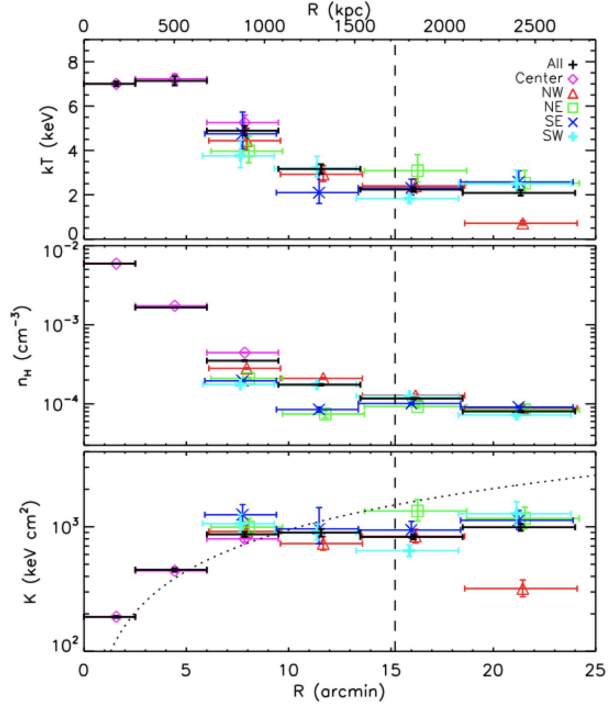}
\end{center}
\caption{Temperature, density and entropy profiles of PKS~0745-191 determined by Suzaku
out to and beyond the virial radius in each direction. The dotted curve at the bottom
shows the expected entropy profile for heating by gravitational collapse
$S \propto r^{1.1}$.  The vertical dashed line shows the estimate of $r_{200}$
\citep{george2009}. } 
\label{fig:george}       
\end{figure}

The low, stable, and well understood instrumental and particle background properties
of the \suzaku\ satellite allowed to perform the first pioneering
studies of clusters out to and beyond their virial radius. \citet{reiprich2009}
determined the temperature profile for the rich, massive, cooling core cluster Abell~2204
from $\sim$10--1800~kpc, close to the estimated $r_{200}$. They find that the temperature
profile between 0.3--1.0$r_{200}$ is consistent with a drop of 0.6, as predicted by
simulations. The first temperature, density, and entropy profiles beyond $r_{200}$ were
reported for another rich, massive cooling core cluster PKS~0745-191 \citep{george2009}.
Between 0.3--1$r_{200}$ they measure an average temperature drop of $\sim 70\%$.  They find
that near to the virial radius the observed entropy profile is lower than that expected
for the heating by gravitational collapse. Fig.~\ref{fig:george} shows the observed profiles
in four directions and the average profiles. The dotted curve at the bottom shows the
expected entropy profile for heating by gravitational collapse, $S \propto r^{1.1}$,
and the vertical dashed line shows the estimated $r_{200}$. The fall of entropy
beyond $r_{200}$ in the NW direction is interpreted by the authors as evidence for
an accretion shock from cooler material falling on the cluster along a filament.

X-ray emission from hot plasma associated with a filament connecting two massive
clusters of galaxies, Abell~222 and Abell~223, was recently detected using \xmm\
\citep{werner2008b}.  The detection of the tenuous gas permeating the filament was
possible because of its favorable orientation approximately along our line-of-sight.
The temperature of the detected gas is $kT=0.91\pm0.25$~keV  and assuming that the
length of the filament along the line-of-sight is $l=15$~Mpc its baryon density is
$\approx 3.4\times10^{-5}$~cm$^{-3}$, which corresponds to a baryon over-density of
$\approx 150$. The entropy of the gas in the filament is $S\approx 870$~keV~cm$^2$~$l^{1/3}$,
which suggests strong preheating. We note that if the detected emission would be associated
with the outer region of radially asymmetric clusters (deriving the gas density by means of
the usual geometric deprojection), the corresponding entropy of
$\approx 420$~keV~cm$^2$ would be much lower than that expected for the virialised
cluster gas ($\approx 1000$~keV~cm$^2$) which has already passed through an accretion shock.

\subsection{Diagnostics of Cluster Mergers}

\begin{figure}
\begin{center}
  \includegraphics[height=8.0cm]{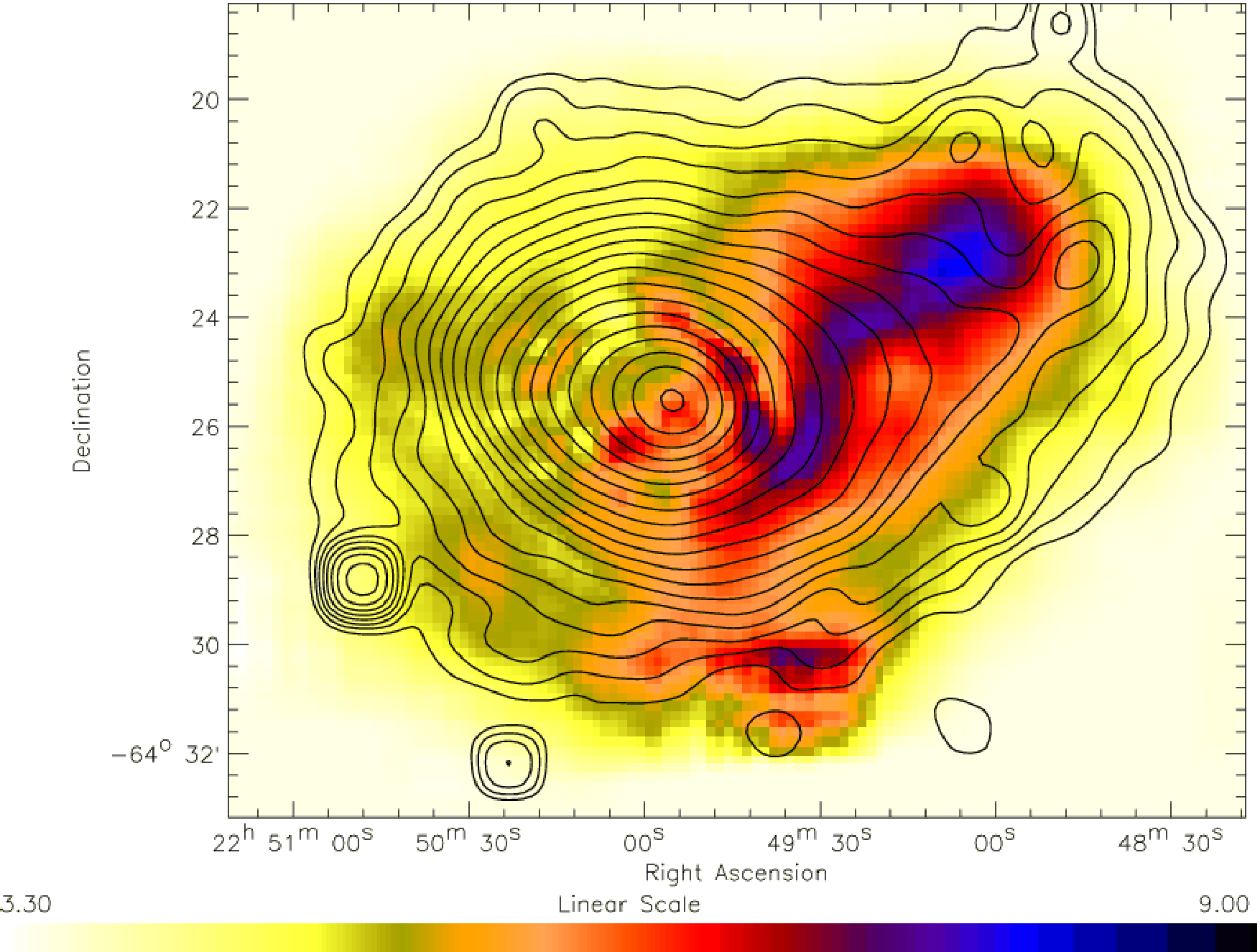}
\end{center}
\caption{Temperature map of the merging cluster A3921 by \citep{belsole2005}. The temperature
map has been produced from \xmm\ observations obtained with the detectors EMOS1 and
EMOS2 with the multi-scale spectro-imaging technique of \citet{bourdin2004}. The barlike hot region
to the NE is interpretated as the signature of an off-axis merger at a stage short after the first
close encounter similar to features observed in simulations \citep{belsole2005,ricker2001}.
Superposed to the temperature map are the contour lines of the surface brightness distribution
in the 0.3 - 10 keV band which was adaptively smoothed. The contours are spaced logarithmicly.}
\label{fig:14a}       
\end{figure}

\begin{figure}
\begin{center}
  \includegraphics[height=11.0cm]{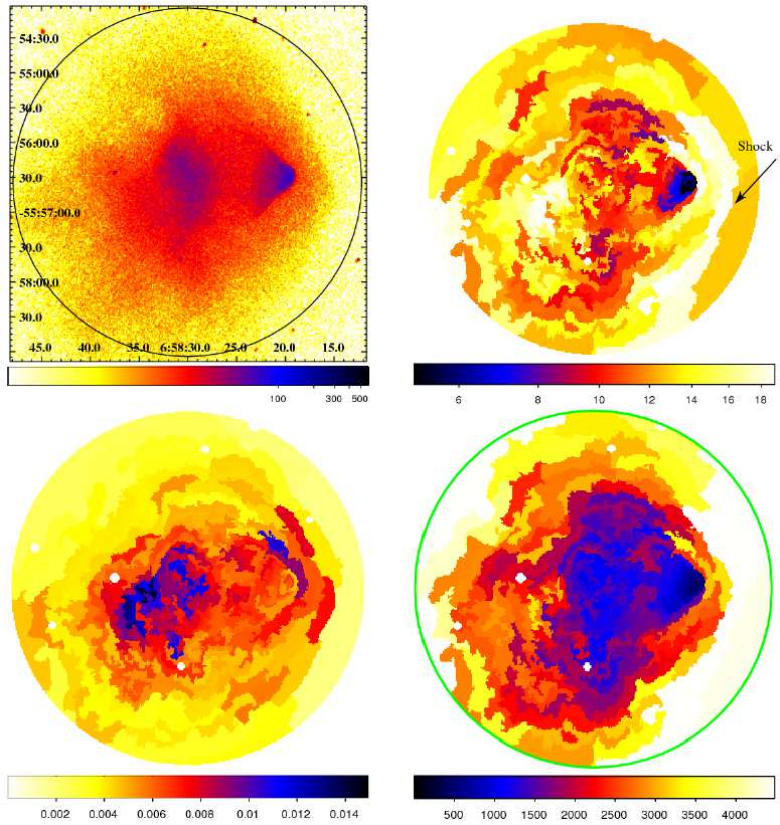}
\end{center}
\caption{Thermodynamic maps for the ICM of the ``bullet cluster'', 1E0657-56 \citep{million2008}
produced from a 500 ksec observation of this cluster with \chandra. The panels show the X-ray surface
brightness in the 0.8 to 7 keV band (upper left), the temperature, $k_B T$, in units of keV
(upper right), the projected pressure in units of keV cm$^{-5/2}$ arcsec$^{-1}$ (lower left) and
projected entropy in units of keV cm$^{5/3}$ arcsec$^{-1}$ (lower right). The shock front preceding the
bullet is marked by an arrow in the temperature map.}
\label{fig:14}       
\end{figure}

Galaxy clusters grow throughout the recent history of our Universe by accreting matter from
their surroundings, preferentially by clumpy accretion from the cosmic network matter 
filaments in the intersections of which they are embedded \citep[e.g.][]{braglia2007}. 
From time to time major merger
events happen in this accretion process, where two larger systems are attracted to each other
and merge violently. These cluster mergers have always been very attractive study
objects \citep[see e.g.][for a review]{feretti2002}. 
The thermal structure of the ICM again offers good perspectives
to unveil the merger configuration and its history as well as to understand the heating processes
of the ICM resulting from the energy release of the merger.

\begin{figure}
\begin{center}
\hbox{
  \includegraphics[height=6.0cm]{Boehringer_f17a.ps}
\hspace{0.01cm}
  \includegraphics[height=6.0cm]{Boehringer_f17b.ps}
}
\end{center}
\caption{{\bf Left:} Surface brightness profile across the shock front preceding the bullet   
\citep{markevitch2002}. The solid line shows the best fit model for the three-dimensional density
distribution with a sharp spherical discontinuity at the shock. {\bf Right :} Projected temperature
profile in a narrow sector across the shock front. The vertical lines show the boundary of the
bullet and the shock. The dashed line shows the average pre-shock temperature \citep{markevitch2006}.
Both data sets have been derived from deep \chandra\ observations.}
\label{fig:15}       
\end{figure}

We will illustrate the diagnostic potential of X-ray spectroscopy in these studies by
concentrating here mostly on the example of the most dramatic merging system, 1E0657-56.
But before we describe an earlier result obtained with \xmm\ for the diagnostics
of the off-axis merger in the cluster A3921 shown in Fig.~\ref{fig:14a} by \citet{belsole2005}.
The temperature map has been obtained by applying the multi-scale spectro-imaging technique
based on a wavelet analysis described by \citet{bourdin2004}. The general result of this analysis
was subsequently confirmed by spectroscopy of the regions highlighted in
the temperature map. While the general gas temperature of the undisturbed Eastern region
of the main cluster has a temperature around 4.9 keV, a hot, bar like region with an extent
of about 160 x 280 kpc  is observed in the highly distorted North-Western side which
has been found to have a temperature around 7.25 keV. The signatures have been interpreted 
as an off-axis merger \citep{belsole2005}. The observed features have been found to be
similar to those in the simulations of an off axis merger with a mass ratio of 1:3
an impact parameter of $b = 5 r_s$ (where $r_s$ is the scale radius of the NFW profile
describing the density distribution of the cluster) and a time of less than about 1 Gyr
after the X-ray luminosity reaches its peak as shown by \citet{ricker2001} in their Fig. 7.
This system seems to show the typical appearance of an intermediate stage off-axis merger
similar to other cases \citep[e.g.][]{reiprich2004}. 

The most detailed and interesting data set on a merging cluster system is that of 1E0657-56
at a redshift of $z = 0.297$. This cluster was observed in a very deep observation with
\chandra\ with an exposure of 500 ksec \citep{markevitch2006} and also with \xmm\ 
(Zhang et al. 2006, Finoguenov et al. 2005). As a first exercise to get an overview 
on the thermodynamic structure of the cluster ICM, it has become standard for such deep, high
photon statistic observations to produce maps of the temperature and density distributions 
of the ICM and to infer the pressure and entropy distribution from these quantities.
Fig.~\ref{fig:14} shows such maps produced by \citep{million2008}. The temperature is
determined from spectroscopy of about 100 image pixels, each containing at least 3000 photons.
The binning of these pixels has been guided by regions of similar surface brightness
\citep[][]{sanders2006b}. The ICM density is derived from the surface brightness
distribution by assuming a certain geometry to deproject the emission distribution along the
line-of-sight. Although the so derived temperature in the temperature map is a projected quantity,
it still gives a good impression about the temperature in the central bin in the line-of-sight
because of the large weight of the innermost bin due to the square density dependence of
the emissivity and the steep density profile. From these maps of approximate temperature and
density distribution in a cross section of the cluster, 
the approximate distribution of pressure, $P = nk_B T_x$, and entropy, 
$S = T/n^{2/3}$, can be constructed. The surface brightness image in Fig.~\ref{fig:14}
shows a disturbed larger cluster component and a compact, cone like structure to the West. 
The latteris identified with a compact subcluster, flying through the main cluster at high relative 
velocity, for which optical images provide further evidence. 
This component has been named ``the bullet'' from which the popular name of the 
whole system, ``the bullet cluster'', originates. In the temperature map the bullet shows
up as a cool core which also has the lowest entropy in the entropy map. The Mach cone
like shape of the bullet suggest that it may fly with supersonic velocity. In the 
temperature map we observe that the region in front of the bullet has a strongly enhanced
temperature that goes hand in hand with high pressure and elevated entropy. This is the
signature of a region heated by a shock preceding and being detached from the bullet.
The high temperature/entropy region shows a sharp edge at the shock. In the pressure map
the region of the bullet shows little enhancement (apart from the narrow band of shock 
compressed ICM), while the highest pressure values are found in the center of the overall
system structure. Thus, despite the disturbances, the pressure maximum is most probably still
indicating the region of the deepest gravitational potential. 

To obtain more quantitative information about the nature of the shock in front of the bullet,
\citep{markevitch2002} has studied the temperature and density distribution around the region
of the shocked ICM in more detail. Fig.~\ref{fig:15} shows two sharp edges in the surface
brightness profile in the left panel. The inner bump is the contact discontinuity that
separates the bullet from the shock heated ICM, the second bump at $90''$ is the shock.
From a density model that fits the projected surface brightness one infers a density jump
that corresponds to Mach number $3 \pm 0.4$ shock. The corresponding temperature jump
as derived from the projected spectra, is shown in the right panel. The relatively high
Mach number implies a relative velocity of the bullet and the ICM of about 4700 km/s. For
an explanation of this high velocity see recent simulations that try to reconstruct the
merger configuration of 1E0657-56 and recover such high velocities which are partly boosted
by the accretion inflow of ICM at the position of the bullet \citep{springel2007, 
mastropietro2008}. In section 6.2 we discuss the use of the observed shock
structure in this cluster to study the processes controlling the thermalization of
the plasma behind the shock. 

So far only two more cases of clear shock signatures in merging clusters are known: 
A~520 \citep{markevitch2005,markevitch2007}; A2219 \citep{million2008}. 

Another type of
interesting structures in the ICM are the so called cold fronts, which also show up in the
surface brightness images of clusters as sharp surface brightness discontinuities. But
thanks to the spectroscopically determined temperatures one can show that the
pressure across the cold fronts is continuous, such that they are boundaries between
colder denser plasma and a more tenuous, hotter environment. \citet{markevitch2007}
provide a nice recent review about the X-ray observations and the physics
of cold fronts and ICM shocks.   

\subsection{Observational Studies of Turbulence of the ICM}

\begin{figure}
\begin{center}
  \includegraphics[height=8.0cm]{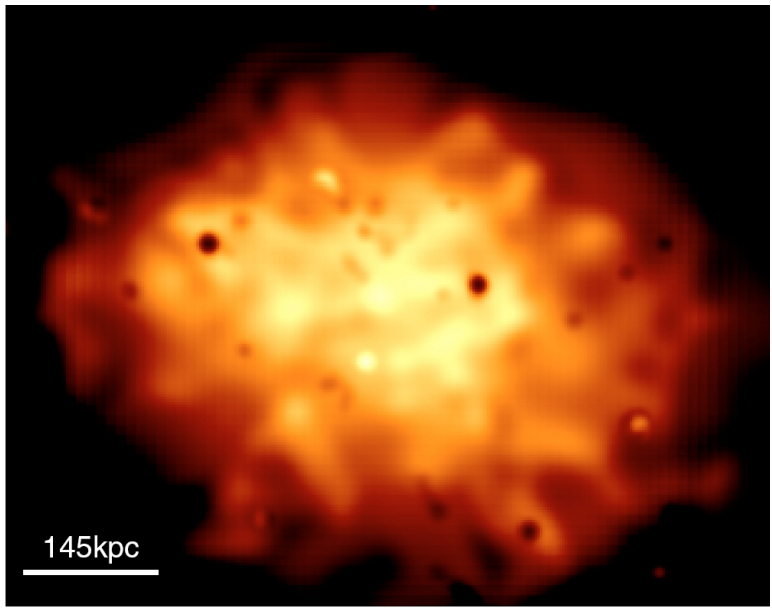}
\end{center}
\caption{Map of the projected pressure distribution of the ICM in the central region
of the Coma galaxy cluster. The scale size of 145 kpc corresponds to the largest 
turbulent eddies traced by the observed pressure fluctuation power spectrum 
\citep{schuecker2004}.} 
\label{fig:15b}       
\end{figure}

Galaxy cluster mergers are expected to induce turbulent motions into the ICM
\citep{sunyaev2003}.
An investigation of the presence of stochastic turbulence has been performed
by means of a spectral reduction of deep \xmm\ observations of the Coma galaxy 
cluster by \citet{schuecker2004}. The turbulence fluctuation spectrum was studied in
the projected pressure map of the ICM in the central region of the Coma cluster.
The Coma cluster has a very flat appearance, characterized by a very large core radius
of the X-ray surface brightness of $\sim 400$ kpc \citep{briel1992}. This enables us
to treat the configuration of the central region of Coma in the first approximation as a 
slab geometry, with corrections to the power spectrum applied later. The analysis
of the fluctuation spectrum testing for a turbulent power law spectrum was performed,
with the pressure rather than with density or temperature fluctuations, not to confuse
the turbulent fluctuations with static entropy fluctuations in pressure equilibrium, 
which would be characterized by contact discontinuities. In fact, Fig.~\ref{fig:15c} (left) 
shows that turbulent pressure fluctuations dominate also the density fluctuations rather 
than contact discontinuities.

\begin{figure}
\begin{center}
\hbox{
  \includegraphics[height=6.0cm]{Boehringer_f19a.ps}
\hspace{0.01cm}
  \includegraphics[height=6.0cm]{Boehringer_f19b.ps}
}
\end{center}
\caption{{\bf Left:} Correlation of the relative fluctuations in squared density, $n^2$,
and temperature, $T$, and their $1\sigma$ error bars. The solid lines give the adiabatic exponents
of $\gamma = 5/3$ (ideal gas) and $\gamma = 4/3$. The third line shows the anticorrelation 
of density and pressure expected for contact discontinuities. The positive correlation not
far from an ideal gas exponent indicates that adiabatic pressure fluctuations, as expected for
the ram pressure effects of turbulent motions dominate also the density and temperature
fluctuation spectrum.
{\bf Right :} Resulting shot noise subtracted, projected power spectrum of the pressure
fluctuations observed in the central region of the Coma galaxy cluster (with $1\sigma$ errors).
Also shown are power spectra for different power law shapes projected in an analoguous way
as the observed spectrum. The model power spectra are labled with the exponents of the
non-projected 3-dimensional power spectra. A Kolmogorov-Obuchov spectrum with an exponent of $n = -7/3$
is not far from the results, which lay in between $n = 5/3$ and $n = 7/3$  \citep{schuecker2004}.} 
\label{fig:15c}       
\end{figure}

The map of the projected pressure distribution in the center of the Coma ICM as shown in 
Fig.~\ref{fig:15b} was obtained by calculating the gas density from the X-ray surface brightness
(with an assumed depth of the ICM in the line of sight) and deriving the temperature
by a spectral analysis of the data in pixels of 20 by 20 arcsec$^2$ and alternatively 40 by 40 
arcsec$^2$, yielding the pressure by means of the ideal gas equation of state. We see the pressure
fluctuations clearly in Fig.~\ref{fig:15b}. A Fourier analysis of these fluctuations seen in
projection results in the power spectrum shown in  Fig.~\ref{fig:15c} (right). This spectrum has 
been corrected for the contribution of Poisson noise and the overall shape of the ICM surface brightness
in Coma. The observed power spectrum is characterized by a shape very close to a power law.
In  Fig.~\ref{fig:15c} we show how different 3-dimensional power law functions compare to the observed
spectrum if they are projected in the same way as the observed spectrum. The observations lay
in between a power law exponent of $5/3$ and $7/3$. An exponent of $7/3$ is the one expected
for the pressure fluctuation spectrum of the classical prediction of \citet{kolmogorov1941}
and \citet{oboukhov1941}. The original work by Kolmogorov was developed for an incompressible
fluid, and in the ICM also magnetic fields play a not fully quantified role, and therefore the
the case of the cluster ICM is not easily comparable to the classical picture. Nevertheless,
a very similar dimensional consideration of turbulence (see e.g. Landau \& Lifshitz Vol. VI) 
is surely applicable to the cluster ICM
where most importantly a scale free spectrum is expected between the driving and dissipation scale.
The observed signature of a nearly scale free power law power spectrum is thus a very interesting
result.  The observations have therefore been interpreted in the way that a nearly
classical turbulence configuration has been established in the Coma cluster ICM 
\citep{schuecker2004}. This can most probably be explained by the fact
that Coma is generally believed to be a post-merger cluster \citep{white1993}. 


\begin{figure}
\begin{center}
\includegraphics[width=9cm,clip=t, angle=0]{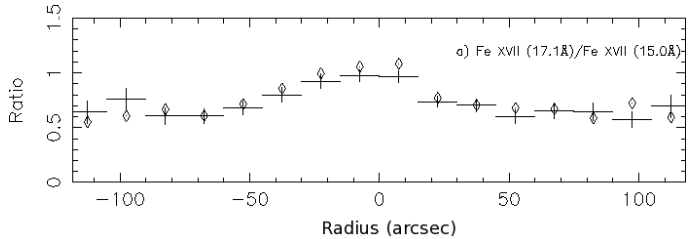}
\caption{Ratio of the $2p$--$3s$ lines of \ion{Fe}{xvii} at 17.05\&17.10~\AA\
to the $2p$--$3d$ line of the same ion at 15.01~\AA. The ratio
has a significant peak at the center of the galaxy, NGC 4636. Since the lines
originate from the same ion, the ratio cannot be due to a spatial
abundance gradient. Given the narrow temperature range and the weak
dependence of the line ratio on temperature it cannot be due to a
temperature gradient either. This trend is the best evidence so far
for resonant scattering of the line at 15.01~\AA\ in a hot galactic halo. }
\label{resscat}
\end{center}
\end{figure}

Important information on the amount of turbulence and turbulent pressure support can
also be obtained by measuring the level of resonant scattering in emission lines observed
in the ICM.
The ICM is generally assumed to be optically thin. This is certainly true for most of the
emitted X-ray photons, but for the strongest resonance lines the ICM can be moderately
optically thick \citep{gilfanov1987}. What happens is that for strong resonance lines the
transition probabilities are large, photons get absorbed, but since the time between the
absorption and emission is extremely short they get very quickly reemitted into a different
direction. Because of the very short time between the absorption and the emission of the
photon the process can be effectively regarded as scattering. \citet{gilfanov1987} pointed
out that because the optical depth of the resonance line depends on the characteristic
velocity of small-scale gas motions, measurements of  this optical depth give important
information about the turbulent velocities in the hot plasma.

The first constraints on turbulent velocities using resonant scattring were obtained
by \citet{churazov2004}, who used \xmm\ EPIC data to compare the relative fluxes of the
He-like Fe~K$_{\alpha}$ and Fe~K$_{\beta}$ lines in the core and in an annulus around the
core of the Perseus cluster. Since the Fe~K$_{\alpha}$ at
6.7~keV has a much larger optical depth than the K$_{\beta}$ line a difference in their
ratios in the two spatial regions with different column densities would be an evidence for
resonance scattering in the core of
Perseus. \citet{churazov2004} found no evidence for resonance scattering in Perseus,
indicating that differential gas motions on scales smaller than $\sim100$~kpc in the
core of the cluster must have a range of velocities of at least half of the sound speed.
Independently, \citet{gastaldello2004} reached similar conclusions.

The first unambiguous evidence for resonant scattering was found using high-resolution
spectra of the hot halo around the giant elliptical galaxy NGC~4636 obtained by
\xmm\ RGS \citep{xu2002}.
The plasma with temperatures below 0.9~keV observed in elliptical
galaxies, groups of galaxies, and in the cool cores of some clusters
emits three strong \ion{Fe}{xvii} lines. While the line at 15.01~\AA\ has a very strong
oscillator strength and is expected to be optically thick, the
blend of lines at 17.05 and 17.1 \AA\ have negligible optical depths.
The radial profile of the ratios of these two lines (\ion{Fe}{xvii}~17.1\AA~/~\ion{Fe}
{xvii}~15.0~\AA) derived using RGS shows a clear gradient with the
peak in the center of the galaxy NGC~4636, proving that many of the 15.01~\AA\
photons get scattered before exiting its core \citep[see Fig.~\ref{resscat}][]{xu2002}.

Following up the work by \citet{xu2002}, \citet{werner2009} analyzed the \xmm\ RGS data of
five nearby bright elliptical galaxies and found that the \ion{Fe}{xvii}
lines in the cores of four galaxies
show evidence for resonance scattering in the innermost region. The data for NGC~4636
in particular allowed the effects of resonant scattering to be studied in detail.
\citet{werner2009} used deprojected density and temperature profiles obtained by \chandra\
to model the radial intensity profiles of the strongest resonance lines, accounting for the
effects of resonant scattering,  for different values of of the characteristic turbulent
velocity. Comparing the model to the data they found that the isotropic turbulent velocities
on spatial scales smaller than $\approx$1~kpc are less than 100~km~s$^{-1}$ and the turbulent
pressure support in the galaxy core is smaller than 5\% of the thermal pressure at the
90\% confidence level, and less than 20\% at 95\% confidence.
Note that the spatial scales of turbulence probed in the cores of elliptical galaxies
by RGS are much smaller than those probed in Perseus \citep{churazov2004} or Coma
\citep{schuecker2004}.

The \ion{Fe}{xvii} lines are the most sensitive probes of turbulence using resonant scattering,
but because they are present only at temperatures less than $\sim$1~keV they cannot be used
to probe turbulence in the much hotter clusters of galaxies. High-resolution spectra obtained
by X-ray calorimeters on the future satellites like {\it Astro-H} and {\it IXO} will allow us
to probe turbulence also in higher mass systems and at larger radii (see Sect.~\ref{sec:7}).

\begin{figure}
\hbox to \hsize{
  \includegraphics[height=6.0cm]{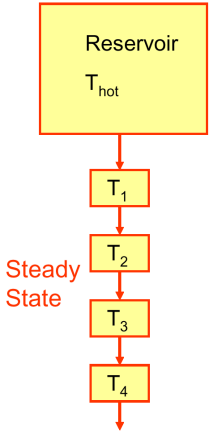}
\hfil\vbox to 5.8cm{\hsize=6.0cm
\caption{Sketch of the temperature evolution of a classical cooling flow model, with
a reservoir of only slowly cooling hot gas at low density and a steady cooling of the
fraction of gas with a mass cooling rate, $\dot M$, down to lower temperatures. The spectrum
of this cooling flow is then determined by the integral of the spectra of the 
transient temperature
phases, where each temperature phase contribution is fixed by the amount of heat that has
to be radiated in the given temperature interval to cool an amount of ICM with mass $\dot M$
(see equations in the text).}
\vfil}}
\label{fig:16x}      
\vskip10pt
\end{figure}

\section{AGN Heated Cluster Cool Cores}
\label{sec:4}

The temperature profiles shown in the right hand panel of Fig.~\ref{fig:10} reveal two types of
clusters: those with temperature profiles falling towards the center are called
cool core (CC) clusters; and clusters with no central drop of temperature are called non-cool core, NCC clusters.
While the former also have high densities in their cores which implies
short cooling times for the central ICM, typically one or two orders of magnitude
smaller than the Hubble time, the NCC clusters have usually central cooling
times exceeding the Hubble time. This observation implies that the ICM plasma
should cool and condense in the CC clusters in the absence of any heat source,
which could balance the cooling. Theoretical considerations focussed on
the consequences of cooling without heating lead to the so-called ``cooling flow'' 
scenario \citep{fabian1977b,fabian1994}, which predicts mass deposition
in the centers of some clusters at a rate of up to hundreds to thousands of solar 
masses a year. In general more than half of the clusters
in the nearby Universe should have cooling flows in this model \citep{peres1998},
where the exact fractional number depends to some degree on the definition
of a cooling flow.

Interestingly, in particular in the context of this review, the cooling flow model
leads to the prediction of very specific and testable spectral signatures of a
steady state cooling flow. Looking at the plasma distribution in this scenario
in temperature space, we have a reservoir of hot plasma and a steady cooling 
of some plasma with a mass deposition rate $\dot M$, as sketched in Fig.~21. 
This simplified picture of a reservoir of hot gas and a constant fraction of gas cooling
is motivated by the fact that cooling is accelerating with decreasing temperature
because the plasma is getting denser while it is cooling in pressure equilibrium
and the cooling radiation power is proportional to the gas density squared.
For the part of the plasma that is steadily cooling to lower temperature, each
temperature interval has to contribute to the cooling radiation with a
power equal to the enthalpy change of the plasma that is cooling across this
temperature interval:
\begin{equation}
L_x~ dT = {d \over dt} {\rm Enthalpy}~ dT = {5k_b~ dT \over 2 \mu m_p} \dot M.
\end{equation}
To calculate the expected spectrum we have to weight this power by the ratio of the
emission at a certain frequency to the bolometric emission:
\begin{equation}
L_x(\nu)~d\nu dT = {{\rm emissivity}(\nu) d\nu \over {\rm emissivity}_{bol} } \times
{5k_b~ dT \over 2 \mu m_p} \dot M.
\end{equation}
To obtain the total spectrum radiated by all steady cooling temperature phases 
we have to integrate the above equation over $dT$:
\begin{equation}
L_x(\nu) d\nu = {5k_b~ dT \over 2 \mu m_p} \dot M \int^{T_{hot}}_{T_{cutoff}}
{ \Lambda_{\nu}(T^\prime) d\nu \over \Lambda_{bol}(T^\prime)}  dT^\prime,
\end{equation}
where $\Lambda_{\nu}(T)$ and $\Lambda_{bol}(T)$ are the emissivity of the
plasma (radiation power per unit emission measure) for radiation at frequency,
$\nu$, and for bolometric radiation, respectively.

\begin{figure}
\begin{center}
\hbox{
  \includegraphics[height=5.9cm]{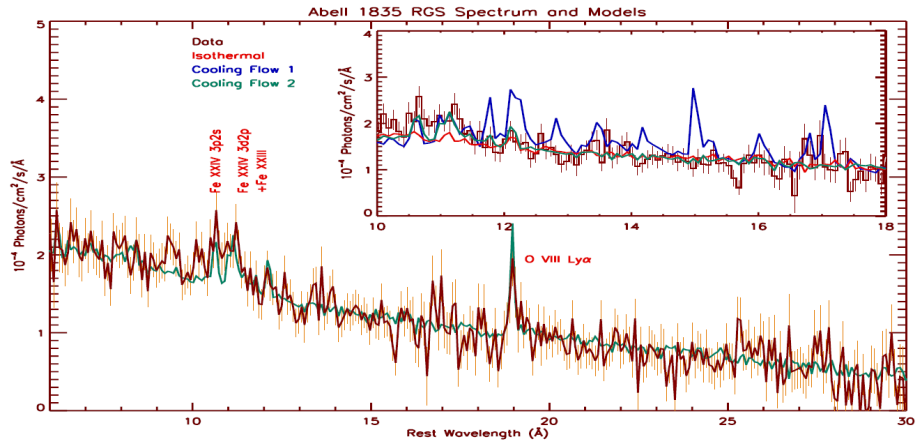}
}
\end{center}
\caption{\xmm\ RGS spectrum of the central region of the prominent cool core cluster, A1835
compared to model predictions \citep{peterson2001}. The red line shows a spectral model
for an isothermal temperature of 8.2 keV. The blue line shows the expectation for 
the classical cooling flow and an ambient temperature of 8.2 keV, the green line shows a cooling 
flow model where temperatures below 2.7 keV have been truncated. 
Clearly, some of the cooling flow predicted lines are missing from the observed
spectrum.}
\label{fig:17}       
\end{figure}

With the advent of high quality spectra provided by \xmm\ and \chandra,
the cooling flow paradigm could be tested by spectroscopy. First evidence that
the cluster cool cores are not consistent with the classical cooling flow model
came from spectra analyzed with the Reflection Grating Spectrometers on \xmm. Some of the cool core
clusters have such highly peaked surface brightness profiles, that they look almost
like blurred point sources to the RGS (which operates analogously to
slitless spectroscopy in optical astronomy). Therefore the RGS can obtain spectra with much higher
resolution than provided by the imaging CCD devices
of the \xmm\ EPIC cameras for these very peaked cool core regions.
Fig.~\ref{fig:17} shows the first published RGS spectrum
of one of the very strong cool core clusters, A1835, at a redshift of $z = 0.2523$
\citep{peterson2001}. In the Figure the observed \xmm\ spectrum (dark red) is compared to the
prediction for the cooling flow model with a mass deposition rate inferred from
the total radiative emission in the cooling flow region determined from the X-ray
images (blue curve). There are clearly some expected lines, notably those of \ion{Fe}{xvii},
missing in the observations. A more detailed analysis of the spectrum shows that for
an ambient ICM temperature of 8.2 keV, the central regions shows temperature phases
down to about 2.7 keV, but the expected intermediate temperature phases below 2.7 keV are
absent in the spectrum. This was generally taken as the first evidence that there is 
something wrong with the classical cooling flow model. Since at the same time high
angular resolution \chandra\ images showed signs of strong interaction of the central
AGN with the ambient ICM in many cool core regions \citep[e.g.][]{david2001,nulsen2002,fabian2005}, 
the most probable solution to the absence of strong cooling
was soon believed to be the heating of cool core regions by the central AGN.

\begin{figure}
\begin{center}
\hbox{
  \includegraphics[height=3.1cm]{Boehringer_f23a.ps}
\hspace{0.01cm}
  \includegraphics[height=3.1cm]{Boehringer_f23b.ps}
\hspace{0.01cm}
  \includegraphics[height=3.1cm]{Boehringer_f23c.ps}
}
\end{center}
\caption{{\bf Left:} Spectrum of hot plasma at various temperatures around the Fe L-shell line blend
near 1 keV as observed with the response function of the \xmm\ EPIC pn instrument.
This feature lends itself as a nice thermometer for the temperature range 0.2 to 2 keV.
{\bf Middle:} Predicted X-ray spectrum around the Fe L-shell line blend for the cooling flow in
the X-ray halo of M87 with a mass deposition rate of  $\dot M \sim 1 $M$_{\odot}$ yr$^{-1}$
compared to the actual spectrum in the radial region $r = 1 - 2$ arcmin as observed with the 
\xmm\ EPIC pn instrument. The observations are clearly inconsistent with the predictions
from the classical cooling flow model. {\bf Right:} Reproduction of the observed spectrum
by assuming that the plasma in this region of the M87 halo occupies the narrow temperature
interval of about 1.44 to 2 keV \citep[for more details see][]{bohringer2002}.}
\label{fig:18}       
\end{figure}

The spectral signature of the absence of cooling was not only obtained from \xmm\
RGS spectra. For example for the cool core region in the X-ray halo of M87, the central
dominant galaxy in the northern part of the Virgo cluster, clear evidence for an 
inconsistency of the spectral signatures with a classical cooling flow model could
also be provided by results with the EPIC pn and MOS cameras of \xmm.
The most prominent feature that can be used for these diagnostics is the blend of
Fe lines in the spectrum from transitions into the L-shell at photon energies around
1 keV. Fig.~\ref{fig:18} shows how the spectral feature of this blend of Fe lines
changes with temperature. The shift is caused by the changing degree of ionization of
the Fe ions that contribute to the line blend. When the temperature is lowered more and 
more electrons in the Fe ions shield the charge of the nucleus decreasing the binding
energy of the L-shell electrons. As illustrated in Fig.~\ref{fig:18} this spectral 
feature can be used as a sensitive thermometer in the temperature range of
0.2--2 keV \citep{bohringer2002}. The middle panel in Fig.~\ref{fig:18} shows
the spectrum implied for the broad temperature distribution expected for the cooling
flow in M87 (in the radial range 1--2 arcmin)
with a mass deposition rate of $\dot M \sim 1 $M$_{\odot}$ yr$^{-1}$.
The predicted spectrum is clearly different from the observed one. A way to modify
the predicted spectrum in the direction of the observed one, is to assume that the
central region of M87 is highly absorbed by cold or warm material to reduce the low
energy part of the broad Fe line feature. The actual absorption can be determined directly, however,
by using the power law spectrum of the M87 AGN with the result that there is no
significant excess absorption above the known value of the galactic absorption
column density of about $1.8 \times 10^{20}$ cm$^{-2}$.
A very good fit to the observed spectrum is then obtained by assuming that the
bulk of the plasma in the central region of the M87 halo only occupies a narrow
temperature interval of about 1.44 to 2 keV \citep{bohringer2002} as shown
in the right panel of Fig.~\ref{fig:18}.

\begin{figure}
\begin{center}
\hbox{
  \includegraphics[height=3.1cm]{Boehringer_f24a.ps}
\hspace{0.01cm}
  \includegraphics[height=3.1cm]{Boehringer_f24b.ps}
\hspace{0.01cm}
  \includegraphics[height=3.1cm]{Boehringer_f24c.ps}
}
\end{center}
\caption{{\bf Left:} Differential emission measure distribution as a function
of temperature of the plasma in the cool core region of the cluster Abell 262
as deduced from a spectral analysis of \xmm\ RGS observations. The
results are derived for the innermost
four shells: 0 - 0.5 armin (triangles), 0.5 - 1 armin (squares), 1 - 2 arcmin
(stars), 2-3 arcmin (circles). For comparison the dashed line shows an approximation
to the slope of the isobaric cooling flow model \citep{kaastra2004}.
{\bf Middle:} Differential emission measure distribution for the isobaric
cooling flow model, for 0.5 solar abundance \citep{kaastra2004}
{\bf Right:} Differential emission measure distribution of the plasma in the 
inner 2 arcmin region in the core of the Centaurus cluster of galaxies. The
results from a spectral analysis of \xmm\ RGS and CHADRA are shown for
comparison. The solid line shows the expected emission measure distribution
for a 10 M$_{\odot}$ yr$^{-1}$ isobaric cooling flow at solar metallicity,
cooling from a temperatue of 4.5 keV to 0.08 keV \citep{sanders2008}.}
\label{fig:18b}       
\end{figure}

It has to be noted that the interpretation of spectra from cooling flows had an interesting
history already before the launch of \chandra\ and \xmm. For example \citet{ikebe1997}
and \citet{makishima1999} already argued on the basis of \asca\ spectroscopy with
similar arguments as those given above for M87, that the prediction of cooling flow models is not met
by the observations. The counterargument to keep the cooling flow model alive at the
time was to postulate internal absorption \citep[see e.g. ][]{allen2001}. The spectral 
resolution and sensitivity of \asca\ was not quite good enough, however, to reject the cooling flow model
in the very compelling way as now done with the \xmm\ results. 
Even earlier, spectroscopic analysis of the X-ray halo of M87 with the Crystal Spectrometer on board of the Einstein satellite
showed the emission lines for the cool phases expected in cooling flows
\citep{canizares1979,canizares1982}, that we are now missing in modern observations.
This evidence for steady state cooling in cooling flows 
gave actually one of the strongest supports to the cooling flow scenario at the time.
It is now clear, however, that the results obtained in these early days were
most probably features of spectral noise. 

Deep observations of cool core regions, in particular with the \xmm\ RGS
instrument, have now also been used to characterize the multi-temperature
structure in cool cores in more detail in terms of the emission measure (or
luminosity) distribution of the plasma as a function of temperature
\citep[e.g.][]{peterson2003,kaastra2004}. This works well only in colder
systems. Fig.~\ref{fig:18b} shows in the left panel the differential emission measure
distribution in the cool core region of the cluster Abell 262 in the innermost
four shells out to 3 arcmin. These observations can be compared to the
expected emission measure distribution for the model of an isobaric cooling
flow, shown in the middle panel of Fig.~\ref{fig:18b}
\citep{kaastra2004}. While the temperature distribution is not isothermal,
we clearly note that the emission measure distribution falls off much more
steeply than the model expectations towards lower temperatures, highlighting
again the fact that less plasma is observed at lower temperatures than
required for the cooling flow model. A similar result is shown 
- in the right panel of the Figure - for the case of 
the Centaurus cluster taken from \citet{sanders2008} based on a very deep
\xmm\ observations of the cluster with the RGS instruments as well as deep
\chandra\ data. Here we can follow the local temperature distribution of the
plasma over a full decade in temperatures from about 0.4 to 4 keV. But again
the contribution of the very low temperature phases is much less than what is
expected for unimpeded cooling.  
  
\begin{figure}
\begin{center}
  \includegraphics[height=8.0cm]{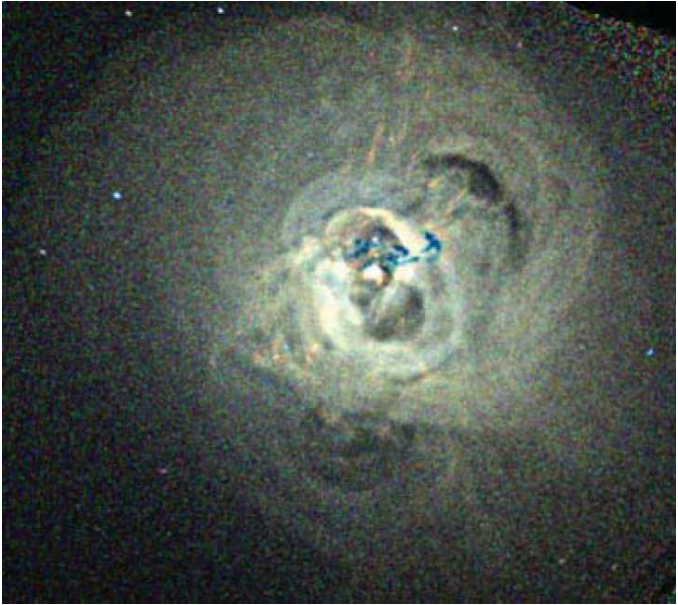}
\end{center}
\caption{False colour image of the central region of the Perseus
cluster produced from images
in three energy bands 0.3--1.2 keV, 1.2--2 keV, and 2--7 keV. An image smoothed on
a scale of 10 arcsec (with 80\% normalization) has been subtracted from the image 
to highlight regions of strong density contrast. The image shows a series of nearly 
concentric ``ripples'' which are interpreted as sound waves or weak shock waves set
off by the activity of the central AGN \citep{fabian2006}.}
\label{fig:19}       
\end{figure}

As the change from the cooling flow paradigm to an AGN feedback scenario
seems to be now widely accepted, the interest
has shifted to the question: how is the ICM actually heated by AGN interaction? One
of the most important earlier arguments in favour of cooling flows in the absence of feedback was that
any possible heating mechanism has to be very well fine tuned, to exactly 
provide the balance to cooling. If too much heat is produced, it disperses
the observed dense gas cores in cooling flows. Secondly, we observe increasing entropy
profiles in the ICM of clusters (Fig.~\ref{fig:13}). It therefore has to be explained how central heating 
can work without inverting the entropy profiles leading to the dissipation of heat by convection. 
Both requirements need to be met and seem almost improbable \citep{fabian1994}. This makes it
even more interesting now to understand how nature achieves this fine tuning.

One of the first, best studied cases is that of the Perseus cluster, which has
now been observed with the \chandra\ observatory for more than 1 Million seconds as proposed
by \citet{fabian2005} providing the most detailed picture of a central cool core region.
Fig.~\ref{fig:19} shows a multi-colour image of the central region of the Perseus
cluster from \citet{fabian2005}. The ``false colour'' image was produced from images
in three energy bands 0.3--1.2 keV, 1.2--2 keV, and 2--7 keV. An image smoothed on
a scale of 10 arcsec (with 80\% normalization) has been subtracted from the image 
to highlight regions of strong density contrast. The central, bright part of the image shows
two features of low surface brightness, interpreted as two cavities of the X-ray emitting
plasma filled with the relativistic plasma from the radio jets of the central AGN. 
This region is surrounded by a series of nearly concentric ``ripples'' which are more clearly
brought out by the unsharp masking processing of the image.

\begin{figure}
\begin{center}
  \includegraphics[height=7.0cm]{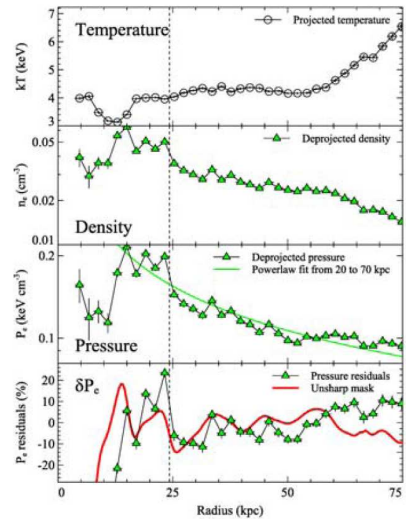}
\end{center}
\caption{Temperature, density, pressure, and pressure residual profiles in the core region ICM
of the Perseus cluster in a sector to the North-East \citet{fabian2006}. The red line shows the
ripples from an unsharp mask image. The dashed line marks the position of a shock front.}
\label{fig:20}       
\end{figure}

A more detailed analysis of the density and temperature variations across the ripples,
shown in Fig.~\ref{fig:20}, implies typical pressure variations associated with the
ripples with amplitudes of about $\pm5$--$10$\%, which are interpreted as sound waves or very weak
shock waves in the innermost region. As detailed in \citet{fabian2005} the dissipation
of the sound waves - if the viscosity is sufficiently high - can balance the radiative cooling 
in the cool core by heating with the realistic assumption that the sound waves cross
the cool core region in about $1/50$ of the cooling time.

\begin{figure}
\begin{center}
  \includegraphics[height=10.0cm]{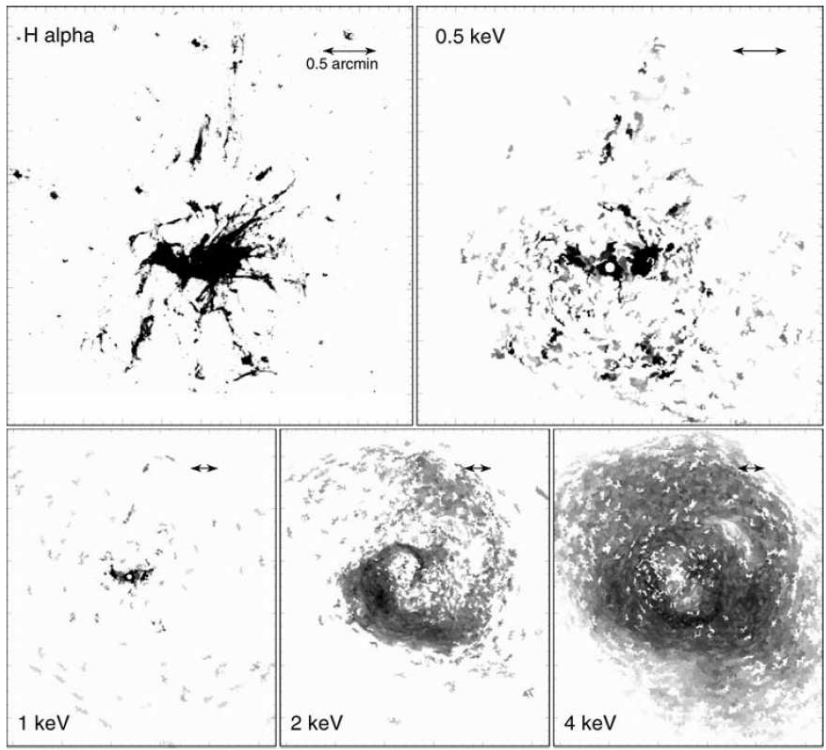}
\end{center}
\caption{H$\alpha$ image in the halo of NGC 1275 in the center of the Perseus cluster (from
\citet{conselice2001}, upper left) shown on the same scale for comparison with maps of the mass distribution
of different temperature components, at 0.5, 1, 2, and 4 keV, of the ICM in the center of the Perseus
cluster \citep{fabian2006}.}
\label{fig:21}       
\end{figure}

\citet{fabian2006} have also carried out multi-temperature plasma modeling in 
various regions of the cluster centre and produced mass distribution maps
of plasma at temperatures around 0.5, 1, 2 and 4 keV. These maps are shown in comparison
to a map of the optical H$\alpha$ emission \citep[observed
by][]{conselice2001} in Fig.~\ref{fig:21}. 
The distribution of the lowest temperature phase around 0.5 keV has a striking similarity
to the H$\alpha$ map. There is relatively little plasma at 1 keV, mostly in the very
center of the cool core. The 2 keV phase coincides mostly with the denser regions of the
ICM. This detailed picture of the central region of the Perseus cluster given in a series
of papers by Fabian et al. and Sanders et al. well illustrate the frontier of the application
of X-ray spectroscopy and thermodynamic analysis of the ICM in the bright central regions
of clusters.

\begin{figure}
\begin{center}
\hbox{
  \includegraphics[height=7.40cm]{Boehringer_f28a.ps}
\hspace{0.01cm}
  \includegraphics[height=7.4cm]{Boehringer_f28b.ps}
}
\end{center}
\caption{{\bf Left:} \chandra\ image of the X-ray halo of M87 in the 0.5--2.5 keV band \citep{forman2005}.
Several features are marked on this image: the surface brightness enhancements coinciding
with the radio lobes (SW arm and E arm) and several sharp surface brightness discontinuities
(14 kpc ring, 17 kpc arc, and 37 kpc arc) which are identified as shock fronts.  
{\bf Right:} Map of the pressure deviations from a smooth spherical symmetric model 
derived from a deep \xmm\ observation of the M87 halo \citep{simionescu2007a}. 
The 14 kpc ring is now clearly revealed as a pressure discontinuity supporting the shock
interpretation. The regions of the radio lobes (SW arm and E arm) are now showing up as zones
of pressure deficit, which is probably compensated for by the unseen pressure of the relativistic radio
plasma.}
\label{fig:22}       
\end{figure}

To show that the physics of the cool core regions seems to be quite diverse and that
we can obviously not easily generalize the behaviour of the Perseus cluster to all cool core
clusters we show two further examples of systems studied in detail, that of M87 and of
Hydra A. M87 was studied in very much detail by a combination of a 500 ks exposure
with \chandra\ \citep{forman2005,forman2006} and a 120 ks exposure with \xmm\ 
\citep{simionescu2007a,simionescu2008a}. 
The X-ray halo of M87 is the closest of all cool core regions in galaxy
clusters and can therefore be studied at the highest spatial resolution. Also 
the AGN in the center of M87 is active and is interacting with the
ambient ICM through relativistic jets \citep[e.g.][]{bohringer1995,churazov2001}. 
The deep \chandra\ exposure shows a surface brightness 
discontinuity which can be almost followed over the entire circumference of a ring
around the nucleus with a radius of 14 kpc \citep[Fig.~\ref{fig:22},][]{forman2005,forman2006}.
Another arc-shaped discontinuity is seen over a smaller sector region at the radius
of 17 kpc. These features are identified as shock waves in  \citet{forman2005} and \citet{forman2006}.
A detailed spectroscopic analysis of the M87 X-ray halo with \xmm\ can
corroborate this interpretation and give enhanced physical constraints.
While the density discontinuity seen with \chandra\
implies a shock Mach number of about 1.2, the temperature jump seen in the projected
spectra is about 5\% at about 2 keV and implies a shock Mach number of 
$> 1.05$ without deprojection correction \citep{simionescu2007a}. Thus the temperature
enhancement identifies the surface brightness discontinuity definitely as a weak shock
front. The shock front has also been visualized in a pressure map of the M87 halo 
constructed from the \xmm\ data shown in Fig.~\ref{fig:22} \citep{simionescu2007a}. 
Similar to the hydrodynamic maps described above this map has been derived
from spectral modeling in many spatial pixel, which in this case have been
constructed by means of a Voronoi tessellation technique developed by \citet{diehl2006}
and originally by \citet{cappellari2003}
such that each pixel has a spectral signal to noise of at least 100. To enhance
the visibility of discontinuities in the presence of the strong pressure gradient,
a smooth, spherically symmetric model has been subtracted from the pressure distribution shown
in Fig.~\ref{fig:22}. The map clearly shows a pressure enhancement at the shock radius 
of 3\arcmin\ (14 kpc) which implies without projection correction a Mach number 
larger than 1.08. There are some further interesting features seen in the pressure
map. Outside the shock front we note two regions of lower pressure in the NE and in
the SW. These are the regions where radio lobes filled with relativistic
plasma emerging from the central AGN are observed. This observation thus
implies that (at least part of) the missing pressure in these regions comes from the pressure of the
relativistic plasma in the lobes, which does not contribute to the X-ray emission
analysed here. Further insight into the physics of this interesting cool core
region in M87 will be described in connection with the analysis of the element
abundance distribution discussed in the next section.

\begin{figure}
\begin{center}
\hbox{
  \includegraphics[height=6.1cm]{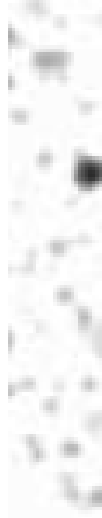}
\hspace{0.01cm}
  \includegraphics[height=6.6cm]{Boehringer_f29b.ps}
}
\end{center}
\caption{{\bf Left:} \chandra\ image of Hydra A with the radio lobes superposed \citep{nulsen2005}.
An oval shaped surface brightness discontinuity at a radius of about 200 - 300 kpc can be noted 
that is identified with a shock wave caused by an energetic radio jet outburst.   
{\bf Right:} Map of the pressure deviations from a smooth azimuthally symmetric model derived
from a deep \xmm\ observation of Hydra A \citep{simionescu2008c}. The location of the X-ray
surface brightness discontinuity observed in the \chandra\ image is approximated by the elliptical
black line. A clear pressure jump can be seen almost all along this line indicating a shock
front with a Mach number of about 1.3.}
\label{fig:23}       
\end{figure}

\begin{figure}
\begin{center}
\hbox{
  \includegraphics[height=5.8cm]{Boehringer_f30a.ps}
\hspace{0.05cm}
  \includegraphics[height=5.8cm]{Boehringer_f30b.ps}
}
\end{center}
\caption{{\bf Left:} Model of the effect of a shock front with different Mach numbers predicting the 
observed surface brightness distribution in the \xmm\ image of Hydra A \citep{simionescu2008c} 
{\bf Right:} Temperature profile around the shock front compared to a smooth model
(blue lines). The observed temperature and density (surface brightness) jumps are consistent
with a shock wave with Mach number $\sim 1.3$ \citep{simionescu2008c}.}
\label{fig:24}       
\end{figure}

In a combined \chandra\ and \xmm\ study of the central region of the 
galaxy cluster associated with the radio source Hydra A, a similar front diagnostics
has also been performed by \citet{nulsen2005} and \citet{simionescu2008c}. 
Here the shock feature has a radial extent of 200 - 300 kpc and hints at an event
which was much more powerful by about two orders of magnitude than that in M87. 
Fig.~\ref{fig:23} shows again the surface brightness discontinuity clearly detected
in the high angular resolution image obtained with the \chandra\ observatory \citep{nulsen2005}.
A deep \xmm\ observation then provides higher photon statistics that allows
us to study this discontinuity spectroscopically. Firstly, we can again
visualize the shock front as a pressure jump in the \xmm\ based pressure map
shown in Fig.~\ref{fig:23} \citep{simionescu2008c}. A more detailed analysis
of the temperature and density distribution around the shock front in several sectors,
shown in Fig.~\ref{fig:24}, provides more detailed information about the shock 
strength. Modeling the discontinuity in three dimensions and fitting the projected model
to the data implies a Mach number of the best fitting model of about 1.2--1.3
\citep{simionescu2008c}. Simple one-dimensional modeling of the evolution of the 
shock driven by the energy input of the radio lobes as well as detailed three-dimensional
hydrodynamical simulations provide good estimates for the age and the total energy 
of the event causing the shock wave, with values for the age of $\sim 2 \times 10^7$ yr 
and the energy of $10^{61}$ erg \citep{nulsen2005,simionescu2008c}.

These few examples show the complex and probably very diverse physics
prevailing in cluster cool core regions. We hope that further detailed studies
of a larger sample of cool core regions will help to generalize the scenario
of cool core heating. A major break through is also expected here from future
high resolution spectroscopy, which will allow us to observe turbulent and bulk motions
in the ICM as described in section 7.      

\section{What Chemical Abundance Measurements Tell Us}
\label{sec:5}

Clusters of galaxies are unique laboratories for the study of the nucleosynthesis and chemical enrichment of the Universe. Their deep gravitational potential wells keep all the metals produced by the stellar 
populations of the member galaxies within the cluster. The dominant fraction of these metals reside within the hot ICM. The chemical abundances measured in the intra-cluster plasma thus provide us with a 
``fossil'' record of the integral yield of all the different stars (releasing metals in supernova explosions and winds) that have left their specific abundance patterns in the gas prior and during cluster evolution. 

X-ray spectroscopy provides an accurate measure of metal abundances in the ICM. These abundances put constrains on nucleosynthesis and on the star formation history of the clusters, and as long as the 
stellar populations where the cluster metals were synthesized are representative, these measurements allow us to put constraints on the chemical evolution of the Universe as a whole. 

\begin{figure}    
\begin{center}
\begin{minipage}{0.45\textwidth}
\includegraphics[width=4.25cm,clip=t,angle=-90]{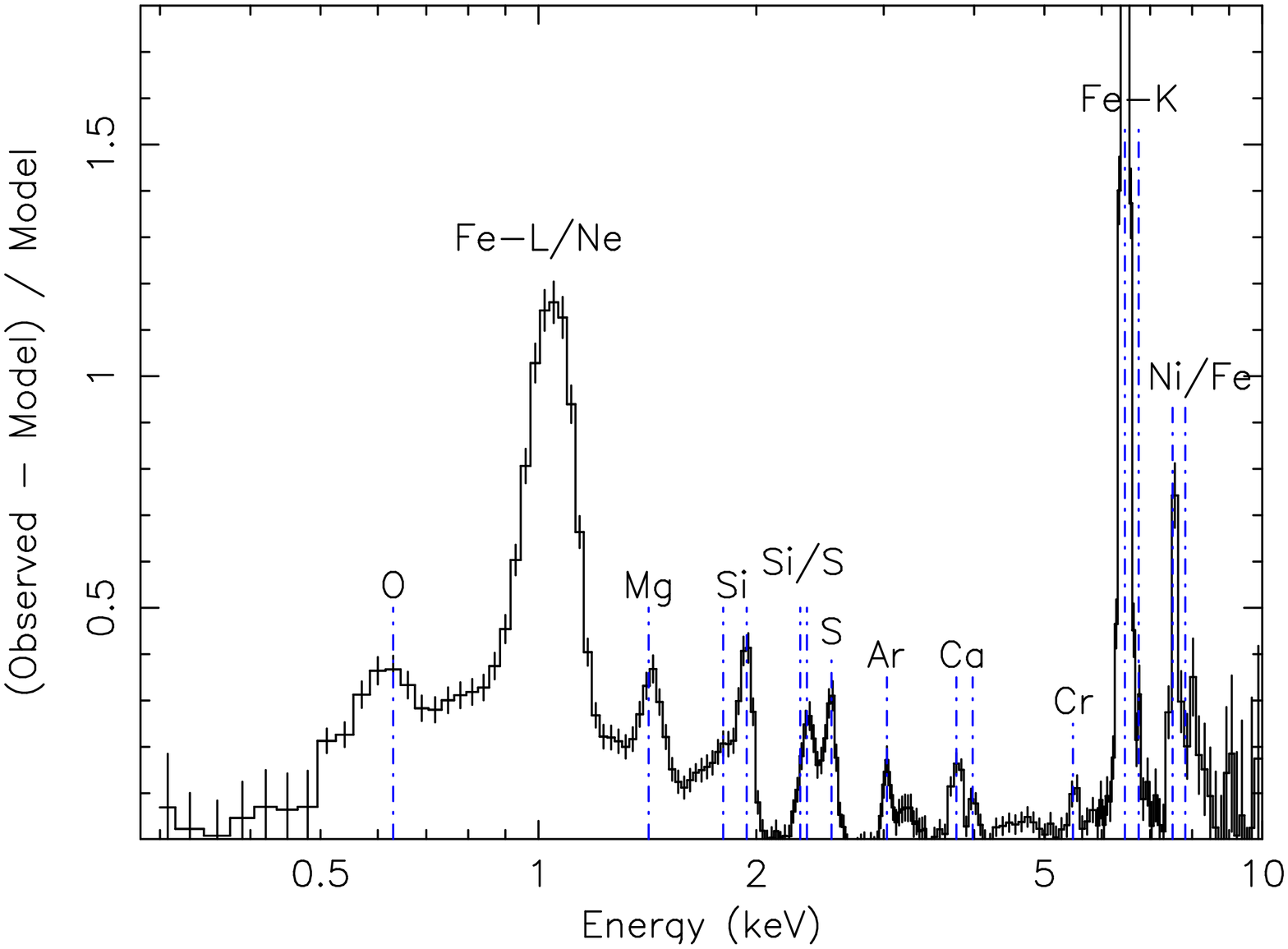}
\end{minipage}
\hspace{0.2cm}
\begin{minipage}{0.45\textwidth}
\includegraphics[width=4.3cm,clip=t,angle=90]{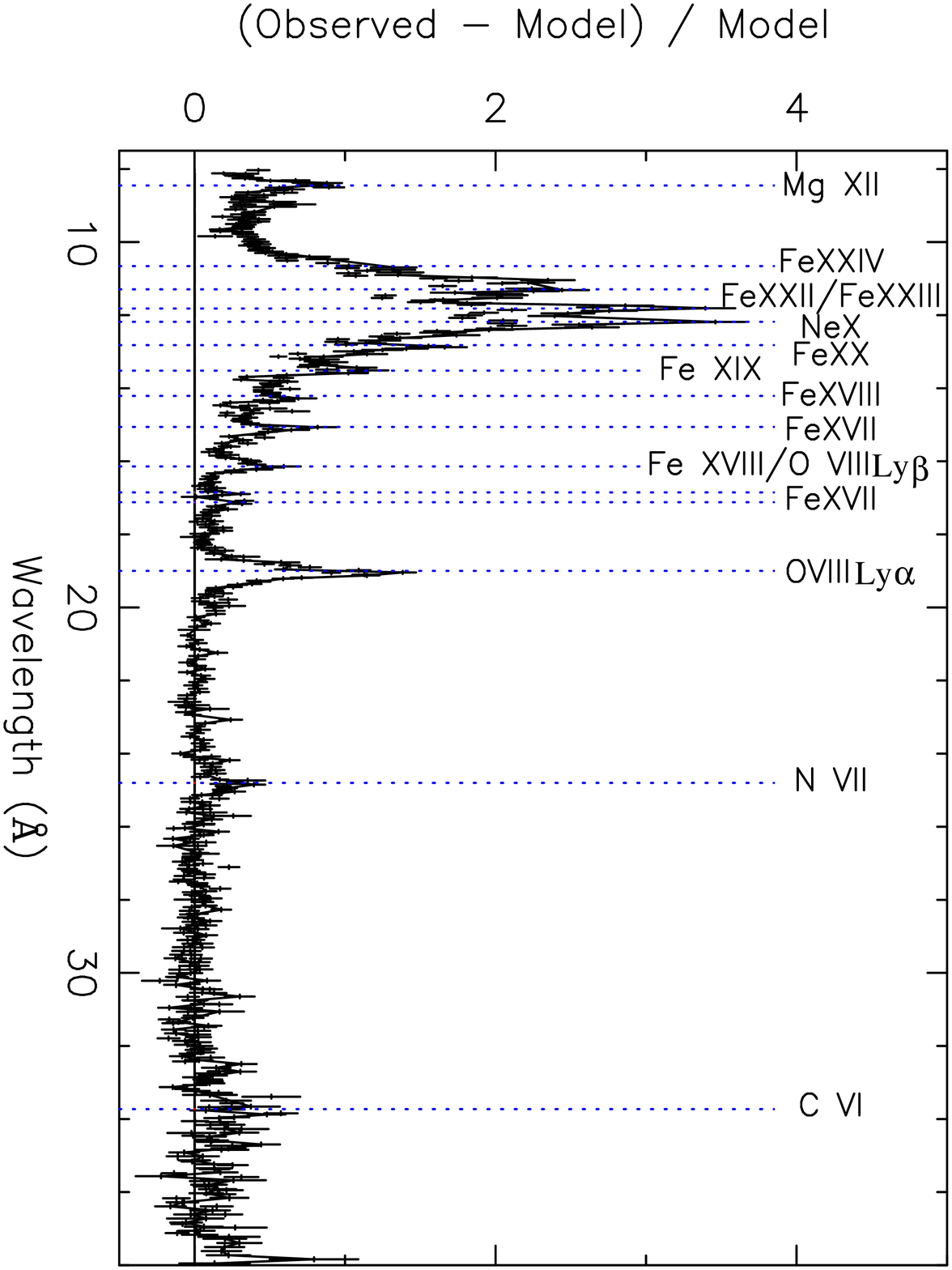}
\end{minipage}
\caption{{\emph{Left panel:}} The line spectrum of the cluster 2A~0335+096, as
  observed with {\emph{\xmm}} EPIC \citep[from][]{werner2006}. 
{\emph{Right panel:}} Line spectrum of M~87, as observed 
with {\emph{\xmm}} RGS \citep[from][]{werner2006b}. This representation of the spectrum 
has been obtained by recalculating
the spectrum with the best fit values without lines, and using the result
to subtract it from the observed spectrum
as indicated on the Y-axis. While the CCD spectra are plotted as 
a function of the observed energy, the grating spectra are shown as a function 
of the observed wavelength.}
\label{lines}
\end{center}  
\end{figure}

The CCD type detectors on \chandra, \xmm, and \suzaku\ allow us to detect the
emission lines of O, Ne, Mg, Si, S, Ar, Ca, Fe, and Ni (see the left panel of Fig.~\ref{lines}). 
The 2p--1s Ne lines at 1.02~keV are in the middle of the Fe-L complex (lying
between 0.8 to 1.4~keV). The energy resolution of the CCD type detectors is 
not sufficient to resolve the individual lines in the Fe-L 
complex, which makes the Ne abundance determination very sensitive to
uncertainties in the temperature modeling. However, the abundance of Ne in the
cores of cooling core clusters, observed with \xmm\ with 
sufficiently deep exposures, can be well determined using the high spectral
resolution of the Reflection Grating Spectrometers (RGS). The RGS has a higher 
spectral resolution than the CCDs. It resolves the Fe-L 
complex into individual lines and allows for accurate measurements of the
equivalent widths of O, Ne, Mg, and Fe. In the case of deep observations of
nearby bright, relatively cool clusters and elliptical galaxies 
even the spectral lines of C and N can be detected (see the right panel of Fig.~\ref{lines}). 
The high resolution transmission grating spectrometers on \chandra\  are because of their 
small effective area unfortunately not well suited for cluster spectroscopy.  

The equivalent widths of the observed lines are under the assumption of collisional 
equilibrium directly converted into abundances of the corresponding elements. Because the 
ICM is a relatively uncomplicated physical environment (ionization non-equilibrium or optical 
depth effects are minimal) these abundances are relatively robust.

Abundances are often shown in the literature with respect to the outdated Solar abundances by
\citet{anders1989} or with respect to newer sets of Solar and proto-solar abundances by \citet{lodders2003}.
The more recent Solar abundance determinations of O, Ne, and
Fe by \citet{lodders2003} are $\sim$30\% lower than those given by \citet{anders1989}.
The Solar abundances of O and Ne reported by \citet{grevesse1998} are higher and
the abundance of Fe is slightly lower than those reported by \citet{lodders2003}.

\subsection{Early pioneering work on chemical abundances}

The first evidence that the ICM is strongly polluted by metals ejected from stars in the cluster galaxies came with the discovery of the Fe-K line emission in the spectrum of the Perseus, Coma, and Virgo clusters by 
the {\it{Ariel~V}} and {\it{OSO-8}} satellites \citep{mitchell1976,serlemitsos1977}. Spectral analysis of samples of clusters observed with {\it{OSO-8}} and {\it{HEAO-1~A2}} showed that the ICM is enriched by Fe to 
0.3--0.5 of the Solar abundance value \citep{mushotzky1978,mushotzky1984}. 

Until the launch of \asca\ in 1993, only the Fe abundance was accurately measured in many clusters. {\it{ASCA}} for the first time detected emission features from many different elements (O, Ne, Mg, Si, S, Ar, Ca, 
Fe, and Ni) in the ICM, opening thus a new chapter in the chemical enrichment studies. While core collapse supernovae (\sncc) produce large amounts of O, Ne, and Mg, Type Ia supernovae (SN~Ia) produce 
large quantities of Fe, Ni, and Si-group elements (Si, S, Ar, and Ca), but only very little O, Ne, and Mg. Early \asca\ measurements of O, Ne, Si, S, and Fe abundances in four clusters suggested that the ICM 
enrichment is dominated by \sncc\ \citep{mushotzky1996}. 
However, few years later analyzing {\it{ASCA}} data, \citet{finoguenov2000} showed that while the \sncc\ products are uniformly distributed in the ICM, the chemical enrichment of the cluster cores is dominated by 
SN~Ia. An interesting result based on \asca\ by \citet{fukazawa1998} and \citet{baumgartner2005} showed that the Si abundance and the Si/Fe ratio increase from the poorer to the richer clusters, 
suggesting that the relative contribution of \sncc\ increases toward the more massive clusters. 
In a pioneering work \citet{dupke2000} used the abundances measured in three clusters of galaxies to put constraints on SN~Ia models. They used the Ni/Fe ratio to discriminate between different SN~Ia 
explosion scenarios available in the literature. Their work preferred the W7 ``deflagration'' models over the ``delayed detonation'' models. 

\begin{figure}    
\begin{center}
\includegraphics[width=9cm,clip=t, angle=0]{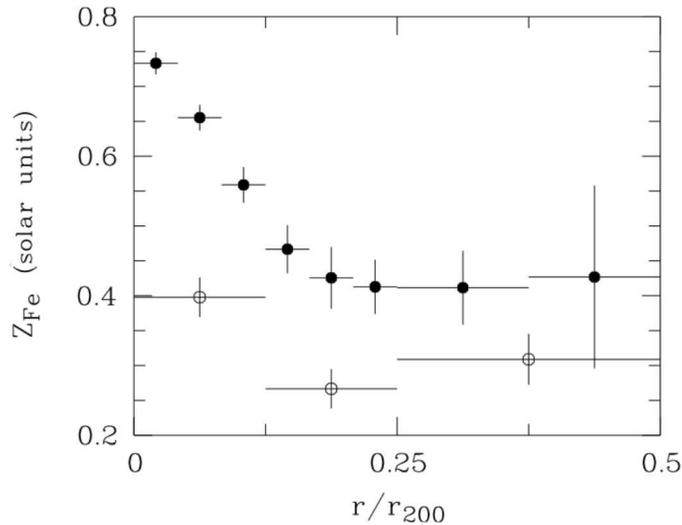}
\caption{The average radial distribution of the Fe abundance for a sample of cooling core (filled circles) and non-cooling core (empty 
circles) clusters observed with \beppo\ \citep[from][]{degrandi2004}.}
\label{degrandi}
\end{center}  
\end{figure}

The \asca\ data of the Centaurus cluster were the first to show a strong Fe abundance peak at the cooling core \citep{fukazawa1994}. \citet{allen1998} used \asca\ spectra and \rosat\ images to show a 
segregation between the metallicities of clusters with and without cooling cores and suggested that it is caused by the presence of metallicity gradients coupled with the sharply peaked surface brightness profiles 
of the cooling core clusters. Unfortunately, the large point-spread function of {\emph{ASCA}} did not allow to investigate the spatial abundance distributions in detail. Using the \beppo\ satellite, which 
had a better spatial resolution than \asca, \citet{degrandi2001} clearly showed that while non-cooling core clusters have flat Fe abundance profiles, the metallicity distribution of cooling core clusters has a 
gradient with a central peak (see Fig.~\ref{degrandi}). Abundance gradients were also found in 9 out of 10 galaxies and groups of galaxies by \citet{buote2000b}, who analyzed ROSAT data of these $\sim1$~keV systems.

\asca\ data were furthermore used for a pioneering search for evolution in cluster metallicities, revealing a lack of evolution out to redshift $z\sim0.4$ \citep{mushotzky1997,rizza1998} and finding no evidence for 
a decrease at higher redshifts \citep{donahue1999}.

\subsection{Chemical enrichment in cooling core clusters}

The emerging picture of the chemical enrichment in clusters of galaxies at the
time of the launch of \xmm\ and \chandra\ was that of an early enrichment of
the ICM by \sncc, the products of which are today well 
mixed and homogeneously distributed, and a subsequent more centrally peaked
enrichment by SN~Ia which continue to explode in the cD galaxy for a long time
after the cluster is formed. 
According to this picture the observed peaked distribution of the 
Fe abundance is largely due to the SN~Ia in the cD galaxy. 

The first abundance studies performed with \xmm\ seemed to confirm this scenario. The radial abundance profiles of the cooling core cluster Abell~496 showed a flat distribution of O, Ne, Mg, which are 
predominantly produced by \sncc, but a centrally peaked distribution of Si, S, Ar, Fe, and Ni \citep{tamura2001b}. However, while most of the Fe and Ni is produced by SN~Ia, Si, S, Ar, and Ca (so called $\alpha$ 
elements) are produced by both supernova types in similar proportions. In many systems, Si has a relatively well determined abundance value and in a cluster with a flat O profile one would expect to see a 
shallower Si abundance gradient than that of Fe. 

However, this has never been observed. In the most nearby cluster cooling core, M87, the Si abundance has a similar gradient as that of Fe, but at the same time it has a low O/Fe ratio in the core that increases 
toward the outer regions \citep{bohringer2001,finoguenov2002}. A similar abundance pattern was observed in the cluster sample analyzed by \citet[][]{tamura2004}, in the Perseus cluster \citep{sanders2004}, in 
Abell~85 \citep{durret2005}, in the Centaurus cluster \citep{matsushita2007b,sanders2006}, in the group NGC~5044 \citep{buote2003}, in Abell~1060 \citep{sato2007a}, and in AWM7 \citep{sato2007b}. 

To solve this discrepancy in M~87, \citet{finoguenov2002} proposed that there are two types of SN~Ia with different Si/Fe yields. Those with longer delay times and higher Si yields are today dominating the 
enrichment of elliptical galaxies, such as the cD galaxies in the centres of cooling core clusters, and those with shorter delay times and lower Si yields dominate the enrichment in the ICM at larger radii, 
because the bulk of them exploded in the time of the strongest star formation period (for more details on SN~Ia models see the Subsect.~\ref{supernovae}). 

Recently, \citet{simionescu2008b} determined radial abundance profiles for O,
Si, S, and Fe in the Hydra~A cluster. They found that all elements have a
centrally peaked distribution. In order to investigate the 
radial abundance distributions in clusters in more detail, they combined their
profiles with those determined for 5 other clusters with high quality X-ray
data (M87, Centaurus, Fornax, 2A~0335+096, S\'ersic~159-03, and Abell~1060). 
In the combined data set they find that the O distribution also peaks in the
cluster core and decreases with radius as $d$O/$d(\log{r}/r_{200}$)=-0.48$\pm$0.07. But the Fe 
abundance still decreases faster: $d$Fe/$d(\log{r}/r_{200}$)=-0.72$\pm$0.04,
and therefore the fraction of O with respect to Fe increases with radius as 
$d$(O/Fe)/$d(\log{r}/r_{200}$)=0.25$\pm$0.09. The Si abundance profile,
however, remains exactly the same as that of Fe, it decreases with radius as: 
$d$Si/$d(\log{r}/r_{200}$)=-0.72$\pm$0.04.

\begin{figure}    
\begin{center}
\includegraphics[width=9cm,clip=t]{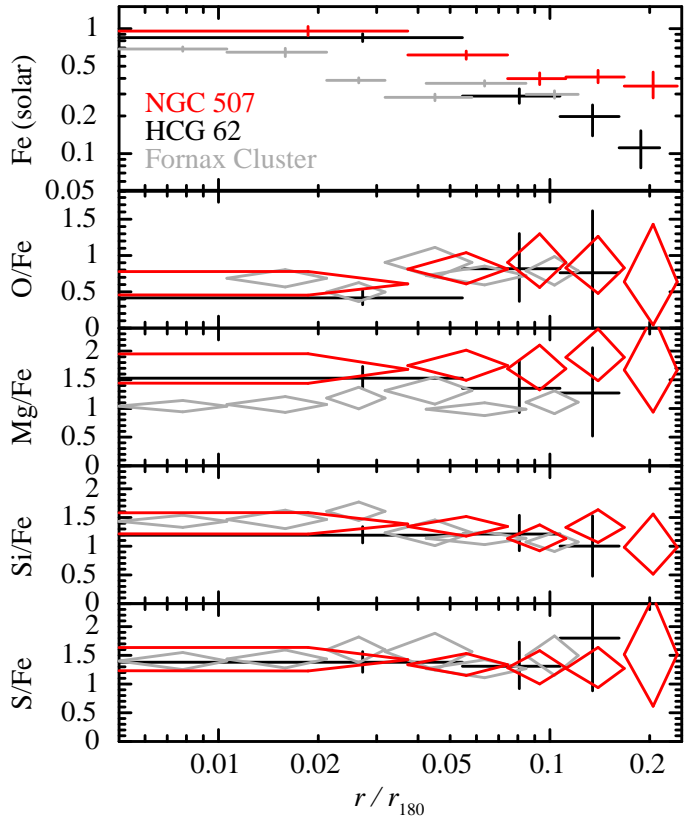}
\caption{Radial profile of abundance ratios for NGC~507 \citep{sato2008b}, HCG~62 \citep{tokoi2008}, and the Fornax cluster \citep{matsushita2007b}. All the abundance ratios are consistent with having a flat 
radial profile, except O/Fe which shows an indication for an increase with the radius. The abundance ratios are in the solar units of \citet{anders1989}. }
\label{sato}
\end{center}  
\end{figure}

Surprisingly, for the group NGC~507, \citet{sato2008b} found steep abundance gradients peaking on the cluster core not only for Si, S, and Fe but also for Mg. They also found a mildly centrally peaked 
abundance distribution for O. The {\it{Suzaku}} satellite has a better sensitivity in the O line energy and it allows for better measurements of the Mg lines in the low surface brightness cluster outskirts than \xmm, 
which has strong instrumental lines and effective area calibration problems at the Mg energy.
As shown in Fig.~\ref{sato} the Mg/Fe ratio as measured with \suzaku\ is consistent with being flat out to 0.2$r_{180}$ in Fornax \citep{matsushita2007b}, HCG~62 \citep{tokoi2008}, and in NGC~507 
\citep{sato2008b}. Since both Mg and O are produced by \sncc\ their radial profiles should follow each other. However, even though the error-bars are large, the O/Fe ratio seems to increase with the radius in 
these three systems. The different abundance profiles for Mg and O might be due to systematic uncertainties in regions with low surface brightness where the confusion with the strong \ion{O}{viii} Galactic 
foreground line becomes an issue. The most straightforward interpretation of these profiles is that the \sncc\ products are also centrally peaked, as indicated by Mg, and the measured peak in the O abundance 
profile is shallower because of the confusion with the Galactic \ion{O}{viii} foreground lines which become important in the faint cluster outskirts.

\begin{figure}    
\begin{center}
\includegraphics[width=9cm,clip=t]{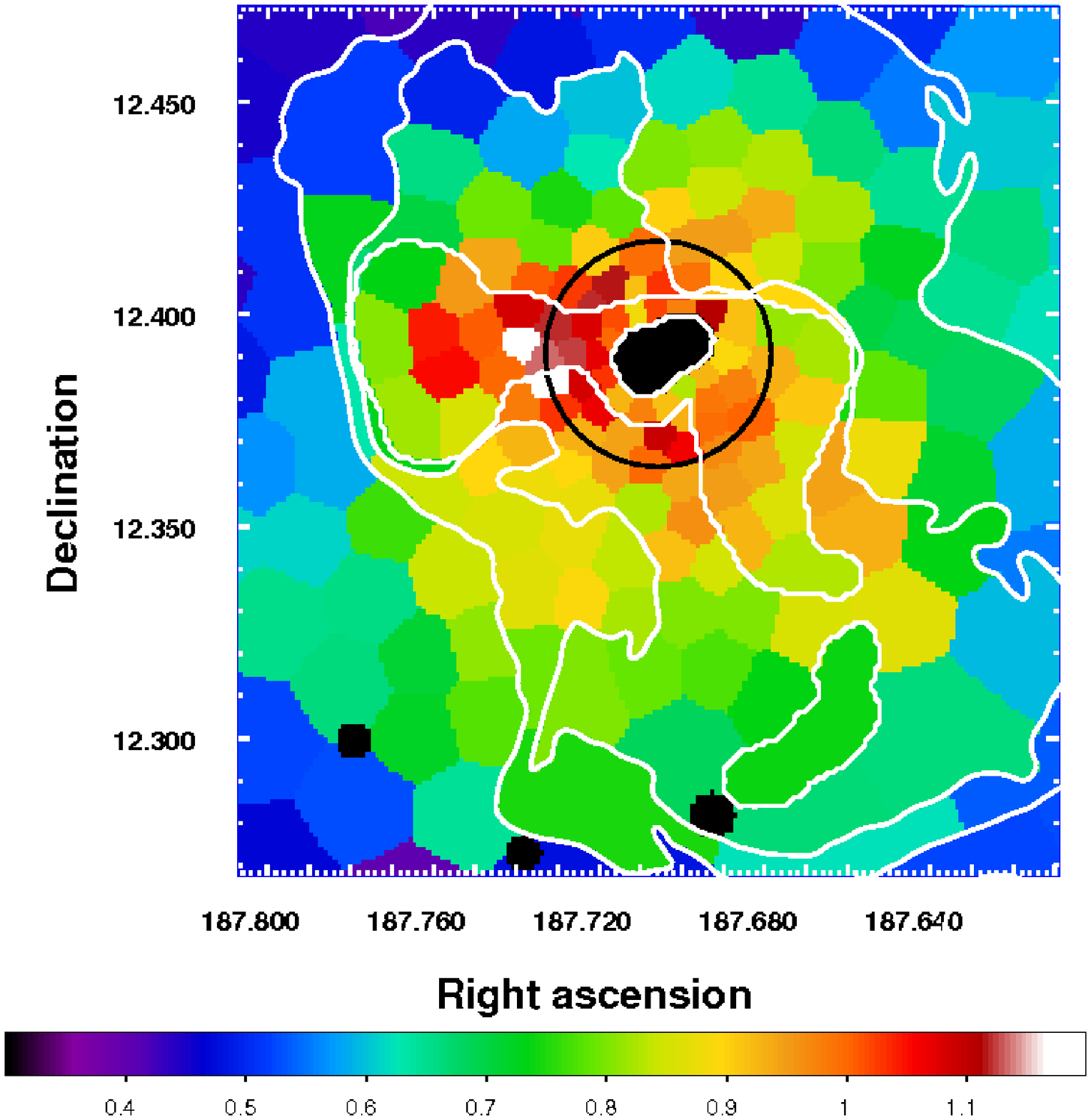}
\caption{Map of the Fe abundance in M~87 in Solar units indicated under the color bar. Contours of the 90 cm
 radio emission \citep{owen2000} are overplotted. Beyond the expected radial gradient, 
one clearly sees the enhanced Fe abundance in the radio arms, especially
within the Eastern arm \citep[from][]{simionescu2008a}.}
\label{M87metals}
\end{center}  
\end{figure}

Based on the Fe abundance profiles, \citet{degrandi2004} and \citet{bohringer2004} concluded that the central abundance peak is due to the metals released by the stellar population of the cD galaxy. To 
produce the observed abundance peaks requires long ($\gtrsim$5~Gyr) enrichment times \citep{bohringer2004}. 
In the absence of mixing the metallicity profiles should follow the optical light profiles of the cD galaxies. But the observed metal mass profiles are much less peaked than the optical light, which implies that the 
metals get mixed and transported out to larger radii, most probably by the AGN/ICM interaction \citep[e.g.][]{rebusco2005,rebusco2006}. 

The role of the AGN in the metal transport was both theoretically and observationally best investigated in M~87. \citet{churazov2001} proposed a scenario where buoyant bubbles of radio emitting relativistic 
plasma, produced by the jets of the AGN, uplift the cool metal rich gas from the central parts of the cD galaxy. This scenario was confirmed by observations of cool metal rich gas, distributed  along the radio lobes 
\citep[see Fig.~\ref{M87metals},][]{simionescu2008a}. The Fe abundance of the uplifted gas (a mixture of ICM, stellar mass loss, and SN~Ia products) is $\sim$2.2 times solar and its mass is $\approx$
$5\times10^8$~$M_{\odot}$. It takes $\approx$30--110~Myr to produce the metals in the cool gas. The relative abundances of the uplifted gas are consistent with those of the ambient ICM, indicating that the 
dominant fraction of metals in the cluster core was uplifted after the last major epoch of star formation and the relative enrichment rate by SN~Ia with respect to that by stellar winds remains fairly constant with time. 
Uplift of metal rich cool gas from the cD galaxy by buoyantly rising bubbles of radio emitting plasma was also observed in the Hydra~A cluster of galaxies \citep{simionescu2008b}. The total mass of the uplifted Fe 
in Hydra~A is $1.7\times10^7$~$M_{\odot}$.

Can the abundance peaks be produced by only stellar winds and SN~Ia in the cD galaxy? To explain the centrally peaked distribution of O, Si, and S in the Hydra~A cluster would require the amount of metals 
produced by stellar winds to be 3--8 times higher than predicted by models or the metals produced by \sncc\ in the proto-cluster phase not to be mixed completely \citep{simionescu2008b}. 
Centrally peaked distribution of all elements can also be produced by ram-pressure stripping of cluster galaxies \citep[e.g.][]{domainko2006}. But both simulations and observations indicate that galaxies are 
getting stripped already at large distances from cluster cores and this mechanism probably does not contribute significantly to the observed abundance peaks on the relatively small spatial scales around the cD 
galaxies. 

The radial metallicity gradients in cooling flow clusters often display an inversion in the center \citep[e.g.][]{sanders2002}. While an apparent drop in metal abundances can be the result of an oversimplified model 
in the spectral analysis of the very center of a cooling core \citep[e.g.][for more discussion about spectral fitting biases see Subsect.~\ref{biases}]{buote1998,molendi2001,werner2006,matsushita2007b} in some cases this inversion seems robust. In particular the metallicity in the Perseus cluster drops very clearly at radii smaller than 40~kpc and the central dip does not disappear when multi-temperature models and extra power-law components are introduced in the spectral fitting, or when projection effects and possible effects of resonant scattering are taken into account \citep{sanders2004,sanders2006,sanders2007}. The metallicity dips in the cluster cores may be related to the AGN/ICM feedback.

\subsection{Constraining supernova models using clusters
\label{supernovae}}

\begin{figure}    
\begin{center}
\includegraphics[width=9cm,clip=t, angle=0]{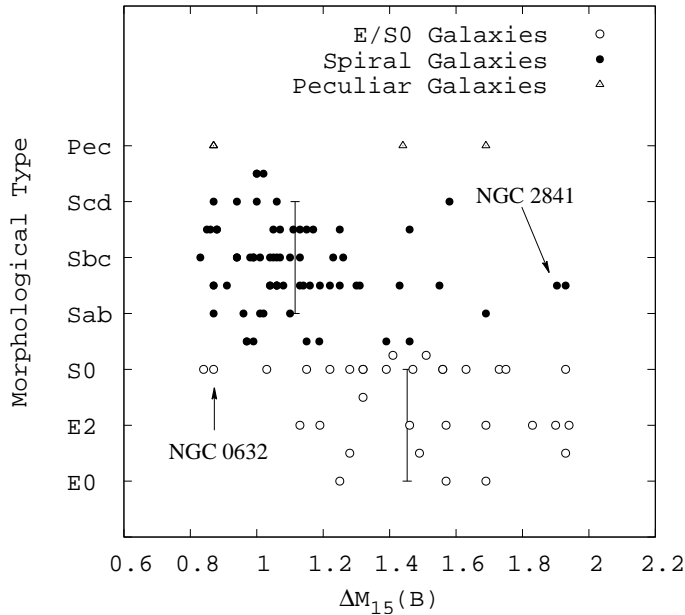}
\caption{A compilation of the morphological type of the host galaxy versus the 
decline rate of SNe~Ia. NGC~2841 has spectral features of an elliptical 
and NGC~0632 has emission lines characteristic for late-type 
galaxies. The vertical markers indicate the average decline rates for 
early- and late-type galaxies. From \citet{gallagher2005}.}
\label{gallagher}
\end{center}  
\end{figure}

As we mentioned before, clusters of galaxies retain all the metals produced by their stellar populations within their gravitational potential. The dominant fraction of these metals resides within the hot ICM 
\citep{finoguenov2003}. All the elements from oxygen up to the iron group are primarily produced by supernovae. Their abundances within the ICM thus provide us with the integral yield of all the supernova 
explosions in the cluster. 

Since the launch of the \asca\ satellite there have been efforts to use the abundance patterns in the ICM to put constraints on the theoretical supernova models. These efforts were mainly focused on constraining 
the Type Ia explosion mechanisms. \citet{dupke2000} and \citet{dupke2001} used the Ni/Fe ratio measured in clusters by \asca\ to distinguish between the SN~Ia models assuming slow deflagration and delayed 
detonation. While the deflagration explosion mechanism of SN~Ia \citep[represented by the W7 and W70 models in the literature,][]{iwamoto1999} predicts high Ni/Fe ratio of 2.18 to 3.22 in the Solar units of 
\citet{grevesse1998}, the delayed-detonation explosion scenarios predict significantly lower Ni/Fe ratios of 0.9--1.4 Solar. They are represented by the WDD1, WDD2, WDD3, CDD1, and CDD2 models in the 
literature, where the last digit indicates the density at which the flame velocity becomes supersonic (deflagration-to-detonation transition density) in units of $10^7$~g~cm$^{-3}$; the ``C'' and ``W'' refer to two 
different central densities (1.37$\times$ and $2.12\times10^9$~g~cm$^{-3}$, respectively) in the model at the onset of the thermonuclear runaway \citep{iwamoto1999}. 
\asca\ measured relatively high Ni/Fe ratios which clearly favored the W7 SN~Ia model \citep[][see also Sect. 5.5.1]{dupke2000,dupke2001}.

The results based on the abundances measured by \xmm, \chandra, and \suzaku\ are not so clear and depending on the relative abundances in the given cluster sometimes deflagration models and sometimes 
delayed detonation models are favored.

An excellent example of the complicated situation is the Virgo cluster with M87 in its cooling core.
The abundance patterns in M~87 cannot be described by any combination of \sncc\ and deflagration SN~Ia models. The low Ni/Fe ratio and very high Si/Fe ratios clearly favor delayed detonation models. The 
high SN~Ia yield of Si-group elements in M87 imply that the Si burning in SN~Ia is incomplete, as predicted by the delayed-detonation models with lower density of deflagration to denotation transition (WDD1 
model). The abundance patterns in the outer region of M~87 are on the other hand characteristic of deflagration supernova models or delayed detonation models with the highest deflagration to detonation 
density (WDD3). \citet{finoguenov2002} conclude that these abundance patterns confirm the diversity of SN~Ia. Today, this conclusion seems to be supported by both optical supernova surveys and more recent 
measurements of cluster abundances.

Optical surveys show a correlation between SN~Ia properties and the morphology of their host galaxy. While a population of brighter SN~Ia with a slow luminosity decline is more common in late-type spiral and 
irregular galaxies with recent star formation (indicating a short delay time between their formation and the explosion), a fainter and more rapidly decaying population of SN~Ia is more common in early-type 
galaxies \citep{hamuy1996,ivanov2000}. This is clearly seen in the Fig.~\ref{gallagher} from \citet{gallagher2005} which shows the morphological type of the parent galaxy versus the decline rate of SNe~Ia. 
SNe~Ia in galaxies whose populations have a characteristic age greater than 5 Gyr are $\sim$1~mag fainter than those found in galaxies with younger populations.
Based on the evolution of the SN~Ia rate with redshift, its correlation with the colour and radio luminosity of the parent galaxies, \citet{mannucci2006} concluded that the delay time distribution of SN~Ia is bimodal. 
About half of the SN~Ia explode $\approx10^{8}$ years after the formation of the progenitor binary system and the delay time distribution of the other half of SN~Ia could be described by an exponential function 
with a decay time of $\sim$3~Gyr. This diversity should be reflected in the abundance yields. The brighter supernovae are expected to produce more Ni and less $\alpha$ elements (Si, S, Ar, Ca) than the fainter 
ones.  

\begin{figure}    
\begin{center}
\begin{minipage}{0.4\columnwidth}
\includegraphics[width=0.81\columnwidth,angle=270]{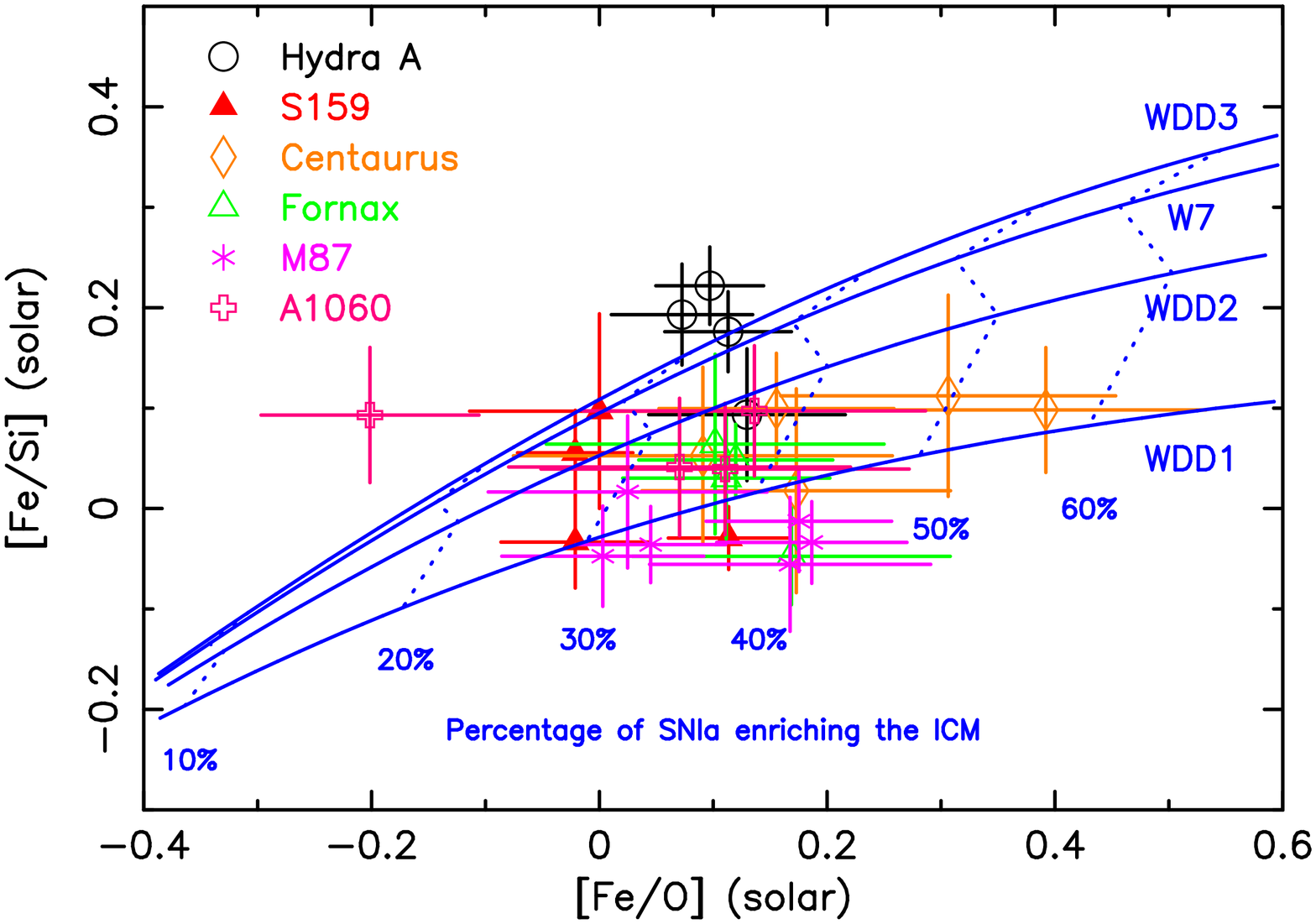}
\end{minipage}
\hspace{1cm}
\begin{minipage}{0.4\columnwidth}
\includegraphics[width=1.1\columnwidth,angle=0]{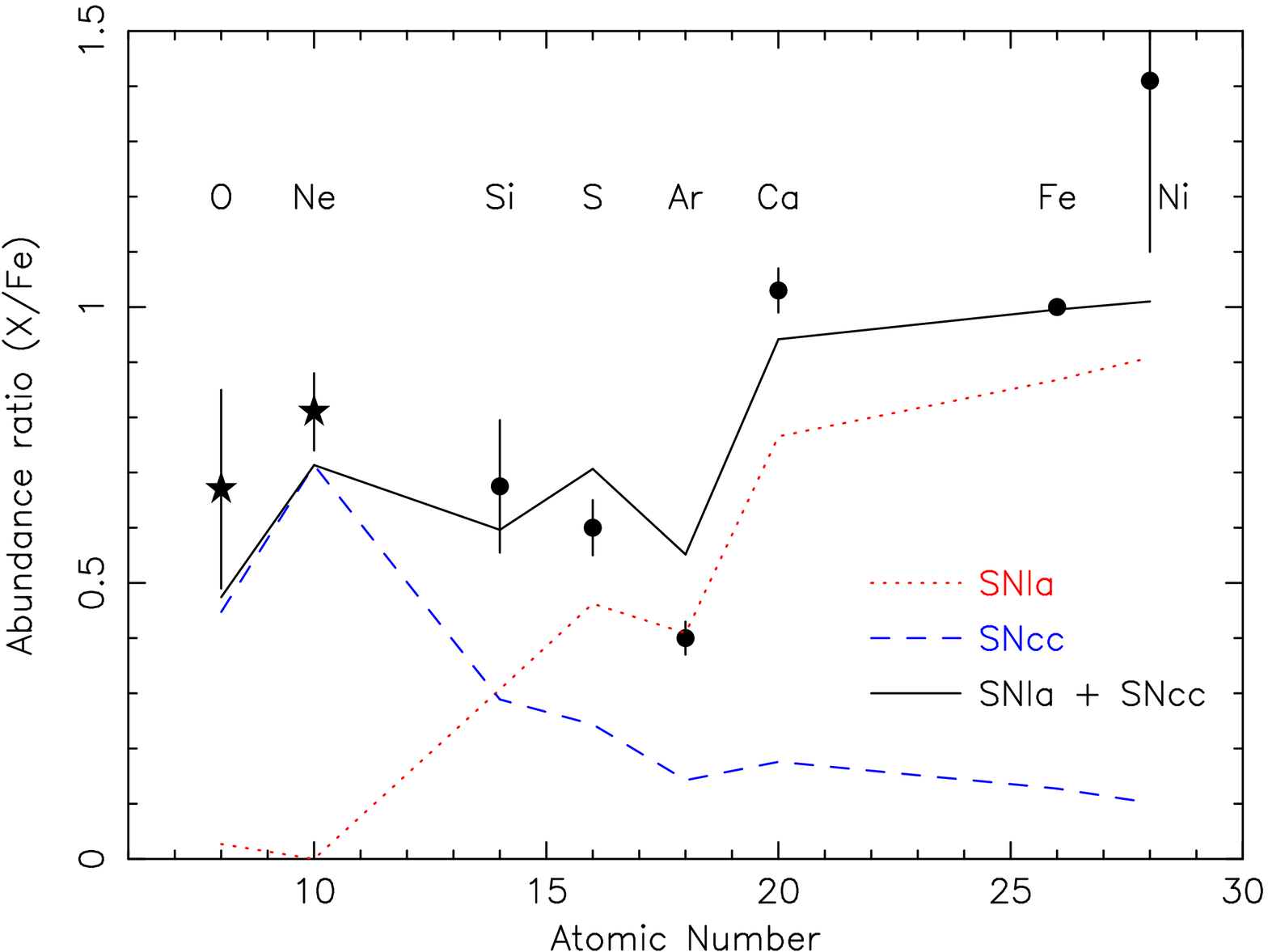}
\end{minipage}
\caption{Intrinsic variation of cluster abundances and their average yields.
{\it{Left panel: }}The observed [Fe/Si] versus [Fe/O] for different spatial 
regions of a sample of six clusters of galaxies. The curves indicate models 
of a mixture of \sncc\ products assuming a Salpeter initial mass 
function \citep{tsujimoto1995} mixed with different SN~Ia products \citep{iwamoto1999}. 
The dotted line connects points of the same number contributions of 
SN~Ia to the enrichment of the ICM. The abundance patterns in the Hydra 
cluster and in M~87 favor very different SN~Ia models. \citep[After][]{simionescu2008b}.
{\it{Right panel: }} Reconstruction of the average abundance patterns 
observed in a sample of 22 clusters of galaxies with theoretical supernova 
yields \citep[from][]{deplaa2007}. The model describing the data 
the best is a combination of SN~Ia yields from a 1D delayed detonation model by  
\citet{badenes2006}  and \sncc\ yields for progenitors with Solar metallicities by 
\citet{nomoto2006} integrated over the Salpeter IMF \citep[from][]{deplaa2007}.}
\label{fig:supernovae}
\end{center}  
\end{figure}

Fig.~\ref{fig:supernovae} shows the log$_{10}$(Fe/Si) against the log$_{10}$(Fe/O) for different spatial regions in clusters of galaxies Hydra~A \citep{simionescu2008b}, S\'ersic~159-03 \citep{deplaa2006}, 
Centaurus \citep{matsushita2007b}, Fornax \citep{matsushita2007}, M~87 \citep{matsushita2003}, and A1060 \citep{sato2007a}. 
The plot shows a relatively large spread in the abundance patterns determined for different clusters. In particular the differences in the Fe/Si ratio between Hydra~A and M~87 are striking. Each overplotted line in 
the figure shows a mixture of yields of a different SN~Ia model with \sncc\ yields by \citet{tsujimoto1995} integrated over the Salpeter IMF. Using \sncc\ yields by different authors, assuming different progenitor pre-
enrichment levels, or integrating the \sncc\ yields over a different IMF would move these curves \citep[see Fig.~11 in][]{simionescu2008b}. But an intrinsic variation in the metallicity or IMF of the \sncc\ progenitors 
alone could not explain the observed variety of abundance patterns. Thus it seems that, indeed, a variety of SN~Ia flavors with different densities at the transition from the subsonic to supersonic flame velocities is 
needed. 

While the ratio of the $\alpha$ elements to Fe can be different for different SNe~Ia, the ratio between the $\alpha$ elements is very similar for all explosion mechanisms, because the fuel (C/O) is always the same 
and the burning conditions where the $\alpha$ elements are synthesized do not vary much in SN~Ia. Therefore the relative abundances of the $\alpha$ elements, especially the Ca/Ar ratio, are the best 
diagnostic of the quality of SN~Ia models. 
\citet{deplaa2007} analysed the abundance patterns of Si, S, Ar, Ca, Fe, and Ni in a sample of 22 clusters observed with \xmm\ within $0.2R_{500}$. Because the SN$_{\mathrm{CC}}$ do not have a significant 
impact on the Ar/Ca ratio, they used the abundances of these elements to find the best fitting SN~Ia model. Delayed detonation models clearly provide a better fit to the data than the deflagration scenarios, but 
none of the supernova models by \citet{iwamoto1999} fit the Ar and Ca abundances. \citet{deplaa2007} obtained a good fit with a one-dimensional delayed-detonation SN~Ia model, calculated on a grid 
introduced in \citet{badenes2003}, with a deflagration-to-detonation density of $2.2\times10^7$~g~cm$^{-3}$ and kinetic energy of $1.16\times10^{51}$~ergs, which was shown to fit best the properties of the Tycho 
supernova remnant \citep{badenes2006}. 

Here it is interesting to note that Tycho seems to be a remnant of a fairly typical SN~Ia with an average luminosity and the data points in the left panel of Fig.~\ref{fig:supernovae} also tend to be centred around the WDD2 model 
which is the closest to the best fitting model of \citet{badenes2006}. Several other observational clues point toward the delayed detonation models. Optical light curves \citep{woosley2007} and spectra 
\citep{mazzali2007}, the X-ray emission of SN~Ia remnants \citep{badenes2006,badenes2008} all show that the structure of the SN~Ia ejecta has to be very close to what one-dimensional delayed detonation 
models predict: most of the white dwarf is burnt, Fe-peak nuclei stay in the inner ejecta and the intermediate mass elements stay in the outer ejecta. The variation of the peak brightness, which correlates with the 
production of $^{56}$Ni and anti-correlates with the production of Si-group elements, can also be explained in the framework of the delayed detonation models by a variation of the deflagration-to-detonation 
transition density (transition from subsonic to supersonic flame velocities).

As shown by \citet{deplaa2007} the relative abundances of chemical elements in the ICM can, in principle, be also used to distinguish between \sncc\ models with different level of pre-enrichment of the progenitors and with different initial-mass functions (IMF). Comparing the measured abundances with the \sncc\ yields predicted by the model of \citet{nomoto2006}, \citet{deplaa2007} found that the dominant fraction of \sncc\ progenitors contributing to the enrichment of the ICM was already enriched and their mass distribution was close to the Salpeter IMF. The main constraints on the \sncc\ progenitors are based on the measured O/Ne ratio in the ICM. 

While elements from O up to the Fe-group are primarily produced in supernovae, carbon and nitrogen are believed to originate from a wide range of sources: winds of short-lived massive metal rich stars, longer lived low- and intermediate-mass stars, and an early generation of massive stars \citep[e.g.][]{gustafsson1999,chiappini2003,meynet2002}. The ICM abundance of C and N was measured only in a few nearby bright elliptical galaxies and clusters \citep{peterson2003,werner2006b}. The small O/Fe ratio and large C/Fe and N/Fe found in M~87 by \citet{werner2006b} suggest that the main sources of C and N are not the massive stars that also produce large quantities of O, but the low- and intermediate-mass asymptotic giant branch stars. 



\subsection{Metallicity evolution and differences between clusters}

As we mentioned before, cluster data obtained by \asca\ were consistent with no evolution in the ICM metallicity out to redshift $z\sim0.4$. More recent observations with \chandra, however, indicate a significant 
evolution \citep{balestra2007,maughan2008}. \citet{balestra2007}, who investigated the Fe abundance evolution on a sample of 56 clusters, found that while in the redshift range of $z\simeq0.3$--$0.5$ the 
average ICM Fe abundance is $\simeq0.4$~Solar, above redshift $z\sim0.5$ the metallicity drops to $\simeq0.25$ Solar. \citet{maughan2008} looked at a larger sample of 116 clusters at $0.1<z<1.3$ in the 
{\emph{Chandra}} archive and essentially confirmed the results of \citet{balestra2007}. They found that the abundances drop by $\sim$50\% between $z\sim0.1$ and $z\sim1$ and the evolution is still present if 
the cluster cores (the inner 0.15R$_{500}$) are excluded from the analysis. This result indicates that the abundance drop is not due to the lack of strong cool cores at large redshifts. 

\citet{balestra2007} also found a trend of the Fe abundance with the cluster temperature. Within (0.15--0.3)$R_{\mathrm{vir}}$ in clusters below 5~keV the Fe abundance is on average a factor of $\sim$2 larger 
than in the hotter clusters. The Fe abundance values measured within 0.2$R_{500}$ for the sample of low redshift 22 clusters analyzed by \citet{deplaa2007} show the same trend with the cluster temperature: 
while in hot massive clusters (k$T\gtrsim5$~keV) the Fe abundance seems to be constant and
equal to $\sim0.3$ solar, for cooler clusters, in the temperature range of $2$--$4$~keV, the Fe abundance shows a range of values between $0.2$--$0.9$ solar. 
This trend is probably linked to the changing stellar mass over gas mass ratio in clusters. While the gas mass fraction was found to, in general, increase with the cluster temperature with a large spread of values 
especially below 5~keV \citep{sanderson2003}, the gas fraction in clusters above $>5$~keV is constant with the ICM temperature \citep{allen2008}. The Fe mass over light ratio seems to be constant as a function 
of temperature \citep[e.g.][]{renzini1997}.  
\citet{lin2003} find that the stellar mass fraction decreases by a factor of 1.8 from low- to high- mass clusters, and the ICM to stellar mass ratio increases from 5.9 to 10.4. This trend suggests a decrease of star 
formation efficiency with increasing cluster mass and provides a natural explanation for the on average lower iron abundance in the more massive clusters. The large range of Fe abundance values in the clusters 
with lower temperatures might be due to the large spread in the gas mass fraction of the lower mass clusters and due to the range of enrichment times and gas densities in the cool cores of clusters. 

\citet{fukazawa1998} and \citet{baumgartner2005} showed based on \asca\ data an increase of the Si/Fe ratio from 0.7 to 3 Solar in the 2--8 keV temperature range. This result was not confirmed by later \xmm\ observations. The abundance ratios of Si/Fe, S/Fe, Ca/Fe, and Ni/Fe measured within 0.2$r_{500}$ in the sample of 22 clusters with temperatures in the range of 1.5--10~keV analyzed by 
\citet{deplaa2007} are consistent with being constant as a function of temperature. The intrinsic spread in the abundance ratios is smaller than 30\%. 
This result shows that the abundance patterns do not change with the observed drop of star formation efficiency with the cluster mass and the ratios of SN~Ia and \sncc\ contributing to the ICM enrichment are 
similar in clusters of all masses \footnote{The increase of the Si/Fe ratio with the cluster temperature observed by \asca\ might have been due to a bias introduced by the strong temperature gradients in the hot 
clusters with cooling cores. If the temperature of the best fit plasma model (the average cluster temperature) is higher than the temperature of the cooler gas with high Si line emissivity, then the model will naturally 
predict a higher Si abundance than the actual value. }.

The observed drop of the Fe abundance with increasing redshift could be at least partly connected to the metallicity-temperature relation. The clusters observed at high redshift are primarily massive clusters 
which eventually evolve into hot clusters like those with lower abundances at low redshifts. The evolution of the metallicity below redshift $z\sim1$ therefore might be substantially weaker than the observations 
suggest.

\subsection{How good are the assumptions behind the abundance measurements?}

The chemical abundances in the ICM are determined assuming a certain plasma model, a certain ICM temperature structure, optically thin plasma, and solar abundances for elements lighter than O or N. How 
good are these assumptions and how can they bias our abundance determinations?

\subsubsection{How much bias due to spectral modeling?}
\label{biases}

Usually two plasma models are used for fitting cluster spectra: MEKAL \citep{mewe1985,mewe1986,kaastra1992,liedahl1995} and APEC \citep{smith2001}. A more updated version of MEKAL compared to the one available in XSPEC is implemented in the SPEX spectral fitting package \citep{kaastra1996}. MEKAL and APEC treat differently the atomic physics and 
differences are also in their line libraries. But despite small differences in the best fit temperature structure, the abundances derived with the two plasma codes are generally consistent within the errors 
\citep{buote2003,sanders2006,deplaa2007}. The two plasma codes thus do not show any systematic inconsistencies. The implemented ionization balance for Ni is inaccurate and can significantly bias the Ni abundance measurements; an update is planned shortly for SPEX (J.S. Kaastra, private communication).

One has to be careful and keep in mind possible biases when interpreting 
the results of measurements of chemical abundances in clusters. The most serious 
biases arise when the temperature structure in the extraction region is 
not modeled properly. Historically, X-ray observations showed significantly 
subsolar abundances in galaxies and groups of galaxies. 
\citet{buote1994} and \citet{buote2000} showed that a low resolution 
spectrum which is composed of intrinsically two temperature components 
with similar emission measures, one below 1~keV and one above 
1~keV, with an average temperature of about 1~keV fitted with a single 
temperature model will seriously bias the Fe abundance determination. 
Since the lower temperature component preferentially excites 
emission lines  in the Fe-L complex below 1~keV (\ion{Fe}{xii--xxi}) 
and the higher temperature component excites the Fe-L lines in the 
range of $\sim$1--1.4~keV (\ion{Fe}{xxi--xxiv}) the shape of the Fe-L 
complex will be flatter than that produced by a single-temperature model 
with the average temperature of the two components (see also Fig.~\ref{fig:18}). The single-temperature 
model can fit the flat shape of the Fe-L complex only with a 
relatively low Fe abundance and higher bremsstrahlung continuum. The 
obtained Fe abundance is thus much lower than the actual value (see the left panel of Fig.~\ref{helium}). 

In the past few years several authors \citep[e.g.][]{buote2003,werner2006,matsushita2007b} 
showed that in the complex cool cores of groups and clusters the obtained 
abundances of essentially all elements increase when two-temperature models 
or more complicated differential emission measure (DEM) models are used instead 
of simple single-temperature models. 

For a multi-temperature plasma with average temperatures of $\sim$2--4~keV, where both Fe-L and Fe-K emission lines are seen with similar statistics, fitted with a single temperature model an ``inverse Fe bias'' 
is observed. \citet{simionescu2008b} point out that in a multi-temperature plasma at these temperatures the cool component produces stronger Fe-L lines and the hotter component emits more Fe-K emission than 
that expected for the single-temperature plasma with the average temperature. Therefore, by fitting a single temperature model to such plasma one will obtain a higher Fe abundance than the actual value \citep[see also][]{rasia2008}. 

These results show that for chemical abundance studies in the ICM 
a good description of the temperature structure is crucial. Accurate 
abundance determinations require deep observations which allow to test 
different differential emission measure distributions to obtain 
the best description of the true temperature structure in the cluster cores.

\subsubsection{Possible biases due to helium sedimentation?}

\begin{figure}    
\begin{center}
\begin{minipage}{0.4\columnwidth}
\hspace{-1.5cm}
\includegraphics[width=1.0\columnwidth,angle=0]{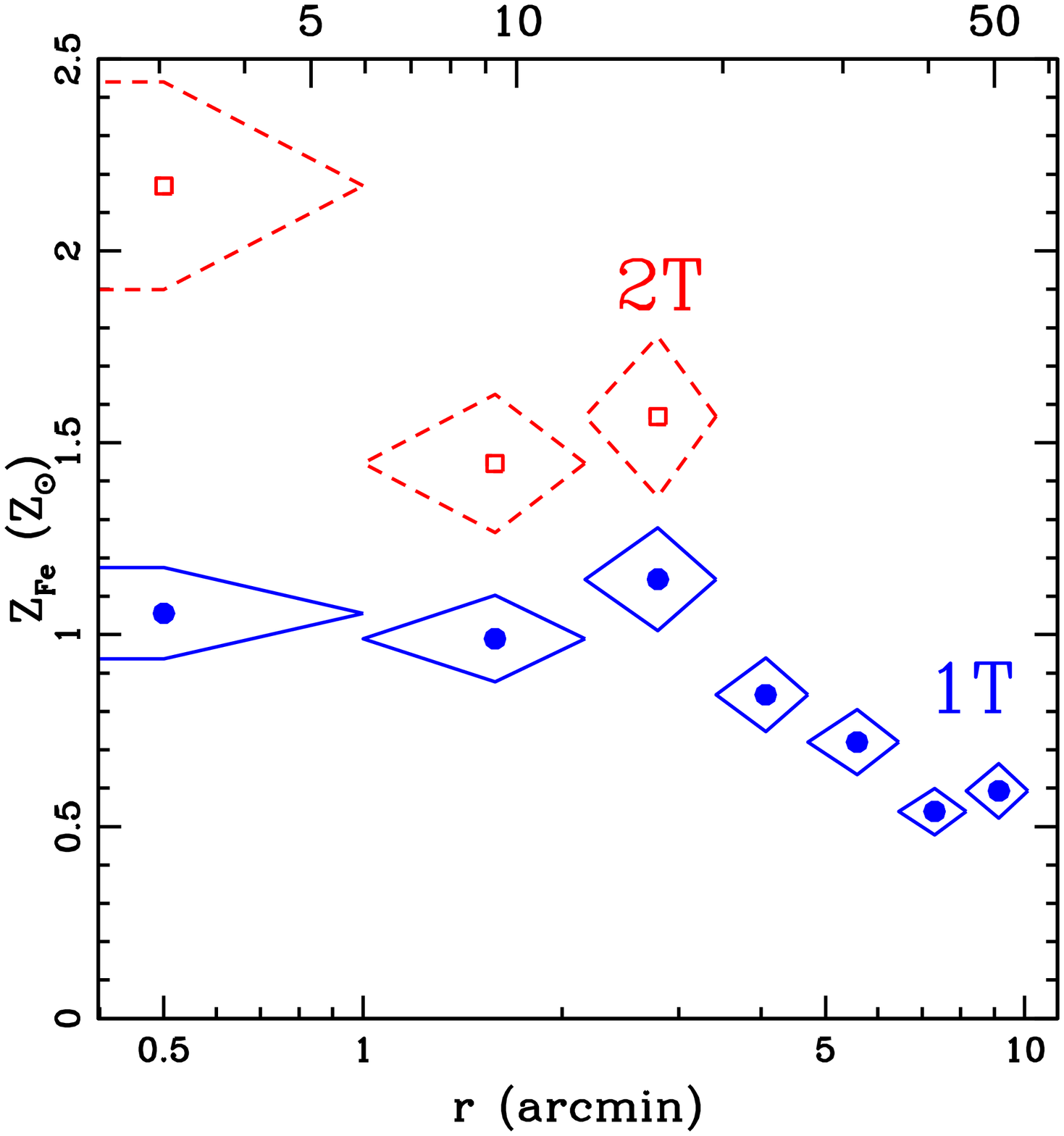}
\end{minipage}
\hspace{-1cm}
\begin{minipage}{0.4\columnwidth}
\includegraphics[width=1.0\columnwidth,angle=270]{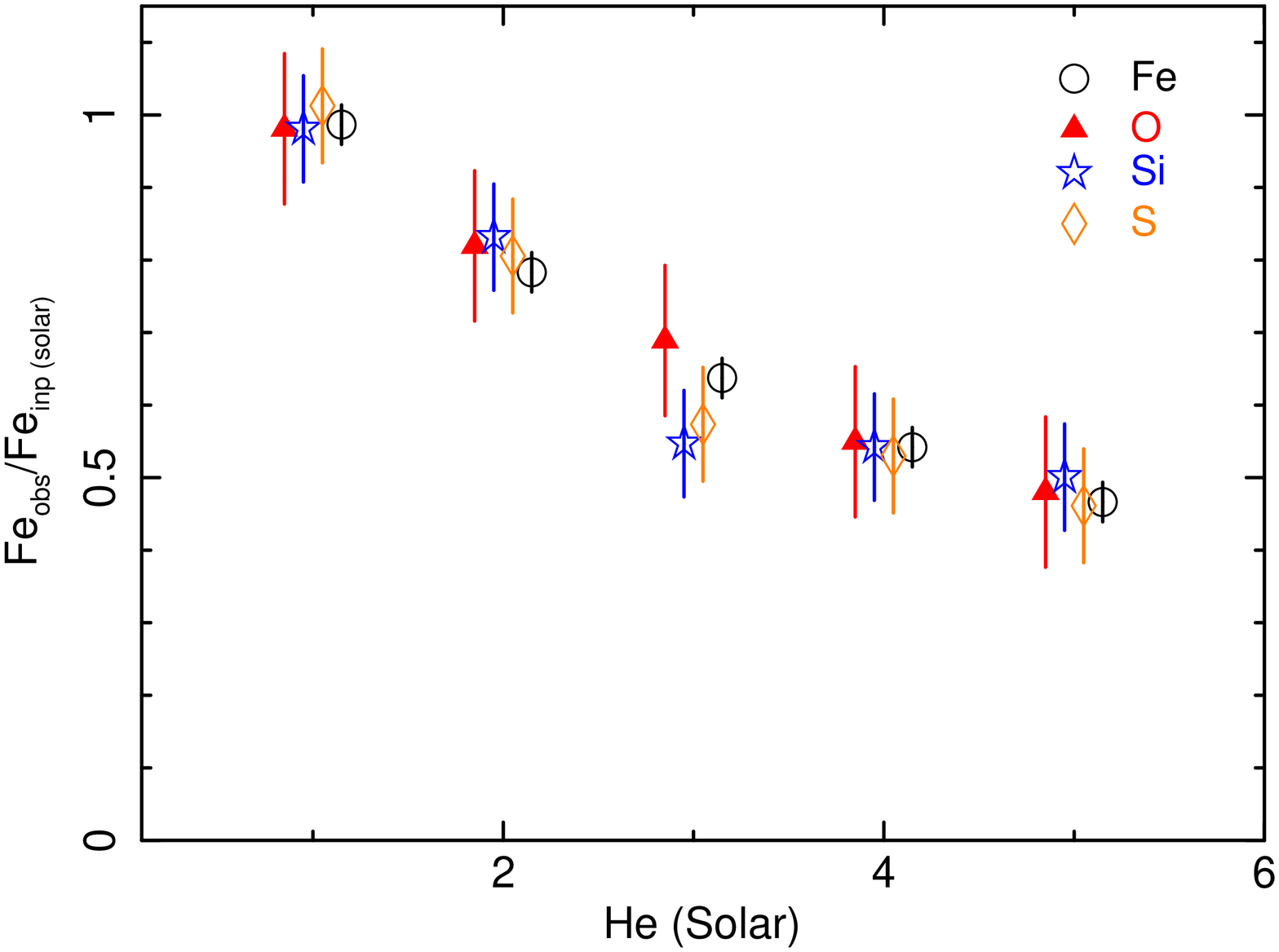}
\end{minipage}
\caption{{\it{Left panel: }}The influence of the temperature model on the best fit Fe abundance as seen in the radial Fe abundance distribution in NGC~1399 modeled with single- and two-temperature models 
\citep[from][]{buote2002}. {\it{Right panel: }}The helium abundance is usually assumed to be Solar in clusters. If the helium abundance in cluster cores is increased by sedimentation and we do not take it into 
account in the spectral modeling, our best fit abundances will be lower than the real values. The effect is indicated in the plot for simulated spectra of a 4~keV cluster. The simulation was made assuming a 100~ks 
observation with \xmm~EPIC and a circular extraction region with a radius of 1\arcmin. }
\label{helium}
\end{center}  
\end{figure}

In the gravitational potential of clusters helium and other metals can diffuse toward the cluster center \citep{fabian1977,abramopoulos1981,gilfanov1984}. The sedimentation is more efficient for the lighter 
elements and in a H-He plasma, the helium is expected to be more centrally peaked than hydrogen. The helium abundance can not be directly measured. The cluster spectra are commonly modeled under the 
assumption of solar or primordial He abundance values in the ICM. In the inner regions of galaxy clusters where the effects of sedimentation are the most important, the underestimation of the amount of helium 
can cause inaccurate determination of gas densities and metallicities. 

\citet{peng2008} solved a set of flow equations for a H-He plasma for clusters assuming 
temperature profiles by \citet{vikhlinin2006}. They showed that at 0.06$r_{500}$ 
of a 11 Gyr massive (10~keV) cluster the 
He/H abundance ratio can be as high as 4.3 Solar. These predictions are, 
however, made for unmagnetized plasma assuming no turbulent suppression of 
particle diffusion and no mixing in the ICM, therefore 
they should be treated as upper limits. \citet{ettori2006} pointed out that 
for an actual helium abundance of $\sim$4 times the value assumed in the 
spectral modeling, would cause an under-determination of the 
Fe abundance by a factor of 2. As we show in the righ panel of Fig.~\ref{helium} the same is 
approximately true for the other elements as well.

\subsubsection{Resonance scattering in cooling cores?}

For the strongest resonant lines the ICM can be moderately optically thick
\citep{gilfanov1987} and as we already mentioned in Sect.~3.5
resonant scattering at the energies of these lines changes the observed line intensities.
This can affect the measurements of elemental abundances in the dense cores of clusters.

\citet{sanders2006} constructed a model that takes into account resonant scattering for
thousands of resonance lines. They tested their model on \chandra\ data of two clusters with
central abundance dips. They conclude that metallicities in cluster cores could be underestimated
due to resonant scattering by at most 10\%.  Resonance scattering can therefore not be
the cause for the observed central abundance dips.

In order to determine the amount of resonant scattering \citet{werner2009} studied the
line ratios of the \ion{Fe}{xvii} ion in 5 elliptical galaxies (see Sect.~3.5).
They concluded that neglecting the effects of resonant scattering in the spectral fitting of
the inner 2~kpc core of NGC~4636 will lead to underestimates of the chemical abundances of Fe
and O by $\sim10$--20\%.

\section{Beyond Thermal Equilibrium}
\label{sec:6}


The intra-cluster medium is usually assumed to be in collisional ionization 
equilibrium (CIE). Plasma in CIE is optically thin for its own radiation and 
external radiation fields do not affect the ionization balance 
which is entirely determined by the temperature of the plasma. The ionization and 
recombination rates in CIE plasma have come to a balance and the electron and 
ion temperatures are in equilibrium. The energy 
distribution of the electrons in the ICM is usually described by a Maxwellian. 

These assumptions are mostly justified but there are certain situations when the plasma deviates from the thermal equilibrium. The shock heated plasma in the cluster outskirts with densities $n_{\mathrm{e}}
\lesssim10^{-4}$~cm$^{-3}$ is likely to be out of ionization balance. The equilibration time scale at those low densities is longer than the age of the cluster 
and therefore the gas may still ionize. At the lowest 
density parts of warm-hot intergalactic medium filaments photoionization by the diffuse radiation field of galaxies and the cosmic background becomes important. Deviations from a Maxwellian electron distribution 
could occur in the cooling cores of clusters where multiple gas phases are present. If electrons diffuse from a hotter into a cooler phase they over-ionize the gas. Plasma could also be out of equilibrium in the 
vicinity of shocks, which may alter the electron/ion equilibrium. Shocks could also accelerate electrons and produce non-Maxwellian tails in the ICM electron distribution. Future observatories with large photon 
collecting area and high spectral resolution will be able to study the spectral signature of these non-equilibrium effects in detail. Moreover, radio observations of clusters reveal the presence of relativistic electrons 
emitting synchrotron radio emission. This relativistic plasma in clusters may also emit Inverse-Compton radiation in the
form of a power law shaped X-ray continuum.

Below we briefly describe what are the spectral signatures of non-equilibrium effects 
in the multiphase plasma and around shocks, and we summarize the state of observations 
of X-ray emission from relativistic plasma.

\subsection{Multiphase plasma}

High signal-to-noise X-ray spectra reveal that multiple spatially unresolved gas phases are present in the cooling cores of clusters \citep{kaastra2004,sanders2007,simionescu2008a}. In M87 and Perseus, the 
most nearby and the brightest cooling cores, images obtained by \chandra\ reveal filaments and blobs in soft X-rays \citep{forman2006,fabian2006}. But the multi-phase extends to much lower temperatures. 
Narrow band optical observations of cooling cores reveal H$\alpha$ emitting warm ($10^4$~K) gas which both in M87 and in Perseus is often associated with soft X-ray filaments \citep{sparks2004,fabian2006}. 
The H$\alpha$ filaments in Perseus are composed of gas with temperatures of $10^4$~K down to 50~K \citep{hatch2005}. \citet{sanders2007} showed that they are surrounded and possibly mixed with soft X-ray 
emitting plasma with temperatures of 0.5--1~keV. 

\begin{figure}    
\begin{center}
\includegraphics[width=7.5cm,clip=t, angle=270]{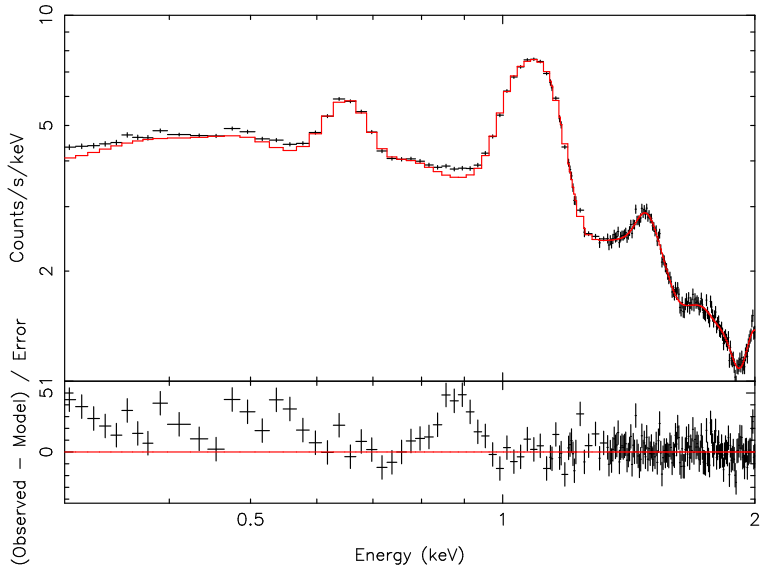}
\caption{Simulated EPIC/pn spectrum of the Eastern X-ray arm of M87 assuming a presence 
of a filament with a diameter of 60~pc and length of 3.5~kpc 
filled with $10^4$~K gas with an average density of 
$n_e=10$~cm$^{-3}$. We assumed that 0.05\% of all the electrons within the 
filament penetrated there from the ambient ICM and have a Maxwellian 
distribution with a temperature of 2~keV. The model plotted 
with the red line shows the emission from the ambient hot ICM. Assuming a 
well calibrated instrument the excess line emission at $\sim$0.85~keV would be clearly revealed. }
\label{multiphase}
\end{center}  
\end{figure}

If electrons penetrate from the hotter to the cooler phase they over-ionize the ``cool'' plasma which may result in observable low energy X-ray line emission. 
In Fig.~\ref{multiphase} we show a simulated EPIC/pn spectrum (100~ks) of the Eastern X-ray arm of M87. Within this extraction region, we assume the presence of a filament with a diameter of 60~pc and length 
of 3.5~kpc filled with $10^4$~K gas with an average density of $n_e=10$~cm$^{-3}$. We assume that 0.05\% of all the electrons within the filament have penetrated there from the ambient ICM and have a 
Maxwellian energy distribution with a temperature of 2~keV. As the spectrum in Fig.~\ref{multiphase} shows, a deep observation with a well calibrated instrument would even at medium spectral resolution reveal 
the line emission from \ion{Ne}{iii-vi} at 0.85--0.9 keV, \ion{O}{iv--vii} at $\sim$0.55~keV, \ion{N}{iii--vi} between 0.40--0.42~keV, \ion{C}{iv--v} at 0.35, 0.37~keV, and at 0.30--0.31~keV. This exercise demonstrates 
the potential of X-ray spectroscopy in constraining the diffusion of electrons from the hot ICM phase into the warm filaments. In reality such attempts are hampered by the complicated multi-temperature structure 
of the ICM around the filaments. Detailed studies of the electron diffusion will be possible with future micro-calorimeters or transition-edge sensors sensitive at low energies.

\subsection{Shocks and non-equilibrium}

In the downstream regions of shock fronts the electron and ion temperatures of the plasma might be different and the ICM is out of ionization equilibrium. 

In a collisional plasma, protons are heated dissipatively at the shock layer which 
has approximately the width of the collisional mean free path, but the faster moving electrons do not feel the shock (for Mach 
numbers smaller than $\sim40$), they get adiabatically compressed and equilibrate with the ions via Coulomb collisions reaching the post-shock temperature predicted by the Rankine-Hugoniot jump condition. 
Astrophysical shocks in a magnetized plasma, like the ICM, are expected to be ``collisionless''. Which means that the electron and ion temperature jump occurs on a spatial scale much smaller than the collisional 
mean free path, but ions and electrons are heated by the shock to different temperatures. In heliospheric shocks with moderate Mach numbers electrons are heated much less than protons, barely above the 
adiabatic compression temperature \citep{schwartz1988}. The plasma behind the shocks is thus expected to be in electron/ion temperature non-equilibrium. Assuming that no kinetic energy goes into acceleration 
of cosmic rays, the post-shock temperature after the Coulomb equilibration time scale reaches the same temperature as the one predicted by the Rankine-Hugoniot jump condition. 

\begin{figure}    
\begin{center}
\begin{minipage}{0.4\columnwidth}
\hspace{-1.5cm}
\includegraphics[width=1.2\columnwidth,angle=0]{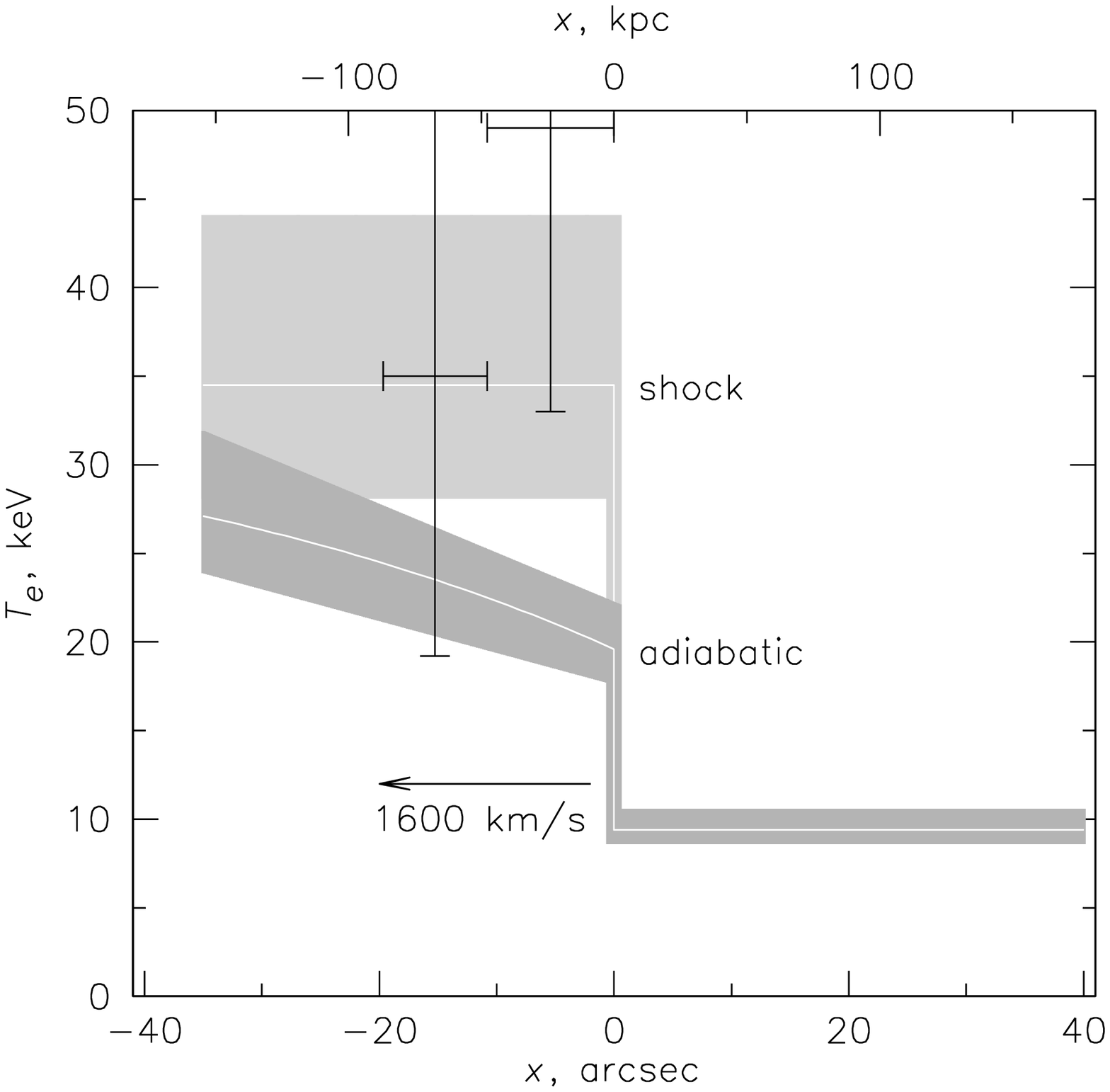}
\end{minipage}
\hspace{-1cm}
\begin{minipage}{0.4\columnwidth}
\includegraphics[width=1.0\columnwidth,angle=270]{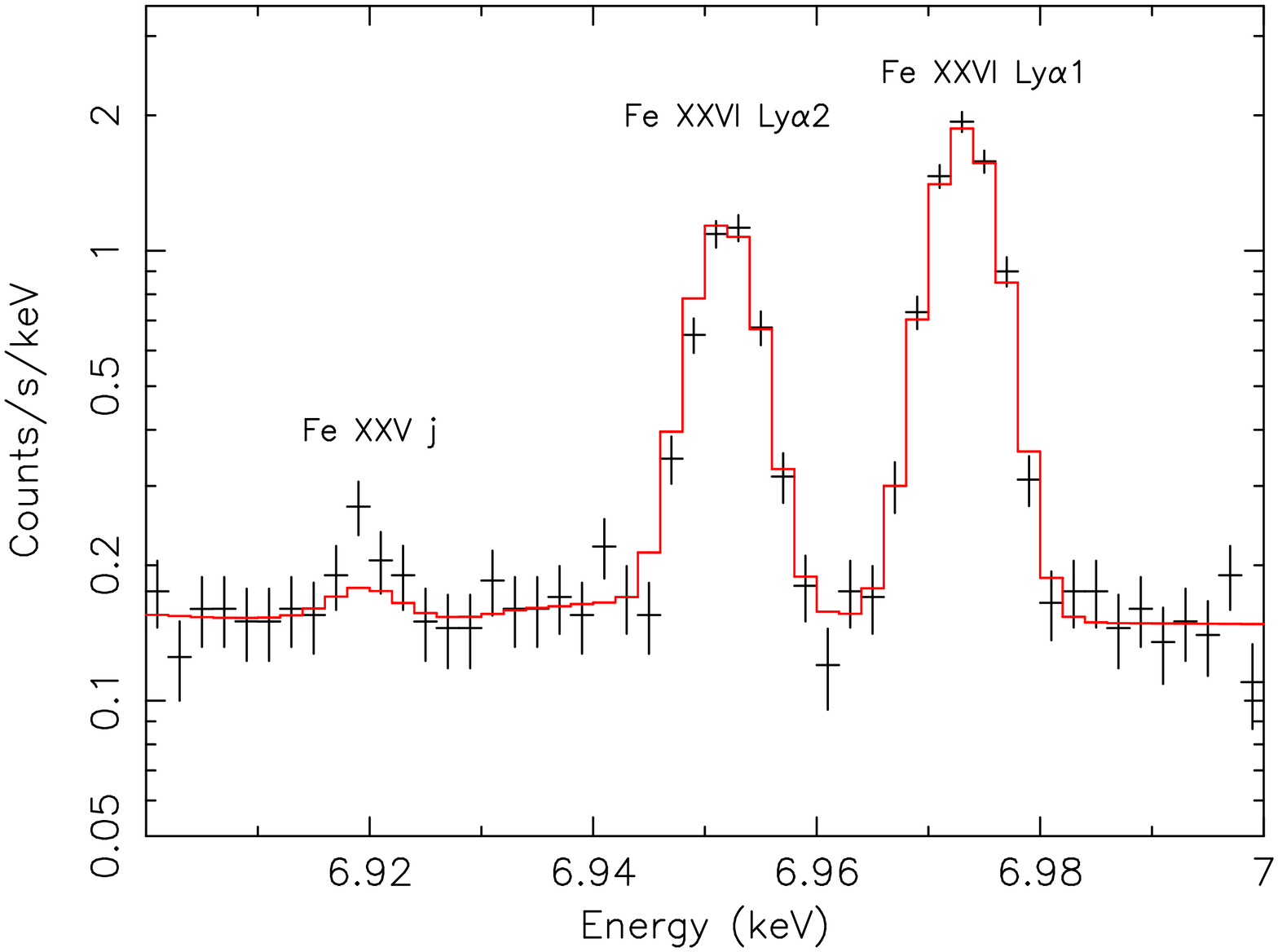}
\end{minipage}
\caption{\leftp\ Electron-ion equilibration in 1E~0657-56 (the ``Bullet cluster'').
The two data points show the deprojected electron temperature for two 
narrow post shock regions. The data points are overlaid on the model 
predictions for instant equilibration (light gray) and for adiabatic 
compression followed by collisional equilibration (dark gray). The velocity 
of the post-shock gas relative to the shock front is $\sim 1600$~km~s
$^{-1}$. \citep[From][]{markevitch2006}. \rightp\ The 6.9--7.0~keV part 
(rest frame energies) of a simulated {\it{Astro-H}} micro-calorimeter spectrum 
of the downstream region of a $M=2.2$ shock propagating through a $10^8$~K plasma. 
The \ion{Fe}{xxvj} satellite line is clearly stronger (with a significance 
of 3$\sigma$ for these simulations) than that predicted by the 
thermal model with a Maxwellian electron distribution indicated by the full line. 
\citep[From][]{kaastra2009c}.}
\label{maxim}
\end{center}  
\end{figure}

\citet{markevitch2006} used the bow shock in the ``Bullet cluster'' (1E~0657-56) to determine whether the electrons in the intra-cluster plasma are heated directly at the shock to the equilibrium temperature 
(instant-equilibration), or whether they equilibrate via Coulomb collisions with protons. They  measure the temperature in front of the shock front seen as a gas density jump, from which, using the adiabatic and 
Rankine-Hugoniot jump conditions, they determine the expected post shock adiabatic and instant-equilibration electron temperatures which they directly compare with observed data. The shock in the ``Bullet 
cluster'' is happening approximately in the plane of the sky, which means that the downstream velocity of the shocked gas spreads out the time dependence of the electron temperature in the plane of the sky. The 
measured temperatures behind the shock in the bullet cluster are consistent with instant heating of electrons. Adiabatic compression followed by equilibration on the collisional time scale is excluded at the 95\% 
confidence level (see Fig.~\ref{maxim}).

As the temperature of the plasma suddenly rises at a shock front, 
it gets out of ionization balance. While the ionization state of ions 
still reflects the pre-shock temperature of the plasma after the instant heating, the 
electron temperature is higher. The plasma is under-ionized compared to 
the equilibrium case and the ionization balance must be recovered by 
collisions. Until the ionization balance recovers there will be more 
ionization in the plasma than recombination. The exact ionization state 
of shocked plasma in non-equilibrium conditions depends on 
its temperature, density, and the time since it has been shocked. Such 
situation is often seen in supernova remnants. 

Shocks and turbulence in the magnetized ICM are believed to be sides of non-thermal 
particle acceleration in clusters \citep[for reviews on particle acceleration 
see e.g.][]{petrosian2008,bykov2008}. Stochastic 
turbulence in the ICM can create a non-thermal tail to the Maxwellian 
electron distribution \citep{bykov1999}. The low energy end 
of the power-law electron distribution enhances the ionization rates and modifies 
the degree of the ionization of the plasma. \citet{porquet2001} 
showed that the plasma is always more ionized for hybrid (Maxwellian plus power-law) 
electron distribution than for a Maxwellian distribution and the 
mean charge of a given element at a given temperature is increased. 
The effect is more pronounced at lower temperatures. In groups of galaxies, 
low mass clusters or in cooling cores a peculiar ionization state of 
Fe which is inconsistent with the temperature determined from continuum 
emission can potentially be a good tool to reveal such 
non-thermal tails in the electron distribution. 
\citet{kaastra2009c} show that a non-Maxwellian tail in 
the electron distribution behind a shock front propagating through 
$10^8$ K plasma cannot be revealed by current detectors. 
They simulate the electron distribution and the resulting 
X-ray spectra of the downstream region of a $M=2.2$ shock in a 
X-ray bright hot cluster and show that a good indicator of hard 
non-thermal electrons is the enhancement of the equivalent widths 
of satellite lines which may be possible to detect with the X-ray 
micro-calorimeters on the future {\it{Astro-H}} satellite or on 
the proposed {\it{International X-ray Observatory (IXO)}}. 
In the right panel of Fig.~\ref{maxim}, we show the 6.9--7.0~keV 
part (rest frame energies) of a simulated 100 ks {\it{Astro-H}} 
spectrum of the downstream region of a $M=2.2$ shock propagating through 
the ICM in a cluster similar to Abell~2029. The \ion{Fe}{xxvi} Ly$\alpha$ lines 
and the \ion{Fe}{xxvj} satellite line are visible in the spectrum and the 
satellite line is clearly stronger than that predicted by the thermal model 
with a Maxwellian electron distribution indicated by the full line.

\subsection{Non-thermal X-ray emission from relativistic plasma}

Direct evidence for the presence of electrons accelerated up to relativistic 
energies comes from observations of radio emission in clusters. 
Large-scale diffuse extended radio emission in the form of halos or relics has 
been observed in about 50 known clusters of galaxies \citep{feretti2007}. 
This radiation is associated with the ICM and has no 
connection to the cluster galaxies. It is clearly produced by synchrotron 
emission by a population of relativistic electrons in the ICM. 
Since the energy density of the cosmic microwave background (CMB) radiation 
($\sim4\times10^{-13}$~erg~cm$^{-3}$) is higher than the energy density in 
the intra-cluster magnetic field ($3\times10^{-14}(B/\mu G)^2$~erg~cm$^{-3}$), 
these relativistic electrons will radiate away most of 
their energy via inverse-Compton scattering of the CMB photons, producing a 
power-law shaped X-ray continuum emission \citep{rephaeli1977}. The detection of 
both the non-thermal X-ray emission and the 
radio emission would be a powerful tool, which would allow us to determine both 
the volume averaged magnetic field in clusters and the energy in the population of 
relativistic electrons. But for the observed radio 
fluxes, detectable hard X-ray emission can only be produced if the magnetic fields 
are of the order of 0.1~$\mu$G. Faraday rotation measurements indicate that the 
intra-cluster magnetic fields are much stronger - 
of the order of $1$--$10$~$\mu$G. The volume averaged magnetic field might, 
however, be weaker than the strong magnetic field measured by Faraday rotation along our line of sight. 

Non-thermal X-ray emission could in principle also originate in non-thermal 
brems\-strahlung of shock accelerated supra-thermal electrons in the ICM. 
However, such non-thermal brems\-strahlung phase must be very short lived. 
\citet{petrosian2001} points out that in order to explain the reported hard 
X-ray luminosity in the Coma cluster (see later) by 
bremsstrahlung emission,  the continuous input of energy into the ICM 
would increase the ICM temperature to $10^{10}$~K in Hubble time. Also 
the high-energy electrons $>50$~keV can not be confined by the 
gravitational potential of the cluster and will escape in a crossing time 
of $<1.5\times10^7$ years, unless they are confined by the intra-cluster 
magnetic field. Alternatively, hard X-ray emission with power-law 
spectra of $\Gamma\sim1.5$ could be produced by synchrotron emission from 
ultra-relativistic electrons and positrons produced by the interaction of 
relativistic protons with the CMB \citep{inoue2005}. 

The nearest hot massive cluster of galaxies with a big radio halo is the Coma 
cluster \citep{feretti1998}. Therefore, Coma has been the primary target to identify 
non-thermal X-rays. Their detection has been 
reported with both \beppo\ \citep{fusco1999} and {\it{RXTE}} \citep{rephaeli2002}. 
More recently, Coma has also been observed with the {\it{INTEGRAL}} satellite \citep{eckert2007}. 
A combined analysis of  \xmm\ 
and {\it{INTEGRAL}} data revealed the presence of hotter ($\sim$12~keV) gas in 
the south-west region overlapping with a radio halo. This hot gas was probably 
heated in a merger \citep{eckert2007}. 
Analysing {\it{INTEGRAL}}, {\it{ROSAT}}, and {\it{RXTE}} data, \citet{lutovinov2008} 
found that the global Coma spectrum is well fitted with a thermal model and the 
evidence for a hard excess is very marginal 
(1.6$\sigma$). \citet{ajello2008} analysed combined \xmm\ and hard X-ray {\it{Swift-BAT}} 
spectra and even though they could not rule out the presence of non-thermal 
hard X-ray emission outside of their 
10\arcmin\ region, within their field of view they found a good fit with two 
thermal models, with no need for a non-thermal component. \citet{wik2009} 
analyzed combined Suzaku Hard X-ray Detector (HXD-PIN) 
and \xmm\ mosaic spectra of a larger 34\arcmin$\times$34\arcmin\  region centered 
on the Coma cluster. They found no statistically significant evidence for non-thermal 
emission implying a lower limit of 
0.15~$\mu$G on the cluster averaged magnetic field. 

Suzaku observations of the bright radio relic in the merging cluster Abell~3667 
show that, at least in some cases, magnetic fields can be high even at distances of 
$\sim1$~Mpc from the cluster core. The upper 
limits on the non-thermal emission from Suzaku put a lower limit of 2.3~$\mu$G 
on the magnetic field in the radio relic. The non-thermal energy density in the 
relics is $>$7\% of the thermal energy density and 
likely near 20\% \citep{nakazawa2009}.  


\citet{ajello2008} analysed the spectra of 10 clusters of galaxies detected above 
15~keV with the {\it{Swift-BAT}}. Except of the Perseus cluster 
the spectrum of which is probably contaminated by hard X-ray 
emission from the central AGN, the spectra of all clusters are well fitted 
by a simple thermal model. Their stacked spectrum of 8 clusters 
(except Perseus and Coma) also confirms the absence of any non-thermal 
high energy component down to a flux of $1.9\times10^{-12}$~erg~cm$^{-2}$s$^{-1}$ 
in the 50--100~keV band. 

Although there have been several other reports claiming the detection of both hard and 
soft X-ray emission in excess to the ICM emission, interpreted as being possibly of 
non-thermal origin \citep[for reviews see][]{rephaeli2008, durret2008}, a solid 
identification of non-thermal X-rays from clusters is still lacking.  
In the relatively near future, X-ray satellites with imaging hard X-ray optics 
sensitive up to 40~keV will be launched: Astro-H, NuSTAR, Simbol-X. They will shed 
more light on the presence of very hot thermal gas, non-thermal and supra-thermal 
electron populations, and on intra-cluster magnetic fields. The {\it{Fermi}} Gamma-
ray Large Array Space Telescope (GLAST) might detect gamma ray emission associated 
with relativistic intra-cluster ions. These missions promise a big 
progress in our understanding of the non-thermal particle 
population in clusters of galaxies.



\section{Future Potential of X-ray Spectroscopy}
\label{sec:7}

One of the most awaited features of the future of X-ray observatories 
is high resolution spectroscopy of the next generation of imaging devices
based on calorimetric measurements of the photon energy in detectors cooled
to few tens of mK. The two missions for which such detectors are envisioned
and actually play a major role in the scope of these missions are Astro-H in Japan
(with a possible launch in 2013\footnote{http://astro-h.isas.jaxa.jp/}) and
the International X-ray Observatory (\ixo) currently discussed by ESA, NASA,
and JAXA as a mission for about 2018--2021\footnote{http://ixo.gsfc.nasa.gov/}.
While Astro-H \citep{takahashi2009} is already being built, 
the \ixo\ mission has not yet been fully approved
and is still in the first phase of technical studies. Nevertheless, to give
an outlook on the coming spectroscopic capabilities of these future X-ray missions, 
we will use the current prospective response files for the \ixo\ mission with the 
planned so called narrow field  detectors (NFI), which provide high resolution spectroscopy 
of about 2--3 eV energy resolution, to illustrate the prospects. 
In addition to providing high resolution spectroscopy
\ixo\ will also significantly increase the photon collecting power compared to 
\xmm\, with a proposed collecting area around 1 keV of about 3 m$^2$ to 
provide a better photon statistics to fully exploit the higher spectral 
resolution statistically. It will also increase
the redshift range of cluster studies providing a detailed picture
of galaxy cluster astrophysics up to redshifts of 2.

\begin{figure}
\begin{center}
  \includegraphics[height=6.6cm]{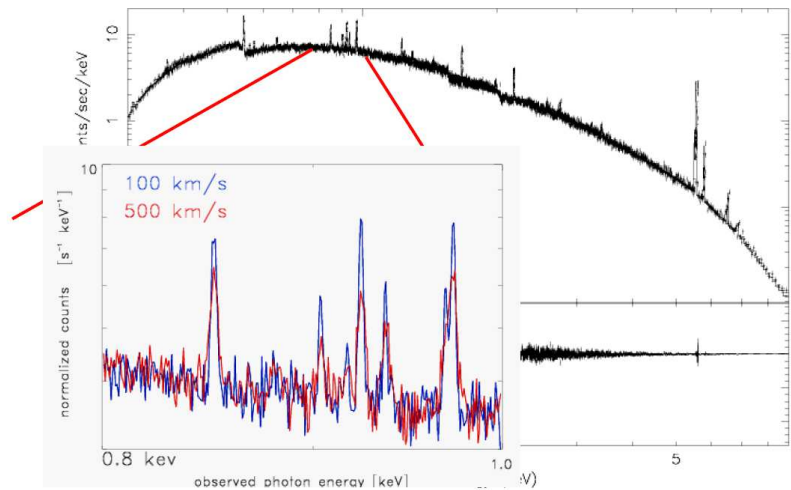}
\end{center}
\caption{Simulated \ixo\ NFI spectrum of cluster emission with a flux of
$3 \cdot 10^{-13}$ erg s$^{-1}$ cm$^{-2}$ in the NFI field-of-view at a redshift $z \sim 0.2$
with an exposure of 100 ksec. The inset shows the spectral structure of major Fe L-shell lines
around 1 keV. Spectral line broadening for plasma with mean turbulent velocities of 100 km/s
and 500 km/s are compared in this inset.} 
\label{fig:k1}       
\end{figure}

With the prospective energy resolution it will be possible to trace gas motions in the 
ICM with an accuracy of better than 100 km~s$^{-1}$ in the Fe lines in the X-ray spectra.
This is made possible due to the fact that the thermal line broadening for lines
from Fe ions is smaller by a factor of about 7 than the sound velocity which corresponds
closely to the thermal velocity of the bulk of the baryons in the ICM, the protons.
Fig.~\ref{fig:k1} shows the simulated spectrum as observed with the \ixo\ NFI of a patch
in a cluster with an aperture radius of 1 arcmin and a flux of $3\times 10^{-13}$ erg s$^{-1}$
cm$^{-2}$ (at 0.5--2 keV).
With an exposure of about 100 ks (40 ks) a Gaussian shape
turbulent velocity broadening can be measured with an accuracy of 40 (70) km~s$^{-1}$.
Even for a cluster at redshift 1 with a flux of $10^{-14}$ erg s$^{-1}$ cm$^{-2}$
velocity broadening accuracy would still be 100~km~s$^{-1}$ in an exposure of 100 ks.
This will provide many new insights into the ICM physics, answering questions such
as for example: how much turbulent energy is injected into the ICM by the
interaction of cool cores with the central AGN? How much of the cool core heating 
comes from the dissipation of this turbulent energy? How is the energy dissipated
in merging clusters and how is this reflected in the turbulent and bulk motion structure
of the ICM? How much can the mass estimate of clusters be improved if we
can measure the energy content of the ICM in turbulent energy?

\begin{figure}
\begin{center}
  \includegraphics[height=6.5cm]{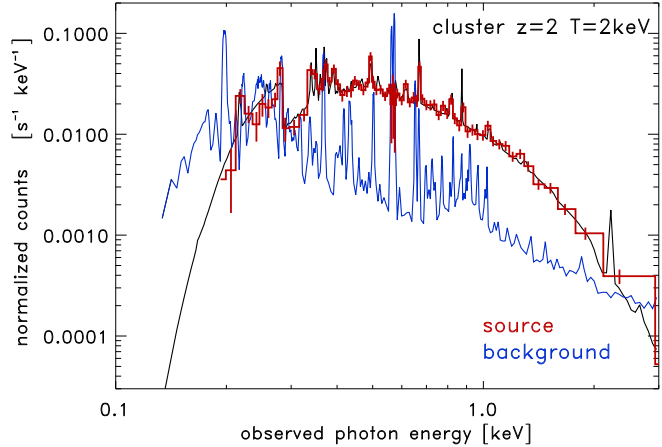}
\end{center}
\caption{Simulated \ixo\ NFI spectrum of a galaxy group with an ICM temperature of 2 keV
at redshift 2. Also shown is the estimated background spectrum. ICM properties like
the temperature and abundance measurements can well be determined from this simulated
exposure of 250 ksec. See text for details.}
\label{fig:k2}       
\end{figure}

Finally Fig.~\ref{fig:k2} illustrates the potential of \ixo\ to provide detailed
information on very distant clusters. Common objects at redshift 2
having a sky density of about 1 per deg$^2$ are poor clusters with masses of about
$3 \times 10^{13}$ M$_{\odot}$ and probably temperatures around 2 keV. Fig.~\ref{fig:k2}
shows the a spectrum of such an object observed with the \ixo\ NFI instrument for 250 ks 
\citep[see also][]{arnaud2009}. 
It will allow the measurement of the bulk temperature
with an accuracy better than 5\% and provide abundance determinations with interesting
precision: Fe ($\pm 11\%$), Si ($\pm 18\%$), O, Mg ($\pm 30\%$). This will allow us
to study the thermal and chemical enrichment history of galaxy clusters over most
of the interesting cosmic epochs in which clusters existed. 
For the more nearby bright clusters \ixo\ will very accurately measure the abundances of 
many chemical elements, including trace elements like Cr, Ti, Mn, or Co. The other elements, 
which we detect in clusters with \xmm\ and \chandra\ (O, Ne, Mg, Si, S, Ar, Ca, Fe, Ni) will be 
traced with an unprecedented accuracy, which will help us to put better constrains on 
supernova yields and thus on their explosion mechanisms. Furthermore, the wide field imager 
on \ixo\ will allow us to resolve the 2-dimensional metal distribution down to the relevant mixing 
physical scales and study in detail the process of metal injection.

The filamentary warm-hot intergalactic medium (WHIM) permeating the cosmic web might 
contain up to 50\% of baryons at redshifts of $z\lesssim2$. This intergalactic gas is 
heated to temperatures between $10^5$ and $10^7$ K as it gathers in regions of 
overdensity 10--100 (with respect to the mean baryon density of the Universe; at the 
current epoch the mean baryon density is $\left<n_b\right> = 2 \times10^{-7}$~cm$^{-3}$). 
Because this gas has such a low density and because it is so highly ionized, it is 
very hard to detect. The current instrumentation only allows us to probe the 
highest overdensity regions of the WHIM, such as the filament between the clusters 
Abell 222 and 223 \citep{werner2008b}, or to look for absorption features toward bright 
continuum sources in sight lines with known large scale structure, such as H2356-309 behind 
the Sculptor Wall \citep{buote2009}.  The combination of the large collecting area of \ixo\ 
and the high spectral resolution of its diffraction grating spectrometers will allow us to 
begin to probe the low-overdensity regions of the intergalactic medium. High-resolution X-ray 
absorption spectroscopy of sight lines towards bright continuum sources may allow us to 
probe overdensities of 10 or more (assuming a metallicity of 0.1 Solar), through the 
absorption features of H- and He-like ions of oxygen.   
Emission line imaging spectroscopy of the WHIM should be pursued with a large 
field-of-view (of order one degree), short focal length, large grasp (effective area times field-of-view) 
soft X-ray telescope, with a high resolution imaging X-ray spectrometer at the focus. 
A successful experiment will need a grasp of order several hundred cm$^2$~deg$^2$, and a 
spectral resolution of order 1--2 eV or better \citep{paerels2008}. 

Apart from these expectations, a new mission like \ixo\, capable of opening a new parameter space
for X-ray spectroscopy and simultaneously increasing the sensitivity substantially   
and thus the reach of the observations, has an enormous potential for new discoveries. 
The most important discoveries of such missions generally come as surprises. Since the
concept of \ixo\ is already realistic - apart from detailed technical problems still to be solved - 
and also not much more costly than previous missions, we can expect its realization.
Therefore we can foresee a bright future for X-ray spectroscopic research also over
the coming two decades, and clusters will surely be among the most interesting objects to study.

\begin{acknowledgements}
We like to thank A. Bykov, G. Pratt, Carles Badenes, and Aurora Simionescu for a critical reading of 
the manuscript and for discussions. NW acknowledges support provided by the National Aeronautics 
and Space Administration through Chandra Postdoctoral Fellowship Award Number PF8-90056 
issued by the Chandra X-ray Observatory Center, which is operated by the Smithsonian Astrophysical 
Observatory for and on behalf of the National Aeronautics and Space Administration under contract NAS8-03060.
HB acknowledges support for the research group through The Cluster of Excellence ``Origin and Structure of the 
Universe'', funded by the Excellence Initiative of the Federal Government of Germany, EXC project number 153
and support from the DfG Transregio Programme TR33.
\end{acknowledgements}



\bibliographystyle{aa}
\bibliography{clusters}

\end{document}